\documentclass[prx,aps,preprintnumbers, floats,floatfix, superscriptaddress,eqsecnum,twocolumn, longbibliography,10pt]{revtex4-2}

\usepackage{amsmath,amssymb}
\usepackage{natbib}
\usepackage{graphicx}
\usepackage{color}
\usepackage{array, enumerate}
\usepackage{bm}
\usepackage{multirow}
\usepackage[breaklinks,colorlinks,citecolor=blue]{hyperref}
\usepackage{braket}
\usepackage{txfonts}
\usepackage{booktabs}
\usepackage{xcolor}
\usepackage{microtype}
\usepackage{relsize}
\usepackage{braket}
\usepackage[normalem]{ulem}
\usepackage{orcidlink}

\renewcommand{\arraystretch}{1.2}

\definecolor{RoyalBlue}{cmyk}{1, 0.50, 0, 0}

\def\be{\begin{equation}}
\def\ee{\end{equation}}
\def\Z{Z}

\newcommand{\AEI}{\affiliation{Max Planck Institute for Gravitational Physics (Albert Einstein Institute) Am M\"{u}hlenberg 1, 14476 Potsdam, Germany}}
\newcommand{\esa}{\affiliation{European Space Agency (ESA), European Space Research and Technology Centre (ESTEC), Keplerlaan 1, 2201 AZ Noordwijk, the Netherlands}}
\newcommand{\PI}{\affiliation{Perimeter Institute for Theoretical Physics, Ontario, N2L 2Y5, Canada}}
\newcommand{\UofG}{\affiliation{University of Guelph, Guelph, Ontario N1G 2W1, Canada}}
\newcommand{\IST}{\affiliation{CENTRA, Departamento de F\'{\i}sica, Instituto Superior T\'ecnico -- IST, Universidade de Lisboa -- UL, Avenida Rovisco Pais 1, 1049-001 Lisboa, Portugal}}
\newcommand{\Tsinghua}{\affiliation{Department of Astronomy, Tsinghua University, Beijing 100084, China}}

\def\mnras{MNRAS}
\def\aap{A\&A}

\begin{document}
\title{
Impact of relativistic waveforms in LISA's science objectives with extreme-mass-ratio inspirals
}
\author{Hassan Khalvati$\,$\orcidlink{0000-0001-5313-9282}}\email{Hkhalvati@perimeterinstitute.ca}  \PI \UofG 
\author{Alessandro Santini$\,$\orcidlink{0000-0001-6936-8581}} \AEI
\author{Francisco Duque} \AEI
\author{Lorenzo Speri} \esa \AEI
\author{Jonathan Gair} \AEI
\author{Huan Yang} \Tsinghua \PI \UofG
\author{Richard Brito} \IST

\begin{abstract}

Extreme-Mass-Ratio Inspirals (EMRIs) are one of the key targets for future space-based gravitational wave detectors, such as LISA. The scientific potential of these sources can only be fully realized with fast and accurate waveform models. In this work, we extend the \textsc{FastEMRIWaveform} (\texttt{FEW}) framework by providing fully relativistic waveforms at adiabatic order for circular, equatorial orbits in Kerr spacetime, for mass ratios up to $10^{-3}$. We investigate the significance of including relativistic corrections in the waveform for both vacuum and non-vacuum environments. Specifically, we develop relativistic non-vacuum EMRI waveforms including two different environmental effects in the EMRI waveforms: power-law migration torques, and superradiance scalar clouds. For EMRIs in vacuum, we find that non-relativistic waveforms incorrectly estimate the predicted source's horizon redshift by approximately  $35\%$ error. Our analysis shows that incorporating relativistic corrections enhances constraints on accretion disks, modeled through power-law torques, and improves the constraints on disk parameter estimates (error $\simeq 8\%$), representing a significant improvement over previous estimates. Additionally, we assess the evidence for models in a scenario where ignoring the accretion disk biases the parameter estimation (PE), reporting a $\log_{10}$ Bayes factor of $1.1$ in favor of the accretion disk model. In a fully relativistic setup, we also estimate the parameters of superradiant scalar clouds with relative errors $\simeq 0.3\%$ for the scalar cloud's mass. These results demonstrate that incorporating relativistic effects is essential for LISA science objectives with EMRIs.
\end{abstract}

\maketitle
\section*{Introduction}
Extreme-Mass-Ratio Inspirals (EMRIs) are systems where a stellar-origin compact object (CO), with mass $\mu \sim  \mathcal{O}( 10) \, M_\odot$, orbits around a supermassive Black Hole (SMBH), typically with mass $M \sim \mathcal{O} (10^6) \, M_\odot$, with small mass ratios, between $10^{-6} - 10^{-4}$. In general, EMRIs follow highly complex, relativistic orbits that evolve over time, generating equally complex gravitational wave (GW) signals~\cite{Hughes_generic_PhysRevD.73.024027}, 
in the $10^{-4} - 10^{-1}$ Hz frequency range, making them ideal targets for space-based detectors such as the Laser Interferometer Space Antenna (LISA)~\cite{Gair_2017,redbook_colpi2024lisadefinitionstudyreport}. 
The expected number of EMRIs detectable by LISA is estimated to be between a few to ten thousand per year over its mission lifetime, depending on factors such as the population of SMBHs, the EMRI formation rate per SMBH, and the detector's sensitivity~\cite{Babak:2017tow}. It has also been suggested that the majority of observed EMRIs could be present in Active Galactic Nuclei (AGNs)~\cite{Pan_2021, Pan2_2021, Broggi:2022udp} with gaseous disks surrounding the system. 

EMRIs perform $\sim 10^4$ orbital cycles in the detector's band, during which the secondary (the smaller object) explores the strong gravitational field regions around the SMBH. For this reason, EMRI observations are expected to be an excellent opportunity to probe the strong-field regime of General Relativity (GR)~\cite{Barack_2007, Tahura_2024}, study the surrounding astrophysical environment~\cite{Barausse_2015, Kocsis_2011, levin2003formationmassivestarsblack}, and explore physics beyond GR~\cite{Gair_2013, Tahura_2022, Speri_fundamental_2024arXiv240607607S}.

Accurate modeling of EMRIs is necessary to avoid systematic errors and achieve their full scientific potential~\cite{khalvati2024flux, Burke:2023lno, Gupta:2024gun}.
These errors can affect the conclusions obtained from EMRI observations about astrophysical processes and the nature of gravity in strong fields. Therefore, it is pivotal to develop robust waveform models that can accurately represent the full scope of EMRI dynamics, while also being generated rapidly for efficient analysis of LISA data~\cite{Chua:2021aah,Speri_2024,Nasipak_2024,lisaconsortiumwaveformworkinggroup2023waveformmodellinglaserinterferometer,lynch2024fastinspiralstreatmentorbital, drummond2023extrememassratioinspiralwaveforms, rink2024gravitationalwavesurrogatemodel}.

In addition to accuracy, the rapid generation of these waveforms is equally critical for successful EMRI searches and parameter estimation (PE). The \textsc{FastEMRIWaveforms} (\texttt{FEW}) framework~\cite{Katz_2021, few_Chua_2021} was developed as a solution for accurate and rapid generation of fully relativistic waveforms on both central processing units (CPUs) and graphics processing units (GPUs). GPUs are particularly well-suited for fast parallel summation of the many harmonics expected in EMRI waveforms. More recently, a frequency-domain EMRI waveform was developed~\cite{Speri_2024}, further enhancing the speed of waveform generation. However, relativistic waveforms in \texttt{FEW} are currently limited to eccentric orbits in Schwarzschild spacetime.

In this work, we present a fully relativistic EMRI waveform for circular, equatorial orbits around a Kerr black hole. Our model incorporates relativistic amplitudes and inspiral based on BH perturbation theory~\cite{Pound:2021qin}, and spans a wider range of mass ratios up to $\eta \gtrsim 10^{-3}$ in comparison to what is currently available in \texttt{FEW} ($\eta \leq 10^{-4}$). Furthermore, we highlight the importance of using fully relativistic waveforms for studying EMRIs in non-vacuum environments, focusing on two distinct scenarios: (1)EMRIs in accretion disks modeled via power law migration torques and (2)EMRIs surrounded by a superradiant scalar clouds in a relativistic framework ~\cite{Speri, Brito_2023}. All models developed in this work are publicly available (\url{https://github.com/Hassankh92/FastEMRIWaveforms_KerrCircNonvac})~\cite{KerrCircNonvac, KerrCircNonvac_Data}.

By comparing the fully relativistic waveform with approximate phenomenological kludge EMRI waveforms, we present the improvements in the waveform and the scientific potential for future EMRI detections. Our findings show that not using fully relativistic waveforms and higher (relativistic) harmonics in the template leads to at least a 20\% error in the SNR and more than 30\% error in the horizon redshift. In addition, we find that the fully relativistic waveform model enhances the ability to detect and characterize environmental effects. We performed a Markov chain Monte Carlo (MCMC) analysis over the full parameter space relevant to our model in both vacuum and nonvacuum to estimate the model parameters and assess the impact of fully relativistic waveforms on the estimation results.

For EMRIs in accretion disk, we also analyze how neglecting this effect would bias the recovery of the model parameters. For the same scenario, our model selection analysis on disk-influenced waveforms, indicates a slight preference for these waveforms over vacuum waveforms. For the second non-vacuum scenario, we study the accumulated orbital phase difference (\textit{phase shift}) of the EMRIs as a simple means to compare inspiral trajectories. Through this approach, we identified the criteria for two parameters describing the superradiant scalar cloud model, namely the cloud's mass and the fundamental field mass to be detectable by LISA. Following this, through a full parameter MCMC analysis, and we could estimate the cloud's mass and the boson's mass
with a remarkably small error.

The paper is organized in two parts as follows: in the first part we cover the methods employed in our study, followed by the second part, where we present the results and analysis.

We begin by briefly summarizing older phenomenological EMRI waveform models in Sec.~\ref{sec:Kludge_models}. In Sec.~\ref{sec:adiabatic_waveforms}, we introduce the basics of adiabatic EMRI waveforms and their connection to Teukolsky formalism. Section~\ref{sec:FEW} outlines the structure of the \texttt{FEW} package, providing the foundation for our study. Finally, we summarize the data analysis setup we use for the next part of the paper in Sec.~\ref{sec:data_analysis}.

The second part begins with Sec.~\ref{sec:traj}, where we focus on the inspiral in circular, equatorial orbits around Kerr BHs, followed by amplitude computations in Sec.~\ref{sec:Rel_amps}. We then introduce the relevant modules of our model in \texttt{FEW} and test it against existing relativistic EMRI waveforms. In Sec.~\ref{sec:rel_corrections}, we present the improvements in the fully relativistic waveforms compared to the legacy kludge models. Finally, in Sec.~\ref{sec:proof of cons}, we incorporate beyond-vacuum effects into our waveform models as a proof of concept, demonstrating the impact of relativistic corrections through PE, focusing on migration torques in accretion disks (Sec.~\ref{sec:disk}),  and superradiant scalar clouds (Sec.~\ref{sec:cloud}).

\section{Methods}
In this section, we begin by reviewing past work on EMRI waveforms and lay out the foundational framework for our study. We also discuss current developments and future directions in this research program.

\subsection{Kludge EMRI waveforms} \label{sec:Kludge_models}

The development of accurate GWs is crucial to do precise science with EMRIs. In general, EMRI waveform models can be described in terms of two key components: phase and amplitude. The phase evolution is determined by the inspiral trajectory, which characterizes the orbital evolution over time, while the amplitude accounts for the harmonics present in the waveform. Various phenomenological models have been constructed,  with the most generic ones being the Analytic Kludge (AK)~\cite{AK_Barack_2004}, Numerical Kludge (NK)~\cite{Kludge}, and Augmented Analytic Kludge (AAK)~\cite{AAK_Chua_2017}. However, these models lack fully relativistic treatments of the inspiral trajectory and waveform amplitude. Below we provide a brief summary of these different ``Kludge" models. 

\begin{itemize}
    \item \textbf{Analytic Kludge (AK)}~\cite{AK_Barack_2004} approximates the trajectory with Keplerian ellipses, incorporating corrections for precession and radiation reaction. The waveform is computed using the Peters–Mathews mode-sum approximation~\cite{Peters_Math_PhysRev.131.435}, which decomposes the mass quadrupole moment into harmonics of the orbital frequency. 

    \item \textbf{Numerical Kludge (NK)}~\cite{Kludge} builds from the AK by employing Kerr geodesics for the orbit with an improved inspiral that incorporates higher-order Post-Newtonian (PN), terms for energy and angular momentum fluxes~\cite{Gair_NK_PhysRevD.73.064037}, along with fitting formulas derived from Teukolsky-generated data.
    The compact object's worldline is associated with flat space coordinates, and the waveform is generated using the quadrupole formula. NK waveforms achieve an overlap of $\sim0.95$ with the more accurate Teukolsky-based waveforms~\cite{Drasco_2006} for pericenter distance of $r_p \sim 5M$, and their accuracy rapidly deteriorates for smaller radius. 
    
    \item \textbf{Augmented Analytic Kludge (AAK)}~\cite{AAK_Chua_2017} model refines the AK by aligning the waveform with Kerr geodesic frequencies, addressing significant dephasing issues. This hybrid model retains AK's computational efficiency while improving accuracy and maintaining phase coherence with NK waveforms for longer times. 
    The accuracy diminishes near plunge due to the divergence of the energy flux beyond the 3PN order in strong-field regions~\cite{Yunes_berti_3PN_PhysRevD.77.124006}.

\end{itemize}

In Sec.~\ref{sec:rel_corrections}, we explore the improvements brought by our fully relativistic EMRI waveforms for circular, equatorial orbits in Kerr spacetime when compared to these approximate, phenomenological kludge models.

\subsection{Adiabatic Waveforms}\label{sec:adiabatic_waveforms}
Adiabatic waveforms~\cite{Hughes_adiabatic_PhysRevLett.94.221101} are an approximate relativistic approach to modeling GWs from EMRIs. They exploit the so-called two-timescale expansion~\cite{Hinderer_2008, Miller:2020bft}, which separates the dynamics into a fast orbital timescale, $T_{\text{orb}} \sim M$, and a slow radiation-reaction timescale, $T_{\text{rad}} \sim M/\eta$, with $\eta = \mu/M \ll 1$ the EMRI's small mass ratio. This separation highly simplifies the problem by treating the system as quasi-stationary on the fast timescale, where the small body effectively follows geodesic motion. The backreaction due to GW emission is averaged over the orbit, and at the adiabatic order, the inspiral is taken to be a ``flow over a succession of geodesics''~\cite{Hughes_2021}. This is equivalent to solving Einstein's equations for the two-body problem perturbatively to the lowest order in the mass ratio $\eta$ and only considering averaged dissipative effects~\cite{Barack:2018yvs, Pound:2021qin}. 

Relativistic waveform computation in the extreme-mass-ratio limit involves solving the Teukolsky equation~\cite{ cutler_eric_PhysRevD.50.3816, Hughes2000_PhysRevD.61.084004, Hughes_generic_PhysRevD.73.024027}.
This is a master equation that governs different types of perturbations, including gravitational ones, around Kerr ~\cite{Teukolsky},
\be \label{eq:teuk_master_short}
\mathcal{D}^2 \Psi = \mathcal{T} \ .
\ee
Here, $\mathcal{D}^2$ is a second-order linear differential operator, $\mathcal{T}$ represents the source term, which is computed of the stress-energy tensor describing a point-particle moving in geodesics of the Kerr spacetime and $\Psi$ is a generic perturbed field quantity. 
The radiative degrees of freedom of the gravitational field are controlled by the complex Weyl scalar $\psi_4$, and can be separated into a radial function, $R_{lm\omega}(r)$, and an angular function, $_{-2} S^{\chi}_{lm}(\theta, \, \phi)$,  in the Fourier domain~\cite{Teuk0_1973ApJ...185..635T,Teukolsky}
\begin{equation}
\Psi = \zeta^{4}  \psi_4 =  \sum_{l,\,m} \ \int_{-\infty}^{+\infty} d\omega\, R_{lm\omega}(r) \  _{-2} S^{\chi}_{lm}(\theta) \ e^{i m \phi} \ e^{ - i \omega t} ,
\end{equation}
with $\zeta=r - i a \cos \theta$, where $a$ is the BH's spin parameter, $\chi = a \omega$, and $_{-2} S^{\chi}_{lm}$ are the spin-weighted spheroidal harmonics, with weight  $s= -2$,  which we henceforth drop to avoid cluttering. The parameter $\omega$ is the gravitational radiation frequency, which will be associated with the harmonics of the source motion. In addition, $(l,m)$ are the multipolar indices.  The GW polarizations can then be obtained using 
\begin{equation}
    \psi_4 = \frac{1}{2} \left( \ddot{h}_{+} - i \ddot{h}_{\times}\right) \quad , \quad r \to \infty.
\end{equation}

One can construct two independent solutions for the radial function in Teukolsky's master equation, one of which describes outgoing waves at \textit{infinity}, $R^{\infty}_{lm\omega}(r)$,
and another describing ingoing waves at the
\textit{horizon} of the BH, $R^{H}_{lm\omega}(r)$. The general radial function can then be expressed as
\be
R_{lm\omega}(r) =  Z_{lm\omega}^{H} \ R^{H}_{lm\omega}(r) + Z_{lm\omega}^{\infty} \ R^{\infty}_{lm\omega}(r) \, ,
\ee
with the Teukolsky amplitudes $Z_{lm\omega}^{\infty,H}$ computed by convoluting the homogeneous solutions with the respective source term (see Eq.~(3.20) in Ref.~\cite{Hughes_generic_PhysRevD.73.024027}).  
A more detailed description of the formalism can be 
found in~\cite{Hughes2000_PhysRevD.61.084004, Hughes_generic_PhysRevD.73.024027,Hughes_2021}. Note that we adopt the modified notation for the Teukolsky amplitudes, as discussed in Appendix D of Ref.~\cite{Hughes_Sullivan_PhysRevD.90.124039}

In general, bound orbits are described by three fundamental frequencies: \(\Omega_r\) (radial), \(\Omega_\theta\) (polar), and \(\Omega_\phi\) (azimuthal)~\cite{Schmidt_2002}, governing the periodic motion in the respective directions. To solve the Teukolsky equations in the frequency domain, the source term is expanded in harmonics of these orbital frequencies
\begin{equation} \label{eq:wave_freq}
\omega_{mkn} = m\Omega_\varphi + k\Omega_\theta + n\Omega_r   \quad , \quad m,\, k, \, n \in \mathbb{Z} \, ,
\end{equation}
%
which allows us to address each harmonic component separately. 
The waveform can then be computed through~\cite{Hughes_generic_PhysRevD.73.024027} 
\begin{equation}\label{eq:main_wave_generic}
    h \equiv h_+ - i h_{\times} = \frac{\mu}{d_L}\sum_{lmkn} A_{lmkn}(t) \ S^{\chi}_{lmkn}(\theta) \ e^{im\phi}e^{-i\omega_{mkn} t},
\end{equation}
where $d_L$ is the luminosity distance to the source, and  \( A_{lmkn} \) are the waveforms's characteristic amplitudes
\begin{equation} \label{eq:amplitudes}
    A_{lmkn} = -\frac{2 Z^{\infty}_{lmkn}}{\omega_{mkn}^2} \, . 
\end{equation}
\\

GWs are radiated to infinity and absorbed at the central BH horizon, causing changes in the orbital constants of motion: energy, angular momentum, and the Carter constant. The orbital averaged fluxes for the first two  can also be computed in terms of Teukolsky amplitudes~\cite{Isaacson_PhysRev.166.1272,Hughes2000_PhysRevD.61.084004}
\begin{align} \label{eq:flux_generic}
    &\left< \dot{E} \right>^{\infty, H}_\text{GW} = \sum_{lmkn} \frac{1}{4\pi\omega_{mkn}^2} \alpha_{lmkn}^{\infty, H}\left| Z_{lmkn}^{\infty, H}\right|^2 \, ,& \\
&\left< \dot{L_z} \right>^{\infty, H}_\text{GW} = \sum_{lmkn}\frac{m}{4\pi\omega_{mkn}^3}\alpha_{lmkn}^{\infty, H} \left| Z_{lmkn}^{\infty, H}\right|^2 \, ,& 
\end{align}
where $\alpha_{lmkn}^{\infty} = 1$ and 
$\alpha_{lmkn}^{H}$ is an horizon factor which can be found explicitly in Sec.~III.D of Ref.~\cite{Drasco_2006}, with fluxes given per $\eta^2$. We do not need to consider the rate of change in the Carter constant since our focus will be on equatorial orbits with no polar motion. At the adiabatic level and in vacuum, the inspiral is obtained by taking the balance-law
\be \label{eq:balance_law}
\mathcal{F}_{\text{orb}} = -\mathcal{F}_{\text{GW}}\, ,
\ee
with $\mathcal{F} \in \{\dot E, \dot L\}$,  
to evolve the orbit from one geodesic to the next on the radiation reaction timescale. As the orbit alters, the Teukolsky amplitudes and the fundamental frequencies change, which leads to a waveform with time-evolving amplitude and frequency. At the adiabatic order, the central BH mass and spin remain constant during the inspiral \cite{Wardell_2023}.

In the rest of this paper, unless otherwise stated, we work in units where the primary mass $M=1$ for simplicity. Additionally, to avoid confusion in the remainder of this paper, we clarify the distinction between the use of $l_{\rm max}$ in amplitude mode summation--Eq.~\eqref{eq:main_wave_generic}, and in flux computations--Eq.~\eqref{eq:flux_generic}, using $l^{\mathcal{A}}_{\rm max}$, and $l^{\mathcal{F}}_{\rm max}$, respectively. Having laid out the basics of the Teukolsky formalism, in the next section, we address how to compute the waveforms in a numerically efficient way with \texttt{FEW}, which is essential for rapid detection and inference of EMRIs with LISA~\cite{redbook_colpi2024lisadefinitionstudyreport, lisaconsortiumwaveformworkinggroup2023waveformmodellinglaserinterferometer}.

\subsection{Fast EMRI Waveforms}\label{sec:FEW}
The \texttt{FEW} package~\cite{michael_l_katz_2020_4005001,few_Chua_2021, Katz_2021} efficiently generates accurate relativistic waveforms for EMRIs. The framework relies on precomputing the Teukolsky flux and amplitude data over a set of grid points and then performs on-the-fly interpolations to ensure detailed and rapid simulation of the EMRI trajectory and subsequent GW signal. The current open-source package, available in the Black Hole Perturbation Toolkit~\cite{BHPToolkit}, provides adiabatic waveforms accurate to $\mathcal{O}(\eta)$ for eccentric orbits in a Schwarzschild background. A 4-year inspiral waveform takes $\sim 10-100 \, \text{ms} $ to compute on a GPU.

As previously discussed, each orbit is characterized by its constants of motion. In \texttt{FEW}, the orbits are parameterized rather by the orbital parameters:  eccentricity \(e\), semi-latus rectum \(p\), and orbit inclination angle \(``I"\) ($x_I = \cos I$). These parameters also fully describe a geodesic and can be transformed into the constants of motion~\cite{cutler_eric_PhysRevD.50.3816,hughes2024parameterizingblackholeorbits}. 
The inspiral trajectory and the phase evolution are determined by solving a system of coupled ordinary differential equations (ODE) for the orbital parameters and phase, utilizing interpolated flux data, $\mathcal{F}_{int} (a, p, e, x)$, across the parameter space. The wave amplitude modes, $A_{lmkn}$, are then interpolated over the inspiral trajectory using the precomputed amplitude data. Amplitude harmonic modes,  $\left\{l,m,k,n\right\}$, can be arbitrarily selected and summed interactively during the waveform computation process in the package.

 The modular design of \texttt{FEW} allows for extensions to more complicated scenarios, including post-adiabatic effects and more generic orbits \cite{Burke:2023lno}. Additionally, the original \texttt{FEW} paper~\cite{Katz_2021} introduced a version of AAK incorporating the PN flux formula using 5PN corrections. The AAK waveform in \texttt{FEW}  extends to more generic scenarios, however as it was mentioned in~\ref{sec:Kludge_models}, the AAK model is less accurate due to its approximate nature. 
 \texttt{FEW} is also capable of transforming the computed waveforms to the detector’s frame, located in the solar system barycenter, with the full parameter space, $\{M, \mu, a, p_0, e_0, x_{I,0}, d_L, \theta_S, \phi_S, \theta_K, \phi_K, \Phi_{\varphi,0}, \Phi_{\theta,0}, \Phi_{r,0} \}$, for a generic orbit scenario, in the detector's frame, referenced to the Solar System barycenter. The indices $S$ refer to the sky position angles, $K$ to the spin angular momentum direction, and subscript $``0"$ indicates the initial value of the quantities.

 In this work, we introduce fully relativistic waveforms for circular, equatorial orbits in a Kerr background, in the sense that we incorporate both Teukolsky-based flux data and Teukolsky-based amplitude data for waveform computation. The details are presented in Sec.~\ref{sec:traj} and \ref{sec:Rel_amps}.

\subsection{Relativistic trajectories in Kerr: Circular, Equatorial} \label{sec:traj}

\begin{figure}[t]
    \centering
    \includegraphics[width = \linewidth]{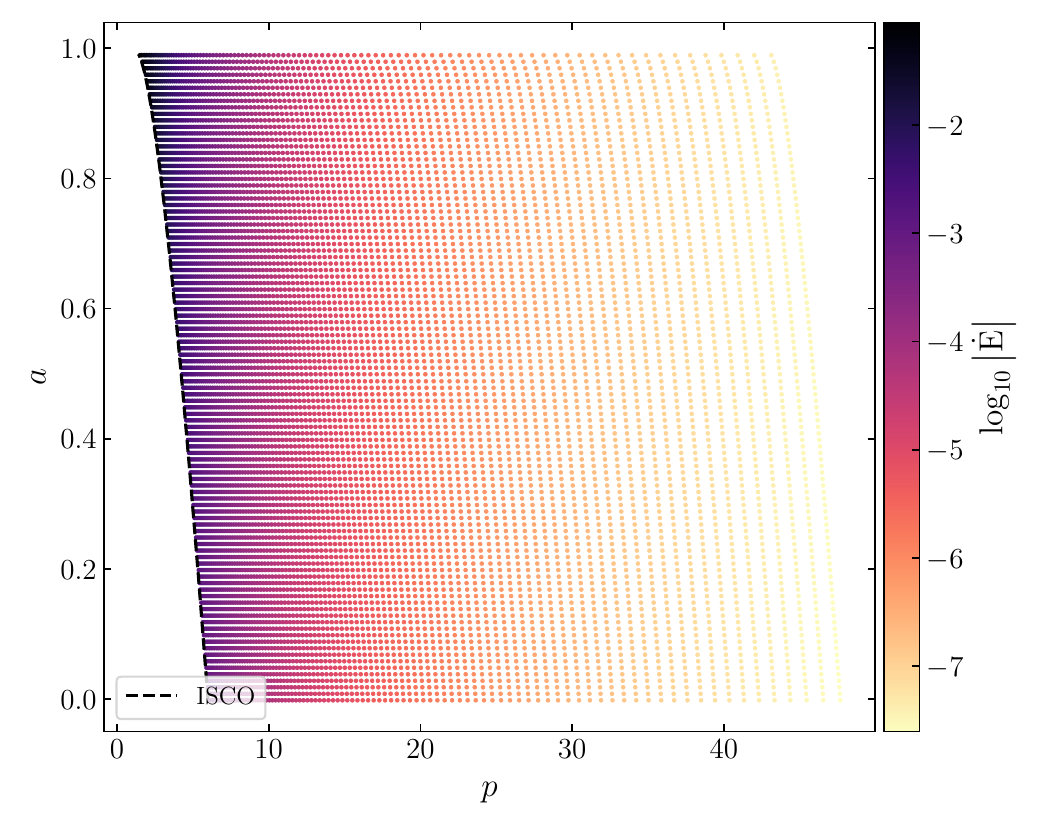}
    \caption{The flux data grid points are shown with the color bar indicating the $\log_{10}$ scale absolute value of the total energy flux in $a-p$ space. We consider uniform grid spacing for the spin parameter with $\delta a = 0.01$ and logarithmic spacing for $p$. The $p$ is scaled by $u = \ln \ (p - p_{\text{ISCO}} + 3.93)$ which is the logarithmic scaled distance to the ISCO  with the grid spacing of $\delta u = 0.025$.
    }
    \label{fig:Edot_grid}
\end{figure}

Circular, equatorial orbits have fewer degrees of freedom than generic orbits. Since the orbit is equatorial, there is no polar motion, and with the eccentricity \(e = 0\), no radial motion exists on the orbital time scale. 
This implies there is no harmonic content in these directions, i.e. the indices \(n\) and \(k\) are omitted in Eqs.~\eqref{eq:wave_freq}--\eqref{eq:flux_generic}. The parameter space reduces to the spin parameter of the central BH \(a\) and the orbital separation $p$ (since we are considering circular orbits, the separation and the semi-lactus rectum represent the same quantity). 
Also, the angular momentum and energy fluxes are related by $\dot L = \dot E \ \Omega_{\varphi}^{-1} $.

To compute the flux, we truncate the summation in Eq.~\eqref{eq:flux_generic} at $l^{\mathcal{F}}_{\rm max}=30$ (with $-l \leq m \leq l$), guaranteeing the truncation error in the flux data is less than $\sim 10^{-7}$ for spins $a\leq0.99$~\cite{khalvati2024flux}. To minimize the error in the flux interpolation, $\dot E_{int}$, similarly to the original procedure in \texttt{FEW}, we subtract the leading PN order flux, $\dot E_{PN}^{(0)}$ from our data, $\dot E_{grid}$ and scale it following
 \begin{align}
     \dot E_\text{int} &= (\dot E_\text{grid} - \dot E^{(0)}_\text{PN} ) \ \Omega_{\varphi}^{-4} \, , \\
     \dot E^{(0)}_\text{PN} &= \frac{32}{5} \ \Omega_{\varphi}^{10/3} \, ,
  \end{align}
 with $\Omega_{\varphi} = (r^{3/2} + a)^{-1}$ for circular orbits in Kerr spacetime. A thorough study of the interpolation-induced error is addressed in a separate work~\cite{khalvati2024flux}. 

The fluxes are then computed on a two-dimensional grid $(a,p)$ with a total of 9900 points. To enhance interpolation accuracy, we require more points in the strong-field region closer to the central BH. To do so, we employ a logarithmic scaling of the distance to the \textit{innermost stable circular orbit} (ISCO), defined as \( u = \ln{(p - p_{\text{ISCO}} + 3.93)} \). This method ensures that our grid is rectangular for different spin values. We use a uniform grid spacing for the spin parameter in the range \(0.0 \leq a \leq 0.99\) with \(\Delta a = 0.01\), and for orbital separation in the range \( p_{\text{ISCO}} + 0.03 \leq p \leq p_{\text{ISCO}} + 42 \) \((1.368 \lesssim u \lesssim 3.818)\) with \(\Delta u = 0.025\).
We show the energy flux values per $\eta^2$ over the grid points in the $(a,p)$ space in Fig.~\ref{fig:Edot_grid}. We take $p_{\text{ISCO}}+0.03$ as the plunge limit. 

With the flux data interpolated over the grid, we can compute the adiabatic inspiral trajectory and phase evolution by solving the following ODEs (equivalent to \eqref{eq:balance_law})
\begin{align} \label{eq:trajectory}
    \frac{dp}{dt} &= \frac{\dot{E}_\text{orb}(a,p)}{E_\text{orb}'(a,p)} \, , \\
    \frac{d \Phi_{\varphi} }{dt} &= \Omega_{\varphi}(a,p) \, ,
\end{align}
where $\Phi_{\varphi}$ is the azimuthal time-varying phase, and the prime denotes the derivative with respect to $p$. 

To obtain the full waveform, we also need to precompute the amplitude data, $A_{lm}$, on the same grid as the flux data above. While the wave strain is simplified for equatorial circular orbits, computing the amplitudes for a Kerr background requires additional care, which we discuss in the following section.

\subsection{Relativistic amplitudes in Kerr} \label{sec:Rel_amps}

As presented in Eq.~\eqref{eq:main_wave_generic}, the amplitudes of  GWs in the adiabatic approximation for EMRIs ($A_{lm}$) are expanded in spin-weighted spheroidal harmonics $S^{\chi}_{lmkn}(\theta)$, with $\chi = a \omega$. This originated from the way perturbations are decomposed in Teukolsky formalism. In a Schwarzschild background ($a=0$), the spin-weighted spheroidal harmonics reduce to the spin-weighted spherical harmonics, $S^{\chi=0}_{lm} =  Y_{lm}$, with no frequency dependence. However, in a Kerr background with a nonzero spin parameter, these bases will vary throughout the inspiral as the orbital frequencies evolve, complicating waveform computations.
The work in Ref.~\cite{Yunes_2011} was the first to address this issue by introducing a new set of amplitudes, $C_{lm}^\chi$, that avoids the evolving basis complication.  This was achieved by expressing the spheroidal harmonics as a series expansion of spherical harmonics
~\cite{Hughes2000_PhysRevD.61.084004}:
\begin{equation}
\label{eq:spheroidal_expan} 
S^{\chi}_{lm}(\theta) = \sum_{j  \, = \,j_{\rm min}}^{\infty} b_{ljm}^{ \ \chi} Y_{jm} (\theta),
\end{equation}
which then allows the amplitudes $C_{lm}^\chi$ to be written in terms of the spheroidal amplitudes, $A_{lm}$:
\begin{equation}\label{eq:new_amps}
C_{lm}^\chi
= \sum_{j = j_{\rm min}}^{\infty}  A_{jm} \  b_{jlm}^{ \ \chi}\,, 
\end{equation}
\begin{figure}[t]
    \centering
    \includegraphics[width =\linewidth]{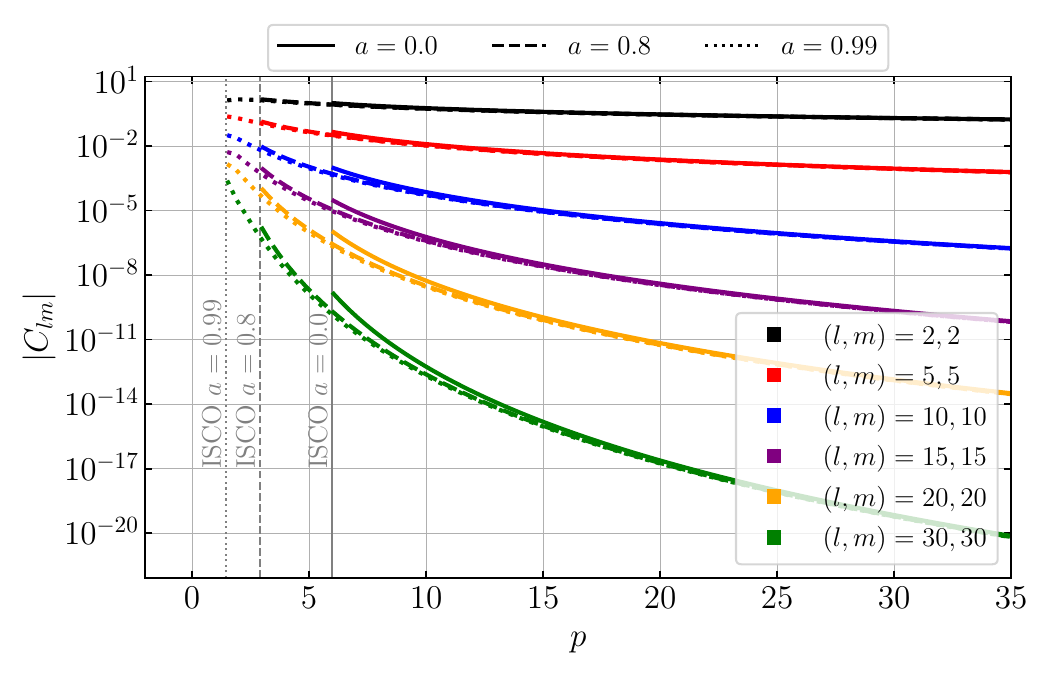}
    \caption{Absolute value of the relativistic wave amplitudes in \textit{spherical harmonics} basis, $C_{lm}^\chi$, as a function of the orbital separation of the secondary, for different spin values and multipoles. We precomputed the amplitude data for up to $l^{\mathcal{A}}_{\rm max}=30$ on the same grid to the fluxes in Fig.~\ref{fig:Edot_grid}.
    }
    \label{fig:amplitudes_Kerr}
\end{figure}
with $j_{\rm min} = \max \{2,|m|\}$ for GWs (tensor perturbations), and we have factored out the azimuthal angle part of the harmonics, $e^{i m \phi}$, from both sides.   The mixing expansion coefficients, $b^{j}_{lm}(\ \chi)$, satisfy an eigenvalue equation derived from the $\theta$ part of the Teukolsky master equation. In our computations, we truncate the series at $j_{max}$, where the contributions to the eigenvalues from higher-order modes fall below an error tolerance of $10^{-8}$ (see Appendix A of~\cite{Hirata2:2010vp}). 
Detailed methods for computing them can be found in Section III of~\cite{press_teuk_1973ApJ...185..649P}, or in Appendix A of~\cite{Hughes2000_PhysRevD.61.084004}. With the new amplitudes, we can rewrite the GW strain Eq.~\eqref{eq:main_wave_generic}, for a circular, equatorial orbit in Kerr background as
\begin{equation}\label{eq:wave_new_amps}
    h = \frac{\mu}{d_L} \sum_{lm} C_{lm}^\chi (a, p) Y_{lm}(\theta) e^{im\phi}e^{-i\omega_{m} t} \, .
\end{equation}
In the course of computing the Teukolsky fluxes, we also compute the new amplitudes $C^\chi_{lm}$ for each $l$ and $m$ on the same $(a,p)$ grid and save them separately for individual modes up to $l^{\mathcal{A}}_{\rm max}=30$. These saved amplitudes are then interpolated over the $(a,p)$ grid, similar to the flux data as part of the overall waveform computation in \texttt{FEW}.
The absolute value of the interpolated $C^\chi_{lm}$ for a set of dominant $l=m$ modes 
 is plotted in Fig.~\ref{fig:amplitudes_Kerr} as a function of the orbital radius for three different spin values, $a\in\left\{0.0, 0.8, 0.99\right\}$ over $p_{\text{ISCO}} + 0.03 \leq p \leq 35$. 

To avoid errors accumulating during the inspiral, the flux computation (Eq.~\ref{eq:flux_generic}) requires summing over a large number of modes. Here we pre-computed the flux data including all modes up to $l^{\mathcal{F}}_{\rm max} = 30$ to ensure accuracy.

In contrast, for the waveform amplitudes (Eq.~\ref{eq:wave_new_amps}), the inclusion of modes is guided by their power or contribution to the total SNR. Figure\ref{fig:amplitudes_Kerr} illustrates how the modes behave in circular, equatorial orbits, highlighting their relative impact on the overall waveform. The selection process for amplitude modes requires careful consideration to balance accuracy and computational cost, ensuring significant information is retained without including unnecessary modes. Using \texttt{FEW}, we can interactively select different sets of modes during waveform computation, providing flexibility in determining the optimal set of modes for a given analysis.

As an example, we consider an EMRI with $\mu = 50 M_{\odot}$ in an orbit around a Kerr BH with spin \(a = 0.99\) and mass $M=10^6 M_{\odot}$, starting at radius $p_0 \simeq 15.35$, set so the EMRI takes 4 years to plunge. In Table~\ref{tab:snr_contributions}, we show the relative SNR contribution from each mode to the dominant mode $(2,2)$ for the modes presented in Fig.~\ref{fig:amplitudes_Kerr}. The SNR contribution from the dominant $(2,2)$ mode is approximately 177. Note that the waveform includes modes projected onto spin-weighted spherical harmonics $Y_{lm}$, Eq.~\eqref{eq:wave_new_amps}, which depend on the polar view angle $\theta$. Thus, the SNR contributions of the modes vary with $\theta$. The values reported here assume $\theta = \pi/4$. 

\begin{table}[ht]
\centering
\renewcommand{\arraystretch}{1.25}
\begin{tabular}{c c}
\toprule
$(l, \, m)$ \hspace{2.5pt} & $\text{SNR}_{(l, \, m)}/\text{SNR}_{(2, \,2)}$ \\ 
\midrule
(5, 5) \hspace{2.5pt}  & $3.8 \times 10^{-2}$ \\ 
(10, 10) \hspace{2.5pt} & $5.4 \times 10^{-4}$ \\ 
(15, 15) \hspace{2.5pt} & $1.4 \times 10^{-5}$ \\ 
(20, 20) \hspace{2.5pt} & $6.5 \times 10^{-7}$ \\ 
(30, 30) \hspace{2.5pt} & $1.5 \times 10^{-9}$ \\ 
\bottomrule
\end{tabular}
\caption{Relative SNR Contributions of different modes to the dominant mode $\left(2,2\right)$, with $\text{SNR}_{(2, \,2)} \simeq 177$, for an EMRI with $M = 10^6 M_{\odot}$, $\mu = 50 M_{\odot}$, and $a = 0.99$, over 4 years inspiral from $p_0 = 15.35$ down to the plunge, with view angles of $\theta, \phi = (\pi/4,0.0)$.
}
\label{tab:snr_contributions}
\end{table}

Based on this analysis, we can truncate the mode summation for the amplitude at lower $l^{\mathcal{A}}_{\rm max}$ values while maintaining $l^{\mathcal{F}}_{\rm max} = 30$ for the precomputed flux data. Using a smaller set of modes can significantly speed up waveform computation, streamlining the mode summation process. In the current version of \texttt{FEW} developed for this work, each additional harmonic mode contributes approximately $~\mathcal{O}(1)$ ms to the total waveform computational time on a GPU for a $4$-year inspiral duration. Further discussion regarding the timing of the waveform computations can be found in Sec.~\ref{sec:test}. 

\subsection{Data Analysis setup} \label{sec:data_analysis}

\texttt{FEW} waveforms are fast enough to allow for full Bayesian inference with standard MCMC methods (e.g.~\cite{Speri, Speri_fundamental_2024arXiv240607607S}). In this work, we run the parallel tempered ensemble sampler \texttt{Eryn}~\cite{2023MNRAS.526.4814K} using 3 different temperatures with 16 walkers each. For each of our PE runs, we monitor the convergence of the samples by making sure that the chains are longer than $50\, \hat{\tau}$, with $\hat{\tau}$ being the average autocorrelation time~\cite{2010CAMCS...5...65G}. 
This MCMC setup allows us to get samples from the posterior distribution $p(\boldsymbol{\theta} \, | \, d)$, which in the standard Bayesian fashion is expressed as
\be \label{eq:posterior}
p(\boldsymbol{\theta} | \, d) = \frac{\mathcal{L}(d \, | \, \boldsymbol{\theta})\, \pi(\boldsymbol{\theta})}{\Z}%
\, , %
\quad 
\Z = \int \text{d} \boldsymbol{\theta} \, \mathcal{L}(d \, | \, \boldsymbol{\theta})\, \pi(\boldsymbol{\theta}) \, ,
\ee
where $\boldsymbol{\theta}$ are the parameters of our model, $d$ the observed data, $\mathcal{L}$ the likelihood and $\pi(\boldsymbol{\theta})$ our prior distributions. $\Z$ is the evidence, which is the integral over the parameter space of the unnormalized posterior $\hat{p}(\boldsymbol{\theta} \, | \, d) = \mathcal{L}(d \, | \, \boldsymbol{\theta})\, \pi(\boldsymbol{\theta})$.  
This quantity is not necessary when performing PE and can safely be disregarded, but it is the key element in Bayesian model selection (see Refs.~\cite{2011PhRvD..83h2002D, 2021PhRvD.104h3027T, 2024PhRvD.109j4019T} for applications in the field of GW astronomy). In fact, if we have two competing models $M_1$ and $M_2$, we can use the Bayes factor $\mathcal{B}_{21} = \Z_2 / \Z_1$ to find the one that better describes the data. While \texttt{Eryn} provides ways to use the temperature ladder to compute $\Z$~\cite{10.1080/10635150500433722, 10.1093/sysbio/syq085}, the number of temperatures required to have faithful estimates would make the overall inference extremely slow. Therefore, for the model selection reported in section~\ref{sec:disk}, we resort to an approach analogous to the one described in~\cite{Srinivasan:2024uax}. Given a set of posterior samples, $\left\{\boldsymbol{\theta}_i\right\}$, with $i=0,\, ..., \, N$, for each sample, we estimate the evidence as the ratio between the unnormalized posterior $\hat{p}(\boldsymbol{\theta}_i \, | \, d)$ and the local probability density function $p(\boldsymbol{\theta}_i)$
~\cite{Srinivasan:2024uax, ivezic}.
We estimate $p(\boldsymbol{\theta}_i)$ using an implementation of masked autoregressive flows~\cite{MAFs} based on the \texttt{nflows} package~\cite{nflows}. With $N$ different estimates of $\Z$, we can compute the associated mean $\hat{\Z}$ and standard deviation $\sigma_{\Z}$.

We assume stationary Gaussian noise, and we sample using the standard Gaussian likelihood $\mathcal{L} \propto \exp\{-1/2 \left< s - h(\boldsymbol{\theta}) \, | \, s - h(\boldsymbol{\theta} ) \right> \}$. We define the data stream $s$, and GW templates
$h(\boldsymbol{\theta} )$ with parameters $\boldsymbol{\theta}$. We denote with $\left< \cdot \, | \, \cdot \right>$ the noise-weighted inner product
\be \label{eq:inner_product}
\left<a \,|\,b \right> = 4\text{Re} \int_{0}^{+\infty} \text{d} f \, \frac{\tilde{a}^*(f) \, \tilde{b}(f)}{S_n(f)} \,  .
\ee
Tildes represent Fourier transforms, while $S_n(f)$ is the one-side LISA power spectral density (PSD). We assume the SciRDv1 curve for the instrumental noise~\cite{scirdv} and include the galactic confusion noise as described in~\cite{PhysRevD.104.043019}. With this definition of inner product, we can introduce other important quantities such as the signal-to-noise ratio $\rho$
\be \label{eq:snr}
\text{SNR} \equiv \rho  = \sqrt{\left< h \, | \, h \right>} \, ,
\ee
the overlap between two waveforms $O(h_i, h_j)$
\begin{equation}
    \label{eq:overlap}
    O(h_i, h_j) = \frac{\braket{h_i|h_j}}{\sqrt{ \braket{h_i|h_i} \ \braket{h_j | h_j} }} \, ,
\end{equation}
and the mismatch $\mathcal{M}$
\begin{equation}
\label{eq:mismatch}
\mathcal{M} = 1 - O(h_i, h_j) \, .
\end{equation}
For the purpose of the mismatch and overlap calculations throughout this work, we set the PSD to 1 for all frequencies, allowing us to focus purely on the discrepancies between the two comparison waveforms.

We inject our signals in noiseless data streams, as this choice does not affect the shape of the posteriors~\cite{Speri}. We evaluate PSDs, inner products, and likelihoods using the \texttt{LISAanalysistools} package~\cite{lisatools}, and we include the full time-domain LISA response as implemented in \texttt{fastlisaresponse}~\cite{2022PhRvD.106j3001K}.
Since searching for EMRI signals in the LISA data stream is an extremely challenging task that is out of the scope of this work, we use priors centered on the injected values for the intrinsic parameters of the systems. We also initialize the walkers of the ensemble sampler with values chosen to be close to the injected ones. All of the posteriors shown in this work have a tighter support than the prior range. The priors used are reported in Tab.~\ref{tab:priors} in Appendix~\ref{sec:appendix}. 

Performing Bayesian inference with this setup is feasible but still expensive. Thus, we focus on three different reference systems, including beyond-vacuum GR effects, which will be discussed later. For all of them, we assume an observation period of $T_{\rm obs}=4$ years and SNR $\rho = 50$ with the initial semi-lactus rectum $p_0$ adjusted to match the fixed observation period and the luminosity distance $d_L$ of the systems varied to maintain the fixed $\rho$.
The injected parameters for different models are reported in Tab.~\ref{tab:injection} in Appendix~\ref{sec:appendix}.

\section{Results}

We now present our findings using the fully relativistic waveforms developed and detailed in the methods section. We begin by outlining the computations, cross-testing, and validation studies. following this, we demonstrate how the inclusion of relativistic corrections impacts LISA science exploitation with EMRIs. In particular, we quantify their impact on the source's SNR and on the detectability of beyond-vacuum GR physics with EMRIs.

\subsection{Waveform computation and model test}\label{sec:test}

As pointed out earlier, our relativistic Kerr waveform model is an extension of the \texttt{FEW} framework, explicitly designed for circular, equatorial orbits around a Kerr BH, and integrated as \verb|KerrCircularFlux| within \texttt{FEW}. This model leverages unique flux and amplitude datasets while maintaining consistency with the existing structure. The inspiral trajectory is accessible through \verb|KerrCircFlux|, and the amplitude interpolation class, \verb|Interp2DAmplitudeKerrCircular|, interpolates all $C_{lm}$ over the spin parameter $a$ and radial separation $p$. The interpolants used for the amplitude and trajectory are exactly the same as those in the \texttt{FEW}~\cite{Katz_2021}.
Both the waveform module and datasets are publicly available and can be accessed in~\cite{KerrCircNonvac, KerrCircNonvac_Data}. 

The new trajectory module computes a 4-year circular inspiral around a Kerr BH in $\lesssim 1 ms$. This performs, on average, about three times faster than the existing \verb|SchwarzEccFlux| in \texttt{FEW}. This is because, in the circular case used for our trajectory, only a single interpolant, $\dot E$, is evaluated. In contrast, the existing FEW framework for eccentric inspirals requires the independent evaluation of two interpolants, $\dot E$ and $\dot L$. For selected EMRI parameters with masses $M = 10^6 M_{\odot}$, $\mu = 10 M_{\odot}$, spin parameter $a = 0.99$, initial phase $\Phi_{\varphi,0} = 0$, and fixed source view angles of $\theta_S = \pi/4$, $\phi_S = 0$, and initial separation of $p_0 \simeq 10.0315$ (corresponding to a 4-year inspiral to the plunge), our waveform module computes the overall waveform in approximately $30-40$ ms on a GPU and $\sim 1.8$ s on a single CPU core for a single selected mode. Including all $(l, m)$ modes up to $l_{\rm max} = 10$ (63 modes), the computation takes approximately $60$ ms on GPU and $65$ s on a single core CPU. The GPU timing was performed on an NVIDIA A100, and the CPU timing was on an Intel Xeon Silver 4214R CPU @ 2.40GHz.

\begin{figure}
\centering
    \includegraphics[width = \linewidth]{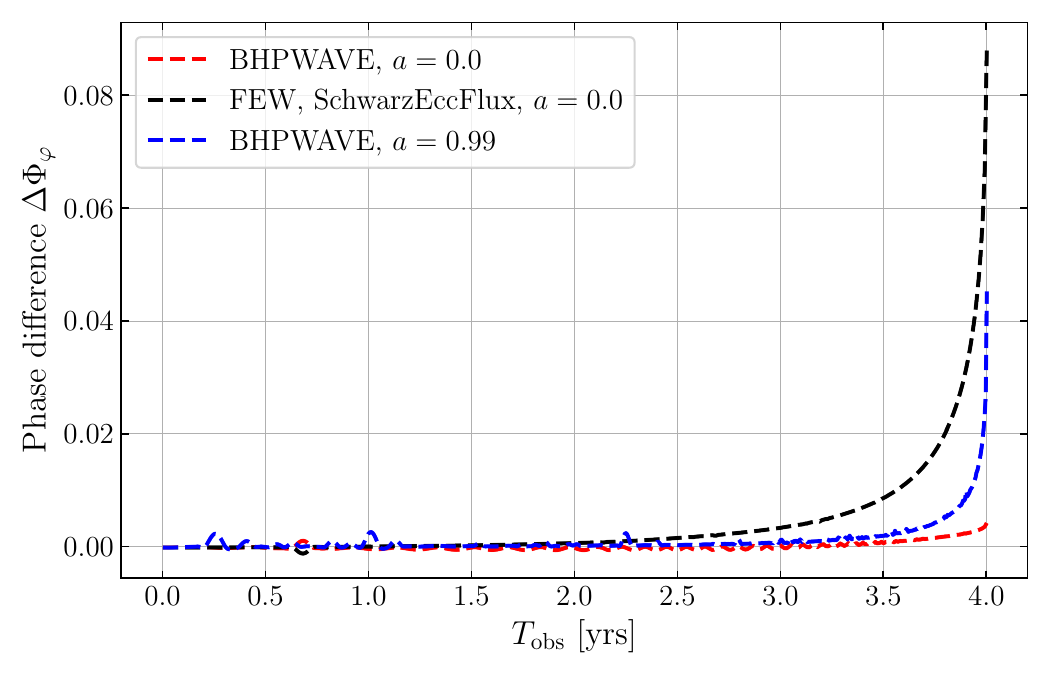}
\caption{Phase shift comparison between the new relativistic circular Kerr waveform and \texttt{BHPWAVE} for $a = 0.0$ (red) and $a = 0.99$ (blue), as well as against the original \texttt{FEW} fast Schwarzschild waveform in the $a=0.0$ limit (black). The inspiral corresponds to a system with $M = 10^6 M_{\odot}$ and $\mu = 10 M_{\odot}$, with an initial radial separation adjusted for a 4-year observation period before the plunge.}
 \label{fig:zach_test}
\end{figure}

To validate the accuracy of our relativistic Kerr circular equatorial waveforms, we tested them against the recently published \verb|BHPWave|~\cite{Nasipak_2024}.
We also compared our results with the original \texttt{FEW} waveform model (Schwarzschild, eccentric). In Fig.~\ref{fig:zach_test}, we illustrate the orbital phase shift between different models for both non-rotating and rotating with spin parameter $a = 0.99$ cases. Initially, all models are perfectly aligned, indicating no dephasing. However, after 4 years of inspiral, a slight dephasing becomes apparent. A more detailed comparison between our model and \texttt{BHPWave} reveals the relative error in interpolated amplitude modes is  \( \lesssim 10^{-5}\) for the majority of the parameter space, increasing up to \(10^{-2}\) near the ISCO of $a=0.99$ for the subdominant modes of $(l,m) \geq (10,10)$. It is important to note that the mode power decreases for higher modes, as depicted in Fig.~\ref{fig:amplitudes_Kerr}.
The total waveform mismatch between our model and \texttt{BHPWAVE} remains $\lesssim 10^{-3}$ for $a=0.99$.

These minor disagreements could stem from either differences in the Teukolsky solvers used for the flux and amplitude data generation or interpolation errors, which may result from variations in the interpolation methods, initial data parameterizations, or data rescaling prior to interpolation. A thorough analysis of these discrepancies is beyond the scope of this paper, and further details can be found in~\cite{khalvati2024flux}. Additionally, Ref.~\cite{Nasipak_2024} considers differences from using different conventions for physical constants, including the solar mass and Newton's constant \(G\). To ensure consistency, we have rescaled both models to the same definitions of these constants for the comparisons above.

\subsection{Impact of relativistic corrections} \label{sec:rel_corrections}
Earlier, we outlined that the GW strain in the adiabatic limit is expressed as a sum of harmonic amplitudes times a complex exponential phase in Eqs.~\eqref{eq:main_wave_generic}, \eqref{eq:wave_new_amps}. The accuracy of waveforms depends on both the phase evolution and the amplitude. For the phase, we account for the fully relativistic flux data from the Teukolsky formalism to handle radiation reaction, as described in Eq.\eqref{eq:trajectory}. For the amplitude, improvements come from using relativistic Teukolsky-based amplitudes and incorporating higher harmonic modes, as shown in Eqs.\eqref{eq:amplitudes} and \eqref{eq:new_amps}. In this Section, we examine their impact on the trajectory, the resulting waveforms, and the implications for LISA's science objectives.

\begin{figure}[t]
    \centering
    \includegraphics[width = \linewidth]{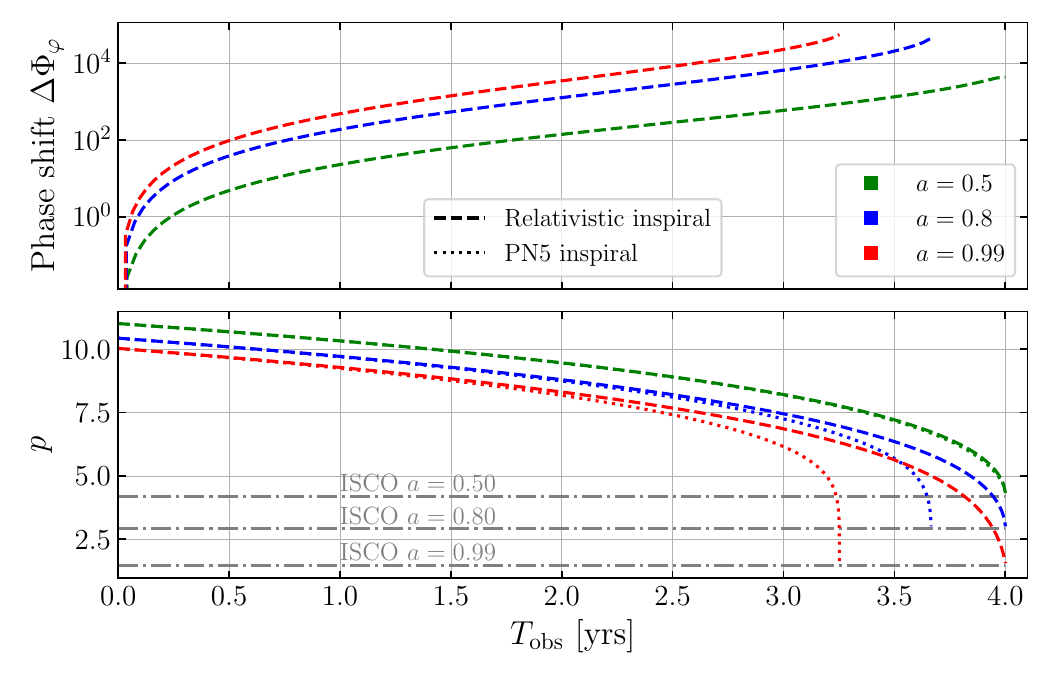}
    \caption{ Phase shift (upper panel), and radial evolution (lower panel) of an EMRI system evolving by relativistic Teukolsky fluxes (dashed), and 5PN fluxes (dotted). We compare the two trajectories for three different spin values $a \in \{0.5, 0.8, 0.99 \}$. The initial separation $p_0$ is adjusted for each spin value to ensure that the EMRI plunges after 4 years of relativistic inspiral.}
    \label{fig:inspiral_kerr_pn5}
\end{figure}

In Fig.~\ref{fig:inspiral_kerr_pn5}, we compare the radial evolution and phase shift between the
inspirals driven by relativistic Teukolsky fluxes, labeled as \textit{Relativistic Inspiral},  and those driven by 5PN fluxes for three spin values: $a = \{0.5, 0.8, 0.99\}$. We set the initial separation $p_0$ such that the EMRI with the relativistic inspiral plunge after 4 years. 
As we increase the spin, we observe that the 5PN expansion overestimates the flux,  leading to greater energy dissipation and, consequently, shorter inspiral. As discussed earlier, this happens because higher spin values allow the EMRI to reach deeper into stronger gravitational regimes before the plunge, where the higher PN order flux terms diverge~\cite{Yunes_berti_3PN_PhysRevD.77.124006}. This discrepancy causes dephasing in the orbital phase of over $10^4$ far exceeding the $\sim 1$ radian threshold often used as a detectability criterion for LISA.

Another approach that has been recently used~\cite{Speri} is to incorporate a relativistic inspiral in the AAK waveform, employing Teukolsky flux data. While in this method the orbital phase evolution is same as the fully relativistic waveform model, the amplitude model is still an approximate model. This introduces errors in the resulting long-duration waveforms' phase and amplitude, leading to discrepancies, as elaborated later. We refer to this waveform model to as the \textit{relativistic} AAK. Given the significant phase shift introduced by the traditional AAK (with 5PN inspiral), we exclude it from further comparisons, focusing solely on the relativistic AAK.

\begin{figure}[t]
    \centering
    \includegraphics[width = \linewidth]{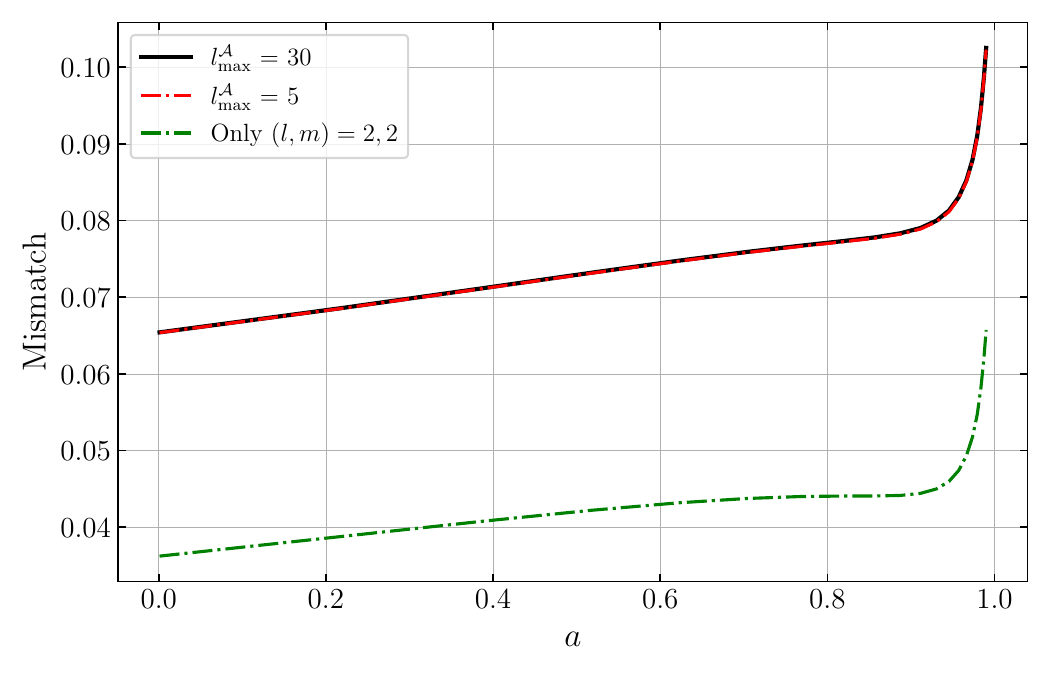}
    \caption{Mismatch between the relativistic AAK and the fully relativistic waveform developed in this work as a function of the BH spin for $4$ years-long waveforms. The mismatch values are maximized on the initial orbital phase with fixed source view angle $\theta_S = 0.7$ and masses of $M = 10^6 M_{\odot}$, and $\mu = 10 M_{\odot}$. 
    }
    \label{fig:mismatch_AAK_vs_Kerr}
\end{figure}

The mismatch $\mathcal{M}$, as defined by Eq.~\eqref{eq:mismatch}, measures the difference between two waveforms.
In Fig.~\ref{fig:mismatch_AAK_vs_Kerr}, we show the mismatch as a function of the spin parameter between the relativistic AAK and our fully relativistic waveform. For the fully relativistic waveform, we select three sets of amplitude $(l,m)$ modes: a single $(2,2)$ mode, all $(l,m)$s with $l^{\mathcal{A}}_{\text{max}} = 5$ and  $l^{\mathcal{A}}_{\text{max}} = 30$.
During the mismatch computation, we fix the sky angles in a way to have the source configured completely in the $x-y$ plane, with the spin, and orbital angular momentum directions perfectly in $z$. The mismatch is maximized over the initial azimuthal phase, with a typical value for the source view angle $\theta_S = 0.7$. Again, we adjust the initial orbital radius $p_0$ for the inspiral so that the EMRI plunges within an observational time of $4$ years. The masses are fixed to typical values of $M = 10^6 M_{\odot}$, and $\mu = 10 M_{\odot}$.

The mismatch with the single \((2,2)\) mode (green curve) illustrates the error due to the relativistic correction by focusing on the only mode present in both the relativistic and the AAK amplitude models. The mismatch with the inclusion of higher modes up to \(l^{\mathcal{A}}_{\text{max}} = 5\) (red curve) demonstrates the improvement due to accounting for additional modes beyond \((2,2)\), while the inclusion of all modes up to \(l^{\mathcal{A}}_{\text{max}} = 30\) shows how including these more modes affect the mismatch (black curve). We assume the most accurate model to be the fully relativistic waveform with \(l^{\mathcal{A}}_{\text{max}} = 30\), so larger mismatches indicate that we are moving further away from the less accurate AAK model and closer to the true model. We observe that including amplitude modes beyond \(l^{\mathcal{A}}_{\text{max}} = 5\) does not significantly change the mismatch curve -- $\sim 10^{-3}$, which is not noticeable in the figure -- and is consistent with expectations from Tab.~\ref{tab:snr_contributions}.

As an illustrative example, we compare the full waveform for three different models in Fig.~\ref{fig:waveforms}: the AAK waveform with 5PN inspiral, the relativistic AAK, and the fully relativistic adiabatic waveform here presented. All three cases involve a Kerr BH with spin parameter \(a = 0.99\) as the central object, starting from an initial orbital separation of $p_0 = 10$ at a fixed luminosity distance $d_L = 1 \, \text{Gpc}$. Both AAK waveforms exhibit noticeably larger amplitudes compared to the fully relativistic waveform due to the neglecting wave absorption at the central BH horizon.  The AAK with 5PN inspiral additionally shows a shorter duration (corresponding to the red curve in Fig.~\ref{fig:inspiral_kerr_pn5}) due to excessive energy dissipation as the 5PN flux diverges closer to the plunge, but even the AAK with relativistic inspiral leads to an overestimation of the total SNR by approximately $20 \%$. 
Notice that, for fixed SNR, the AAK waveform underestimates the amplitude during the early inspiral stage and overestimates it near the plunge when compared to a fully relativistic model. In Sec.~\ref{sec:proof of cons}, we study how this impacts the detectability of beyond-vacuum GR physics with EMRIs, in particular environmental effects.

\subsection{Impact on EMRIs Horizon redshift}

The error in the amplitude of the waveform would also affect the predicted
``redshift of horizon", which is the maximum redshift (or distance) at which the instrument can detect a source. The horizon redshift depends on the amplitude of the emitted GWs, the sensitivity of the detector, and cosmological parameters.
We now quantify how the relativistic corrections in the waveform amplitude influence the prediction for the horizon redshift. Specifically, we compare the relativistic AAK and the fully relativistic model we presented, which differ solely in the ways that their mode amplitudes are computed. 

\begin{figure}[t]
    \centering
    \includegraphics[width = \linewidth]{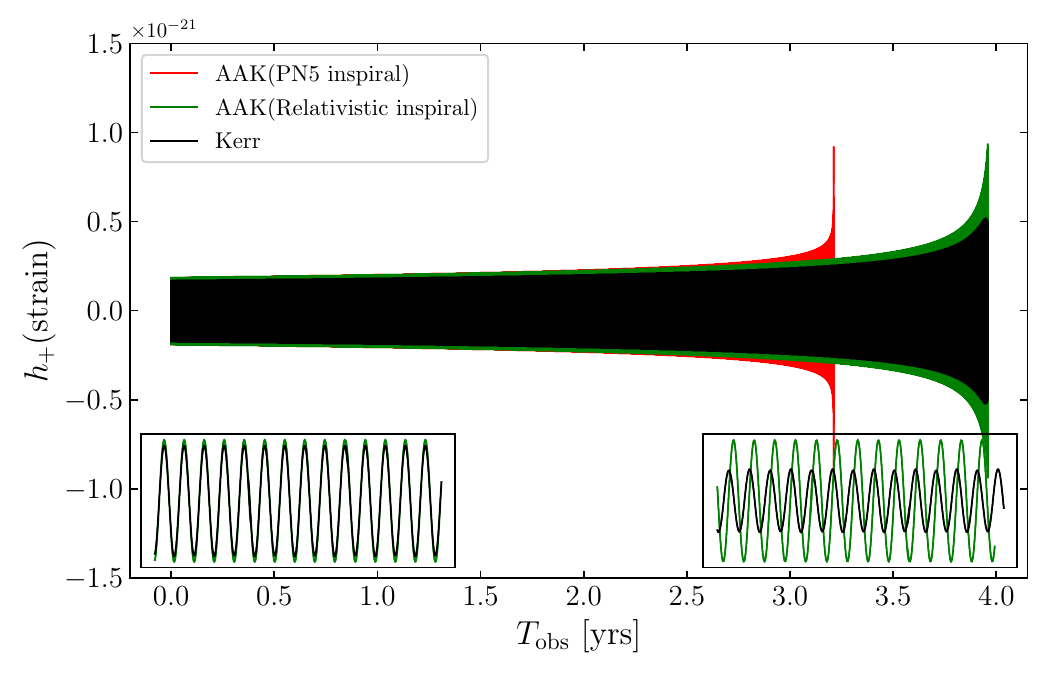}
    \caption{Three different waveforms for the same circular, equatorial orbit around a Kerr BH. The EMRI has $M=10^6\, M_\odot$, $\mu = 10M_\odot$, $a=0.99$, $p_0 = 10$ and fixed luminosity distance $d_L = 1 \, \text{Gpc}$. We used the fully relativistic Kerr waveform (black, $\text{SNR} \simeq 112$), the AAK waveform incorporating relativistic inspiral (green, $\text{SNR} \simeq 134$), and the AAK waveform with 5PN flux-based inspiral (red, $\text{SNR} \simeq 89$). The insets show the phase alignment of the Kerr and relativistic AAK waveforms for the early inspiral (left) and just before the plunge (right).} 
    \label{fig:waveforms}
\end{figure}

In our analysis, we determine the redshift at which the sky-averaged total SNR meets the threshold $\bar \rho = 20$ for various primary source masses $M$ while keeping the mass ratio constant.
We convert luminosity distances in redshift assuming the \textsc{Planck18} cosmology~\cite{2020A&A...641A...6P}.

As discussed in Sec.~\ref{sec:adiabatic_waveforms}, the radiation reaction time scale in EMRIs is inversely proportional to the mass ratio, $\propto \eta^{-1}$. Therefore, EMRIs with larger mass ratios will undergo a faster inspiral. Once again, to ensure that the secondary compact object plunges within a fixed typical value for observation time $T_\text{obs}$, we adjust the initial orbital radius, $p_0$.
Our waveform model has an extended validity up to approximately $p \gtrsim 43$, allowing us to investigate mass ratios up to $\eta \sim 10^{-3}$ (mostly over two-year-long inspirals). The applicability of adiabatic waveforms for such a large mass ratio may be uncertain~\cite{Geoffrey_PhysRevLett.126.241106, K_chler_2024}. Previous studies found that the detection of EMRIs with $\eta \sim 10^{-3}$ and adiabatic waveforms is possible without significant loss of SNR in the case of Schwarzschild spacetimes (see Table II of~\cite{Burke:2023lno}).

\begin{figure}[t]
    \centering
    \includegraphics[width = \linewidth]{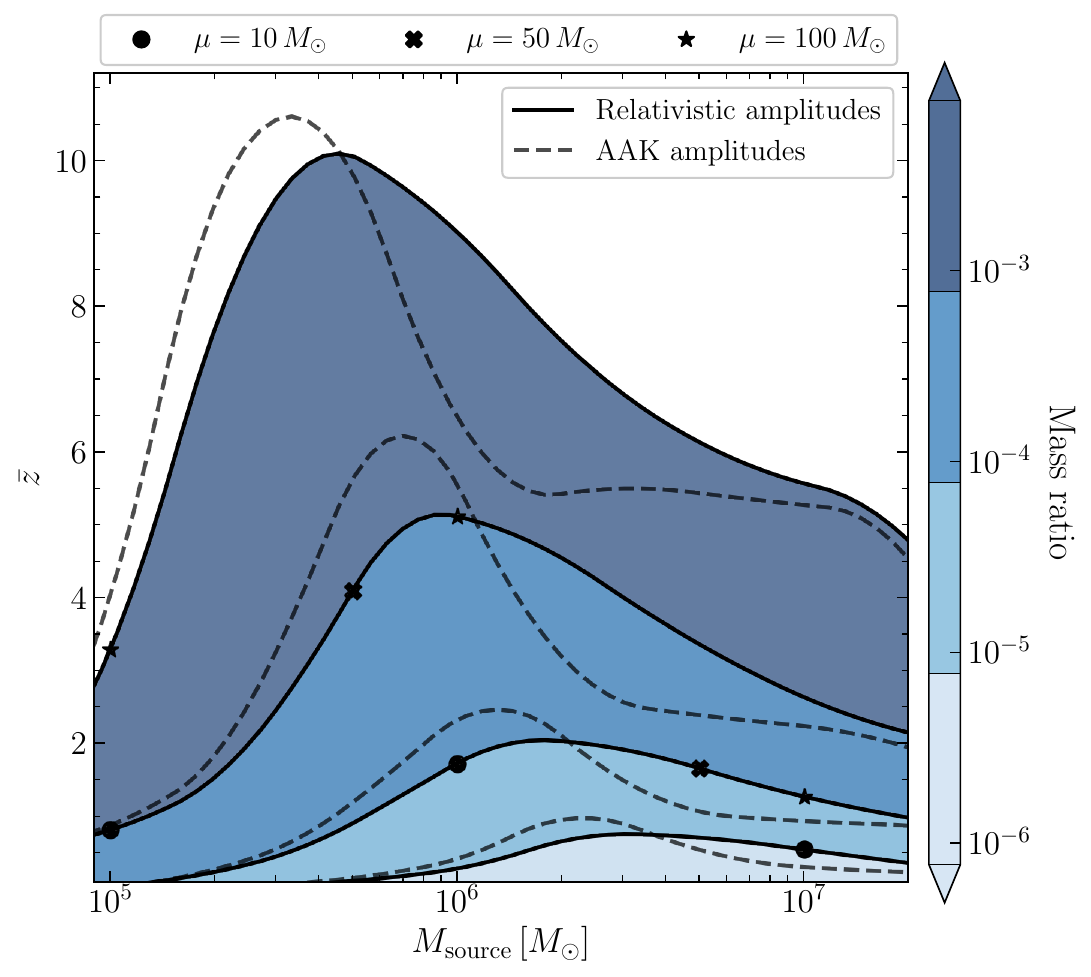}
    \caption{ Horizon redshift $\bar{z}$ at which the sky-averaged SNR of a circular, equatorial, plunging EMRI into a (prograde) MBH with spin $a=0.99$ reaches a threshold SNR $\bar{\rho}=20$, as a function of the source-frame primary mass. Solid (dashed) lines refer to relativistic (AAK) amplitudes, while different colors represent different mass ratios. For all systems, the observation time is assumed to be $T_{\rm obs} = 2$ years, and the trajectory is evolved using relativistic fluxes. The different markers represent the position of three fiducial secondary masses $\mu = 10, \, 50, \, 100\, M_\odot$. For a typical EMRI with mass ratio $\eta = 10^{-5}$, the AAK model overestimates the horizon redshift by about $34\%$ for $\mu = 10\, M_\odot$, whereas for  $\mu = 50\, M_\odot$, it underestimates the horizon redshift by about $35\%$.
}
        
    \label{fig:snr_waterfall}
\end{figure}

We show the curves of horizon redshift $\bar{z}$ against the primary source mass $M$ for different mass ratios $\eta \in \{10^{-3}, 10^{-5}, 10^{-4}, 10^{-6}\}$ in Fig.~\ref{fig:snr_waterfall}. Each curve represents a specific mass ratio, and accordingly, each point on the curve corresponds to a different value of the secondary mass $\mu$. There are regions on the plot where the secondary mass is relatively massive, depending on the primary mass.

We observe the horizon redshift in both models peaks at specific values of \( M_{\text{source}} \), though the peak sensitivity occurs at different mass values for each model.
At the lower end of \( M_{\text{source}} \), the AAK model tends to overestimate the horizon redshift, which aligns with the expectations discussed earlier in relation to Fig.~\ref{fig:waveforms} (neglecting the perturbations falling into the BH's horizon). However, for larger values of \( M_{\text{source}} \), the AAK model significantly underestimates the horizon redshift $\bar{z}$, particularly for the smallest mass ratios, such as $10^{-3}$.
For each mass ratio, as the primary source mass $M_{\text{source}}$ increases, the overall wave frequency decreases. In this regime, the higher harmonics in the relativistic amplitude play a crucial role, as they are present at frequencies higher than the dominant quadrupole mode, $C_{22}$. Since confusion noise (Sec.~\ref{sec:data_analysis}) is stronger at lower frequencies, these higher harmonics enhance the estimated SNR by contributing power where noise is weaker. Therefore, the transition from overestimation to underestimation in the horizon redshift values for the AAK occurs when these modes are included in the relativistic amplitude, a behavior not observed in previous studies \cite{Babak:2017tow}.
We consider two representative secondary masses, $10 M_{\odot}$ and $50 M_{\odot}$, with fixed mass ratio $\eta = 10^{-5}$ to illustrate the significant differences between the relativistic and AAK amplitude models. For the $10 M_{\odot}$ case, the AAK model overestimates the horizon redshift by approximately 34\%, whereas for the $50 M_{\odot}$ case, it underestimates the redshift by approximately 35\%. These secondary masses are of particular astrophysical relevance, corresponding to typical EMRI systems, and provide valuable benchmarks for comparing model predictions given the substantial discrepancies in horizon redshift.

Finally, we note that we assume a sampling rate of $dt=10$ seconds for all the generated waveforms. Although this value is too large to fully resolve the high-frequency region of the spectrum, where the majority of the SNR is concentrated for low-mass systems, we have checked that using a sampling rate as small as $2$ seconds does not significantly affect our results for the low mass tail of Fig.~\ref{fig:snr_waterfall}. We do not use such a small sampling rate for the whole plot to avoid dealing with the high-frequency zeros of the noise PSD. 

\subsection{Impact on beyond-vacuum GR effects}\label{sec:proof of cons}
In the previous section, we quantified the impact of relativistic corrections on the waveform amplitudes in vacuum. In this final section, we use the newly built fully relativistic Kerr waveform for the study of two beyond-vacuum-GR EMRI systems. This serves as a proof-of-concept for the modality and flexibility of the package and also highlights the importance of relativistic contributions to EMRI physics.

\subsubsection{\textbf{Migration Torques in Accretion disks}}\label{sec:disk}

A fraction of the population of SMBHs is expected to be actively accreting matter surrounding it, forming dense-gas accretion disk~\cite{Dittmann:2019sbm, Yang_2021, 2019ApJ...874...54M, Madau:2014bja}. EMRIs formed and evolving immersed in this environment will dynamically interact with it, which can alter its trajectory and subsequent GW signature~\cite{Kocsis_2011, Yunes_disk, Speri, Derdzinski:2020wlw, Zwick:2021dlg, Garg:2022nko, Levin_2007MNRAS.374..515L}.
These interactions were first considered in the context of EMRIs more than a decade ago~\cite{Kocsis_2011, Yunes_disk}.  Effects like accretion onto the secondary and, in particular, migration torques were found to lead to observable imprints in the waveform, which could potentially be used to study the disk properties. 
\textit{Migration} results from the density wake generated by the secondary in the surrounding gas, which then assisted by the differential rotation of the disk, forms a spiral arm structure that can resonantly exchange angular momentum with the EMRI, accelerating or stalling the inspiral~\cite{Tanaka_2002, Tanaka_2004, Hirata:2010vn, Hirata2:2010vp}. 

Migration torques can be added as a correction to the balance-law equations governing the evolution of the orbital and angular momentum energy (Eq.~\eqref{eq:balance_law})

\begin{align} \label{eq:disk_balance}
\dot{L}_{\text{orb}} &= -\dot{L}_{\text{GW}} - \dot{L}_\text{mig} \, .
\end{align}

The simplest model for the gas corrections is a power-law~\cite{Kocsis_2011} 
\begin{align}
    \Dot{L}_\text{mig} = A  \left(\frac{p}{10 M} \right)^{n_r}  \Dot{L}^{(0)}_{PN} \, ,  \label{eq:MigTorques}
\end{align}
where $\Dot{L}^{(0)}_{PN} = \Omega_{\varphi}^{-1} \Dot{E}^{(0)}_{PN}$ is the leading PN order angular momentum flux. The amplitude $A$ and slope $n_r$ depend on the internal properties of the disk, e.g. viscosity, opacity-law, and the primary accretion rate.  For example, in the inner-region of the canonical Shakura-Sunayev $\alpha$-disk~\footnote{for a  short but more detailed description of this model, we refer the reader to\cite{Shakura_1973A&A....24..337S}}, the slope is $n_r = 8$ and the amplitude~\cite{Kocsis_2011}
\begin{equation}
    A \sim 7 \times 10^{-10}  \left(\frac{0.1}{\alpha} \right)\left( \frac{f_\text{Edd}}{0.1}\frac{0.1} {\epsilon} \right)^{-3}\left( \frac{M}{10^6 M_\odot}\right) \, , 
\end{equation}
where $\alpha$ is the viscosity parameter, $f_\text{Edd}$ the accretion ratio with respect to the Eddington limit, and $\epsilon$ the conversion efficiency of mass-energy to luminosity in the disk. The factors introduced correspond to their typical values: $\alpha = f_{Edd} = \epsilon = 0.1$, though variations of one order of magnitude can occur. In the waveform modeling presented here, these factors are all encoded into the amplitude parameter, $A$. For further discussion, refer to the Ref.~\cite{Speri}. 
\begin{figure}[t]
    \centering
    \includegraphics[width = \linewidth]{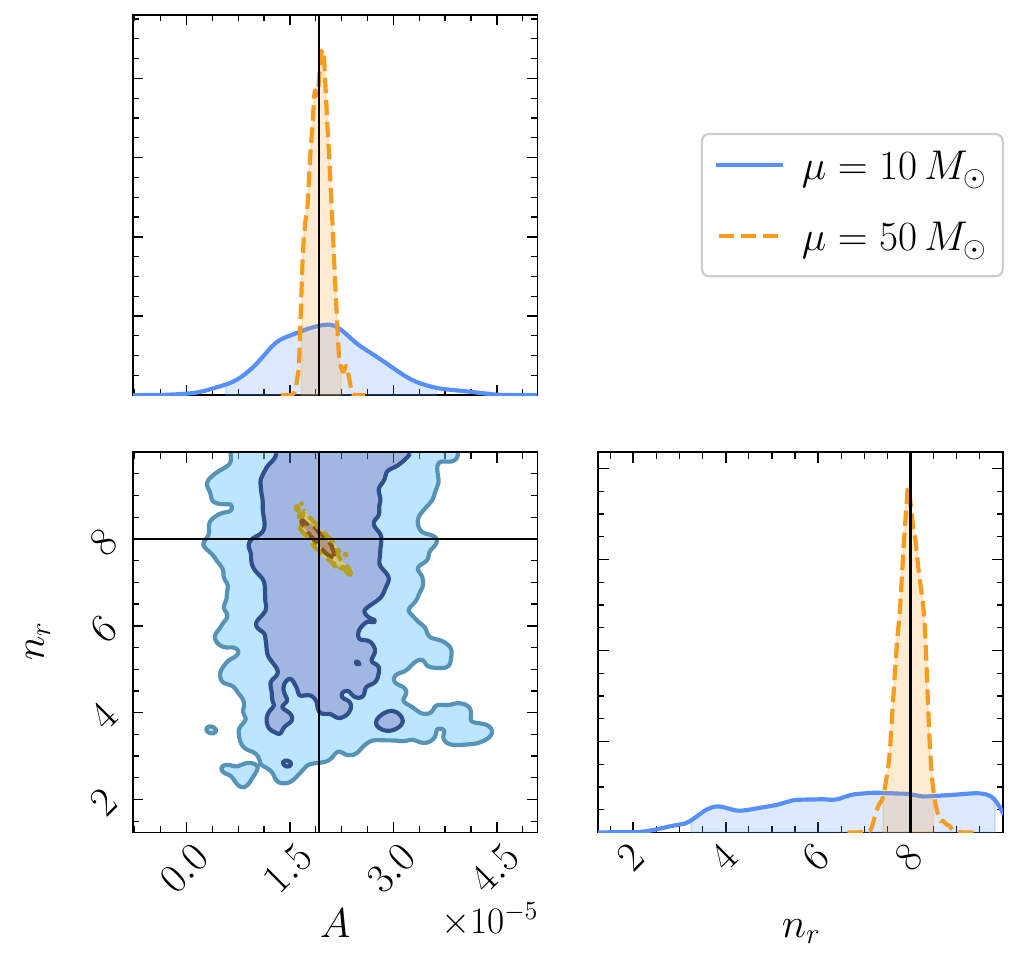}
\caption{Marginalized posterior distributions for the migration torque~\eqref{eq:MigTorques} amplitude $A$ and slope $n_r$ for two similar EMRIs with different secondary mass (the full injected parameters are listed in Table~\ref{tab:injection}). Shaded areas represent the $1-$ and $2-\sigma$ regions ($95\%$ credible regions). 
We use the fully relativistic waveforms developed in this work, which for the heavier secondary case $\mu = 50 M_\odot$, lead to a significant improvement on the constraints of the torque parameters with respect to Ref.~\cite{Speri}. The medians and $95\%$ credible intervals of the inferred amplitude and slope are $A = 2.0^{+1.7}_{-1.4} \times 10^{-5}$, $n_r = 7.0^{+2.8}_{-3.7}$
 for the $\mu = 10 M_\odot$ case, and $A = 1.93^{+0.31}_{-0.27}\times 10^{-5}$, $n_r = 7.99^{+0.50}_{-0.58}$
 for the $\mu = 50 M_\odot$ case.
}
\label{fig:disk_vs_disk}
\end{figure}
\begin{figure*}[ht]
	\centering
 \includegraphics[width=\textwidth]{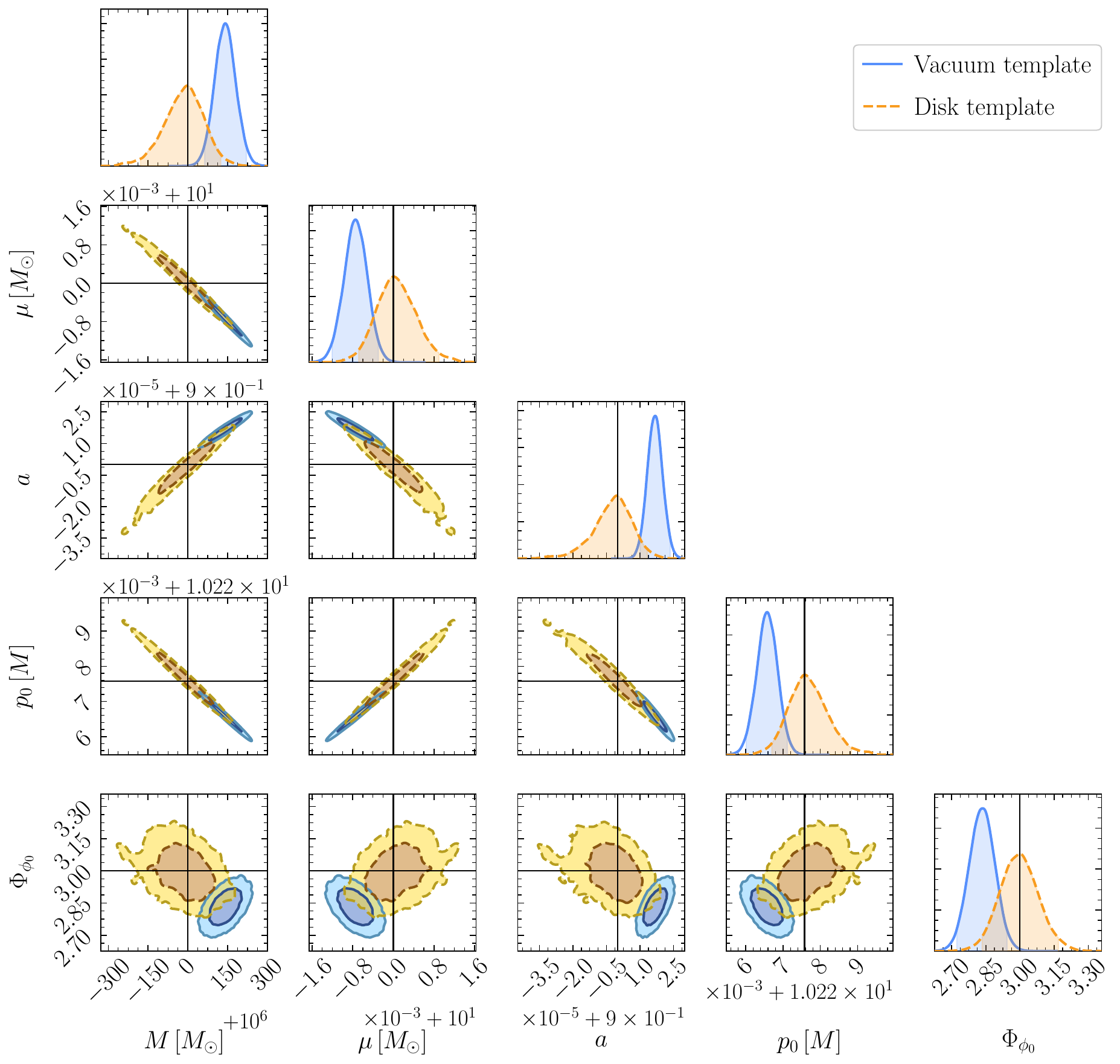}
	\caption{Comparison between the posteriors on the EMRI intrinsic parameters recovered when using a vacuum (solid blue) and disk (dashed orange) template to analyze the same injection produced with the ``Disk'' template.   
For each parameter, we show the difference with respect to the injected value (solid black lines). Shaded areas in the 2D (1D) histograms represent the $1-$ and $2-\sigma$ regions ($95\%$ credible regions).}
	\label{fig:mcmc_disk_circ_compare}
\end{figure*}
In the same work, this model was incorporated in \texttt{FEW}  and used to study, within a Bayesian framework, the ability of LISA to measure migration torques in EMRIs. It was concluded that for the $\alpha$-disk, torque amplitudes as small as $A \sim 2 \times 10^{-6}$ could be detected for a fiducial circular, equatorial EMRI system with $M=10^6\, M_\odot$, $\mu = 50 \, M_\odot $, $a = 0.9$ and $\text{SNR} = 50$ after 4 years of observation prior to plunge (initial orbital separation $p_0 \approx 15.5M$). For an amplitude $A = 1.92 \times 10^{-5}$, the amplitude and slope could be constrained simultaneously with a relative error ($\frac{\sigma_x}{\bar x} \times 100$) of $\lesssim 24\, \%$ while maintaining the typical uncertainties for the other binary parameters, suggesting the possibility of distinguishing between different disk models.

However, the results listed above were obtained using the relativistic AAK model. As discussed in Sec.~\ref{sec:rel_corrections}, these approximate waveforms can underestimate the relative amplitude, i.e. power or SNR, of the early-inspiral with respect to the close-to-plunge region. Since typically environmental effects are more significant at larger orbital separations, and in a PN language correspond to negative PN corrections~\eqref{eq:MigTorques}, we expect the use of AAK amplitudes in Ref.~\cite{Speri} underestimates how well migration torques can be constrained. 

We therefore repeat their study with the fully relativistic waveform here presented, using the accurate amplitude model (Sec.~\ref{sec:Rel_amps}), for two different values of secondary mass, $\mu = 10\, M_\odot$ and $\mu = 50\, M_\odot$.  Our findings are illustrated in Fig.~\ref{fig:disk_vs_disk}, where we show the marginalized 2D posteriors for the migration torque parameters obtained from a full Bayesian analysis over the entire parameter space, using standard MCMC methods. The details of our MCMC setup were
discussed in Sec.~\ref{sec:data_analysis}. Additional details are provided in Appendix~\ref{sec:appendix}, where we list the injection parameters in Tab.~\ref{tab:injection} and show the full posteriors for the case with the heavier secondary in Fig.~\ref{fig:mcmc_disk50}. The waveform and trajectory for the accretion disk's migration torque effect can be accessed under the modules \verb|MigrationTorqueKerrCircularFlux| and \verb|MigTorqKerrCircFlux|.

The EMRI with $\mu = 50 M_\odot$, similar to the one considered in~\cite{Speri}, is analyzed using the fully relativistic waveform.  For a fixed averaged SNR of $\bar \rho = 50$, the early inspiral amplitude is about 10\% higher than that of the AAK amplitude. We find relative errors on the recovered parameters of the migration torque of only $\sim 8\, \%$, a significant improvement with respect to the results obtained with the AAK amplitude waveform~\cite{Speri} (which were $\sim 24\, \%$). 
This decrease corroborates our expectation and highlights the importance of relativistic corrections to study beyond-vacuum GR with EMRIs, which inherently probe the strong-field regime. Even corrections applied only to the vacuum sector, e.g. including relativistic amplitudes in the waveform, have a significant impact on how well we can probe accretion physics with EMRIs. The other intrinsic parameters are the same as in~\cite{Speri}. 

The EMRI with a lighter secondary, $\mu = 10 M_\odot$, exhibits broader posteriors, a consequence of the smaller initial orbital separation, $p_0 \simeq 10.03 M$, required to maintain the plunge within 4 years. In comparison, the EMRI with $\mu = 50 M_\odot$ starts its inspiral at a larger separation, $p_0 \simeq 15.36 M$, where the disk-induced angular momentum loss is approximately an order of magnitude larger for the same injected disk parameters, $A = 1.92 \times 10^{-5}$ and $n_r = 8$. This explains why we can better constrain the disk parameters for the heavier EMRI, as the migration torques at the beginning of the inspiral are significantly stronger.

Additionally, we investigated the bias in PE by injecting disk waveforms and recovering with vacuum templates. Our focus here is on the $\mu = 10, M_\odot$ case, where the disk's influence is weaker. Nonetheless, this case remains important for exploring potential biases in parameter recovery. To assess the bias, we compare the results to the vacuum posterior for the intrinsic parameters in Fig.~\ref{fig:mcmc_disk_circ_compare}. We find that the posterior medians are shifted away from the true values by $2 \sigma$ or more, specifically leading to an overestimation of the primary's mass and spin. A similar behavior was observed in \cite{Speri}, though in a different EMRI scenario, with a heavier secondary mass, primarily due to the use of less accurate waveform modeling.
However, we caution that these observed biases should be considered in the context of potentially larger systematic biases present in the waveform models, as ignoring the environmental phenomena alone may not be the predominant source of the total bias in PE~\cite{Burke:2023lno, Dhani:2024jja, Gupta:2024gun}.

To quantify the support for the disk model in the lighter EMRI scenario, we compute the Bayes factor between the two models (disk vs vacuum) following the approach presented in Sec.~\ref{sec:data_analysis}. Labeling with ``D" and ``V" the disk and vacuum template, respectively, we find $\log_{10}\mathcal{B_{\rm DV}} = 1.1$ and $\sigma_{\log_{10}\mathcal{B}_{\rm DV}} = 0.7$. Following the criterion in~\cite{doi:10.1080/01621459.1995.10476572}, this value translates into positive but not conclusive evidence in support of the disk model. Not finding strong support for the same template used for the injection poses extra attention on how sensible our search and PE pipelines will have to be to fully exploit the scientific potential of these signals~\cite{redbook_colpi2024lisadefinitionstudyreport}. Notably, for the heavier EMRI, the environmental effects are so significant that our MCMC algorithm does not converge within the assumed prior volume 
when using the vacuum template, posing an interesting problem for EMRI searches with beyond-vacuum GR effects~\cite{Chua:2021aah}.

\subsubsection{\textbf{Superradiant scalar clouds}}\label{sec:cloud}

Finally, we consider an EMRI evolving in a different type of environment: a superradiant bosonic cloud composed of an ultralight scalar field. New ultralight fundamental fields, whether scalar or vector ones are ubiquitous in extensions of the Standard Model and high-energy physics~\cite{PhysRevLett.38.1440, PhysRevLett.40.223,Arvanitaki_2010,Fabbrichesi:2020wbt} and have been proposed as a component of dark matter~\cite{Arvanitaki_2010, Ferreira:2020fam,Hui:2021tkt,Antypas:2022asj}. The canonical example is the QCD axion but in the context of astrophysical BHs, boson masses ranging from $m_b \sim 10^{-23}-10^{-11}\, \text{eV}$ have been considered. Bosons in the lower end of this mass range can form self-gravitating structures, known as \textit{boson stars}~\cite{Seidel:1993zk, Liebling:2012fv}, which describe well the flatter profile of the inner core of DM halos suggested by observations~\cite{Schive:2014hza, Schive:2014dra, Cardoso:2022nzc}. On the other hand, if the Compton wavelength of the boson is comparable to the BH size, i.e. $\alpha_{b} = M\mu_{b} \lesssim 1/2$, where $\mu_b = m_{b} c / \hbar $ is the field's reduced inverse Compton wavelength, the field can effectively extract rotational energy from the spinning BH through superradiance (the wave analogous to the Penrose process) and condensate into  \textit{boson clouds} (see Ref.~\cite{Brito:2015oca} for a review).  These have a structure similar to the hydrogen atom with $\alpha_b$ playing the role of a ``gravitational fine-structure constant''. Boson clouds can attain densities much larger than other astrophysical environments, like accretion disks. At the end of the superradiant growth, the cloud's mass, $M_c$, can reach up to $\sim 10 \%$ of the BH host mass, with densities of $\rho \lesssim 5\times 10^2 \, \text{g}/\text{cm}^3 $. Since the clouds are non-spherical, they decay via emission of continuous, quasi-monochromatic GWs with frequency $\omega \sim 2\mu_b$~\cite{Yoshino:2013ofa,Arvanitaki:2014wva,Brito:2017zvb,Siemonsen:2019ebd,Siemonsen:2022yyf}. However, this decay is polynomial and on timescales that are extremely larger than the superradiant growth timescale~\cite{Yoshino:2013ofa,Arvanitaki:2014wva,Brito:2017zvb,Siemonsen:2019ebd,Siemonsen:2022yyf}. 

Similarly to the accretion disk scenario, if the EMRI is inspiralling while immersed in a superradiant cloud, it interacts with the scalar environment. The secondary will accrete the scalar field, generate density wakes that exert dynamical friction~\cite{Zhang_2020,Baumann:2021fkf,Tomaselli:2023ysb}, and that can also induce resonant transitions between different cloud states when the orbital frequency matches the energy difference between states, akin to atomic resonances between orbitals with different energy~\cite{Baumann:2018vus,Baumann:2019ztm}. The impact of these effects on an EMRI's evolution has been studied extensively under this Quantum Mechanics analogy~\cite{Baumann:2018vus, Zhang:2018kib,Baumann:2019ztm,Zhang_2020, Baumann:2021fkf, Baumann:2022pkl, Tomaselli:2023ysb, Tomaselli:2024dbw, Tomaselli:2024bdd}, and integrated into a Bayesian framework in~\cite{Cole:2022yzw}. 
However, these works resorted to Newtonian models for the waveforms and the interaction with the cloud. More recently, Refs.~\cite{Brito_2023, Duque:2023cac} addressed this problem for circular EMRIs around non-rotating BHs in a fully relativistic setup, employing techniques from BH perturbation theory~\cite{Barack:2018yvs, Pound:2021qin}. In their framework, one can compute gravitational radiation for a given scalar configuration but also obtain the energy and angular momentum carried by scalar waves both to infinity and absorbed at the BH horizon. In the Newtonian limit, these quantities can be directly related to the force felt by the secondary and the energy imparted to the scalar configuration~\cite{Zhang_2020, Tomaselli:2024dbw}. In General Relativity, this equivalence is lost; both scalar and GW emission and accretion onto the BHs imply the cloud is no longer completely stationary. Nevertheless, for a sufficiently light secondary, these effects occur on timescales much longer than the inspiral/observation time of an EMRI. We can, therefore, still write the balance-law equation for the orbital evolution (see also~\cite{Clough:2021qlv} for a derivation of relativistic continuity equations)
\begin{equation}
\dot{L}_{\text{orb}} = -\dot{L}_{\text{GW}} - \dot{L}_\text{s} \, . \label{eq:BalanceLawScalar}
\end{equation}
Note that $\dot{L}_\text{s}$ is not necessarily positive~\cite{Brito_2023}. In fact, for the system we study below, $\dot{L}_\text{s}$ is negative, and the presence of the cloud slows down the inspiral. 
With this,  we revisit the detectability of the environmental effects for a circular EMRI immersed in a superradiant cloud in the fundamental dipolar state (the one with the fastest superradiant growth time), but using the relativistic results in Ref.~\cite{Brito_2023} for the interaction with the cloud. Using the same framework presented there, we computed the scalar angular momentum fluxes at infinity, $\dot{L}_\text{s}^\infty$, and at the BH horizon, $\dot{L}_\text{s}^H$, up to the $l_\text{max}^s=5$ mode, on the same grid described in Section~\ref{sec:traj} (we checked higher scalar modes contribute less than $1\%$ in the parameter space region we are going to study). 
The cloud's effect is described by two parameters: the ``gravitational fine-structure'' (coupling) constant $\alpha_b = M \mu_b$; and the total mass of the cloud, $M_c$, which depends on the difference between the initial and final BH spin at which the superradiant instability saturates. We assume the cloud has saturated the superradiant growth and it is in a (quasi-)stationary configuration. This condition allows setting a relation between the BH spin and the gravitational coupling given by $\alpha_b = \Omega_H$,  leading to the expression,  
\be \label{eq:saturation}
a \simeq 4\alpha_b / \left(1 + 4 \alpha_b ^2 \right), 
\ee
for the spin of the BH (see Eq. 25 in~\cite{Brito:2017zvb}). This effect is integrated in a new trajectory module, \verb|CloudKerrCircFlux|, and the waveform module, \verb|ScalarCloudKerrCircularFlux| , with the balance-law in Eq.~\eqref{eq:BalanceLawScalar} governing the orbital evolution, where $\dot{L}_\text{s} = \dot{L}_\text{s}^\infty + \dot{L}_\text{s}^H$. 

Although the values of the scalar flux are obtained in Schwarzschild, we are going to use them for EMRIs with rotating primary BHs. As illustrated in Fig.~5 in Ref.~\cite{Brito_2023}, the relative contribution of the scalar sector to the total flux with respect to GWs is more relevant at large distances, where spin effects are less important. As a reference, the relative difference between the GW energy flux in the dominant quadrupole mode for circular EMRIs between Schwarzschild and a Kerr BH with $a=0.6$ (which we will use below) is ~$ \lesssim 10 \%$ for $p_0 \geq 10$. Since the effect of the cloud is already a small (but potentially relevant) correction to the overall trajectory, a $\sim 10 \%$ error on the scalar fluxes should not impact our results qualitatively. Additionally, note that in the region of the parameter space we explored, the scalar fluxes are always at least 2 orders of magnitude smaller than the gravitational fluxes and reach up to 6 orders of magnitude of relative difference at $p=6$. A recent fully relativistic calculation of scalar fluxes in Kerr spacetime further confirms the robustness of our approach, supporting our initial estimate that using the scalar flux from Schwarzschild background in our setup does not introduce significant errors in the parameter region that we consider (see Fig.~2 in Ref.\cite{dyson2025environmentaleffectsextrememass}).

\begin{figure}[t]
    \centering
    \includegraphics[width = \linewidth]{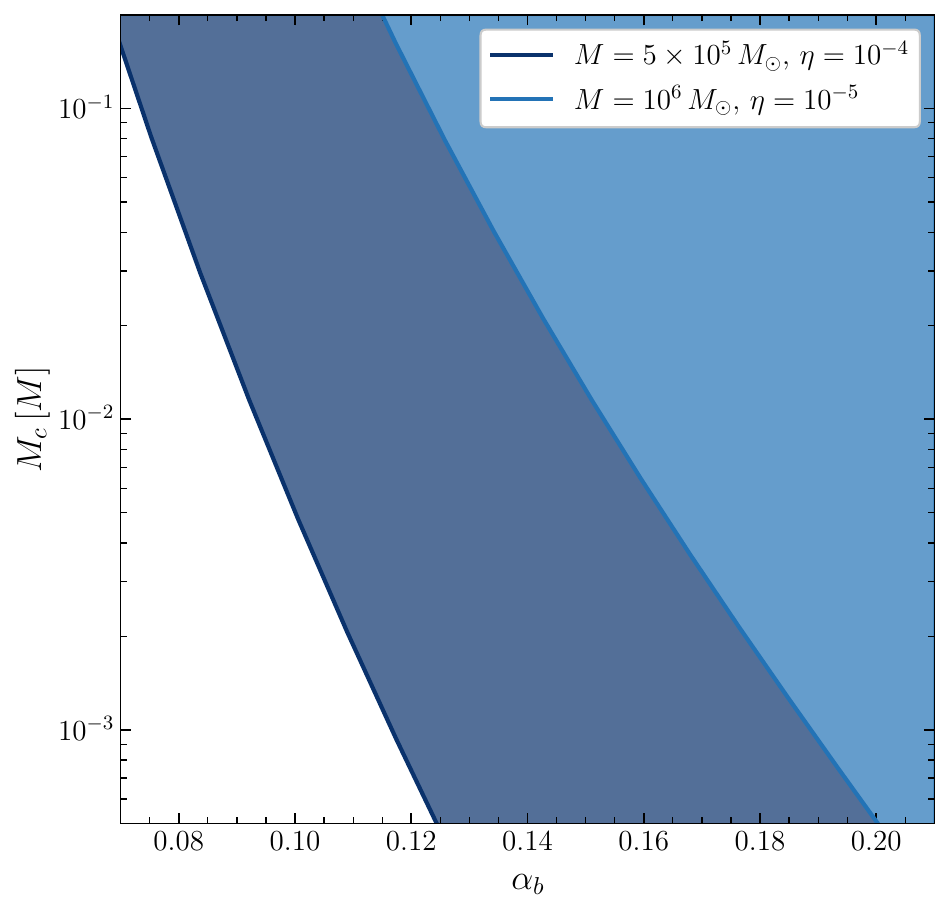}
    \caption{Minimum mass of the scalar cloud, $M_c$, which for a given mass of the scalar, $\alpha_b = M \mu_b$, leads to a 1 cycle of orbital dephasing after 4 years of observation prior to plunge. For cloud masses above the lines, the dephasing is bigger. We show two representative EMRIs with different mass ratios, corresponding to different initial radii ($p_0 \sim 22$ for $\eta = 10^{-4}$ and $p_0 \sim 11$ for $\eta = 10^{-5}$). At larger separations, the relative ratio between the scalar and GW flux is larger, explaining the larger dephasing obtained for the EMRI with a larger mass-ratio.
    }
    \label{fig:Min_Dephase_Cloud}
\end{figure}

\begin{figure}[t]
    \centering
    \includegraphics[width = \linewidth]{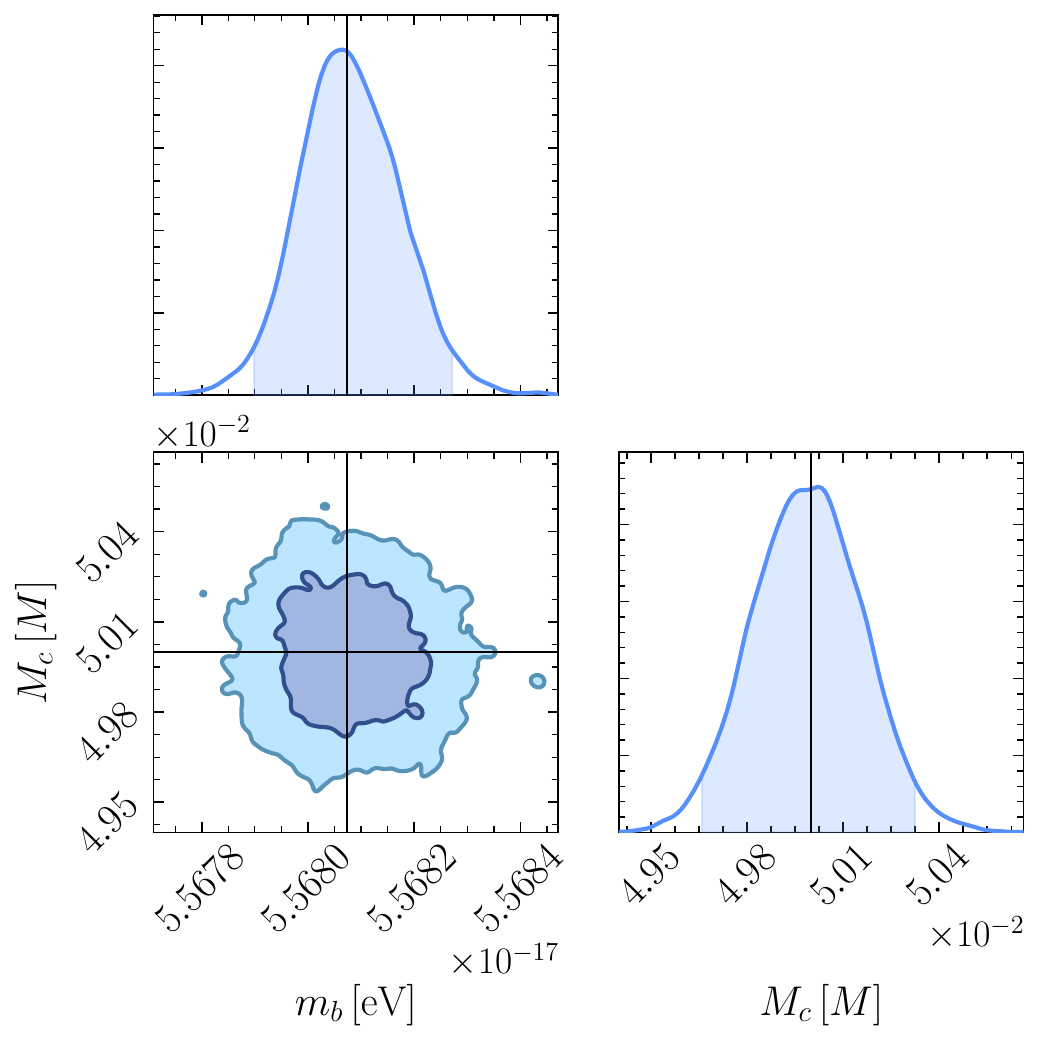}
    \caption{
    Marginal 2D posterior distribution for the cloud parameters $m_b$ and $M_c$. Black solid lines represent the injected parameters, while shaded areas in the 2D (1D) histograms represent the $1-$ and $2-\sigma$ regions ($95\%$ credible regions). The full posterior distribution can be found in Fig.~\ref{fig:mcmc_axion}. We stress that $m_b$ is not directly sampled but is instead derived from other parameters' posterior distribution under the assumption of (quasi-)stationary configuration for the cloud. The medians and $95\%$ credible intervals of the inferred masses are $m_b = \left(5.5681 \pm 2 \times 10^{-4} \right) \times 10^{-17}\, \rm eV$, $M_c = \left( 499.9\pm 3.3 \right) \times 10^{-4}\, M$.
    }
    \label{fig:corner_Cloud}
\end{figure}

We start by using this corrected trajectory to compute which scalar cloud configurations lead to one cycle of dephasing after 4 years of observation prior to the plunge. This is a benchmark value for the detectability of a beyond-vacuum GR effect, albeit not a sufficient one. 
As mentioned above, the cloud's mass can reach up to $~ 10 \%$ of the primary BH mass. Our results are displayed in Fig.~\ref{fig:Min_Dephase_Cloud} for two different EMRIs. We only show values for $\alpha_b \lesssim 0.2$ because larger couplings dissipate too quickly via GW emission~\cite{Brito:2015oca}. As in the accretion disk case, the EMRI with a larger mass ratio ($\eta   = 10^{-4}$) covers larger orbital separations and, thus, for the same coupling $\alpha_b$, requires less massive clouds for the environmental effect to be detectable. From the plot, we read EMRI observations with LISA could potentially probe superradiant scalar clouds for $\alpha_b \geq 0.07$, corresponding to masses $\mu_b \sim 10^{-16}-10^{-17} \, \text{eV}$ for SMBHs with $M\sim 10^{5}-10^{6} \, M_\odot$. This range of masses is virtually unprobed, with current constraints based on the observation of highly-spinning SMBHs, which are sensitive to possible systematic errors in the measurements of the BH spin and to the unknown formation and growth history of the SMBHs~\cite{Arvanitaki:2014wva,Cardoso:2018tly}. If EMRIs around massive BHs with $M\lesssim 10^{5} \, M_\odot$ are also observed, one could also potentially probe the mass range $\mu_b \gtrsim 10^{-16} \, \text{eV}$, which is where constraints based on GW emission from the superradiant cloud (either directly or via a stochastic background) are likely to be less effective (see Figs. 2 and 3 of~\cite{Brito:2017wnc}).

We then repeat the analysis of the previous section and perform a Bayesian inference with MCMC over the full parameter space of a circular equatorial EMRI in Kerr spacetime and the two cloud parameters. We focus on an EMRI with primary mass $M = 4 \times 10^{5} \, M_\odot$, spin parameter $a=0.6$, and secondary mass $\mu = 20\, M_\odot$, again with 4 years of observation time prior to plunge (initial orbital separation $p_0\approx 19.94$) and $\text{SNR} = 50$.  The BH's spin was picked corresponding to the superradiant saturation spin for the cloud's coupling $\alpha_b \approx 0.16667$, while for its mass, we chose a moderate value of $M_c=0.05M$. This configuration leads to $1388$ cycles of dephasing ($8.7 \times 10^3$ radians) with respect to vacuum after the 4 years of evolution. The full posterior is displayed in Fig.~\ref{fig:mcmc_axion} in the Appendix~\ref{sec:appendix}, while the marginalized 2D posterior for the cloud parameters is shown in Fig.~\ref{fig:corner_Cloud}. 

Note that among the two model parameters here, the cloud's coupling, $\alpha_b$, is not directly sampled in our analysis; instead, it is derived from the BH's spin at each step assuming the superradiant saturation condition, Eq.~\eqref{eq:saturation}. Then we get the posterior on the boson mass $m_b$, using the $m_b = \hbar \ \alpha_b / M $. Therefore, the constraints for $m_b$ are obtained from the BH spin measurement and the BH's mass. For this reason, $\alpha_b$ does not appear in Fig.~\ref{fig:mcmc_axion}, and in Tabs.~\ref{tab:priors},~\ref{tab:injection}, and~\ref{tab:recovery}, which refer only to the sampling parameters. Remarkably, the cloud's mass can be constrained with a relative error of $\sim 0.3 \, \%$. The mass of the ultralight field can be tightly constrained, as it is derived from intrinsic parameters (relative error $\sim 10^{-3} \ \%$; see Fig.~\ref{fig:mcmc_axion}).

Our results confirm the early exploration of Ref.~\cite{Cole:2022yzw} using Newtonian models. We did not pursue a direct comparison with their results since there are substantial differences in the models used for both the waveforms and the effects of the environment. Also, they considered larger mass ratios and lighter primary masses, which can reach even larger orbital distances, where the environmental effects are typically stronger.
Nonetheless, it is important to quantify the impact of astrophysical systematics in GW astronomy, particularly how improved modeling of environmental effects affects their detectability and biases in the vacuum parameters if the environment is ignored. We leave this study for future work. Also, as for the heavier EMRI in the accretion disk, the effect of the environment is so strong that we cannot recover the injected signal with the vacuum template within our prior volume, and therefore we do not pursue model selection for this system. 
Note that for the case of superradiant clouds, the dephasing due to dynamical friction and energy transfer between the EMRI and the cloud could be combined with a direct search of the cloud's monochromatic GW emission for a consistency check on the presence of this environment.

\section{Discussion and Future Directions}

We implemented fully relativistic fast adiabatic EMRI waveforms, based on the Teukolsky formalism, for circular equatorial orbits in Kerr spacetime as an extension to the \texttt{FEW} package. This extension involves computing both Teukolsky-based flux data and Teukolsky-based amplitude data for the waveform evaluation. The waveforms are fast—between 10 and 100 ms—making them suitable for data analysis investigations.  Additionally, we publicly release waveform modules that include beyond-vacuum effects alongside this work at \url{https://github.com/Hassankh92/FastEMRIWaveforms_KerrCircNonvac}~\cite{KerrCircNonvac,KerrCircNonvac_Data}. The effects include: (1) power-law effects, such as migration torques, Eq.~\eqref{eq:MigTorques}, which offer flexibility for various applications, including those PN tests of general relativity, as demonstrated in Ref.~\cite{Speri}; and (2) a fully relativistic waveform accounting for interactions with relativistic superradiant scalar clouds. The module and accompanying dataset are available for broader research applications, ensuring the reproducibility of the results presented in this paper.

We observed that using non-relativistic amplitudes or neglecting higher harmonics individually introduces errors in the waveforms despite using a relativistic inspiral. For instance, the relativistic AAK model -- which applies the kludge model for the amplitude -- has mismatches up to $\sim 6\times 10^{-2}$ against the fully relativistic $(l,m) = (2, 2)$ mode alone (the green curve in Fig.~\ref{fig:mismatch_AAK_vs_Kerr}). Including higher harmonic modes further improves the waveform amplitudes. However, these improvements are only evident up to $l_{\text{max}}^{\mathcal{A}} = 5$. Beyond that, the mismatch curves between the waveforms stabilize (red and black curves in Fig.~\ref{fig:mismatch_AAK_vs_Kerr}). We should note that a larger mismatch here indicates greater improvement over the AAK amplitude model. Nevertheless, our model can include harmonics up to $l_{\text{max}}^{\mathcal{A}} = 30$ for circular orbits around Kerr, making it well-suited for future analysis that accounts for higher modes.

The implementation of fully relativistic waveforms allowed us to examine their impact on LISA's science objectives. In particular, we observed that for a typical EMRI with a 4-year inspiral time (Fig.~\ref{fig:waveforms}), the discrepancies in the predicted SNR can be as high as 20\% compared to approximate waveform models. Similarly, in Fig.~\ref{fig:snr_waterfall}, we show deviations as large as 35\% in the predicted EMRI source's horizon redshift compared to relativistic AAK waveforms.

Our studies of beyond-vacuum effects further underscore the importance of accurate relativistic amplitudes and higher harmonics in waveform templates. Specifically for migration torques in accretion disks, this led to a significant reduction in uncertainty from $\sim 24\%$ to $\sim 8\%$ in the disk's PE for an EMRI with $\mu = 50 M_\odot$ when compared to relativistic AAK waveforms (Fig.~\ref{fig:disk_vs_disk}). In contrast, for a lighter EMRI with $\mu = 10 M_\odot$, the effect was less pronounced due to the inspiral starting from a shorter orbital radius -- where the torques are weaker. Nevertheless, our analysis revealed at least $2\sigma$ deviations in the intrinsic parameters when the disk effect was ignored for $\mu = 10 M_\odot$. The Bayes factor, $\log_{10}\mathcal{B_{\rm DV}} = 1.1$, favored the disk effect, showing positive but modest support in this case.

We extended our study to another beyond-vacuum phenomenon: the presence of a superradiant scalar cloud surrounding the primary SMBH. For the first time, we incorporated this effect into EMRI waveforms within a fully relativistic framework, both for the cloud structure and its interaction with the binary. We investigated the potential of EMRI observations to probe scalar clouds in Fig.~\ref{fig:Min_Dephase_Cloud}. Through a full MCMC analysis, we further showed our ability to measure the cloud's properties, namely its total mass and the fundamental field's mass, with an error of $\sim 0.3\%$ and $\sim 0.002\%$, respectively (Fig.~\ref{fig:corner_Cloud})
\newline

While our primary focus has been on highlighting the need for the most accurate relativistic adiabatic templates in EMRI studies, potential systematic errors remain a concern, particularly those stemming from computational errors. It is crucial to exercise caution when discussing bias in PE, especially biases attributed to neglecting beyond-vacuum effects, as systematics could introduce larger biases~\cite{khalvati2024flux,Shubham_PhysRevD.110.084060}. Furthermore, we have only focused on adiabatic waveforms, which miss certain physical effects and are another source of systematic errors~\cite{Burke:2023lno}. 

In our analysis of the horizon redshift curves, we cover a wide range of binary masses and mass ratios in an EMRI system. While our model accommodates those, the validity of adiabatic waveforms in large mass ratio regimes needs to be addressed.

Additionally, in the beyond-vacuum model, it is important to note that the power-law behavior in the migration torque is an approximation, which may not fully capture the complexities of the underlying physics of the accretion disk. This limitation should be considered when interpreting results related to the impact of disk migration torques on EMRI waveforms. 
Regarding the scalar cloud model, our scalar flux calculations were performed in a Schwarzschild background. Extending these results to a Kerr background is not expected to drastically alter the overall effect (discussed in Sec.~\ref{sec:cloud}) .

Our effort in this work marks a step forward towards the generalization of fast EMRI waveforms, incorporating both fully relativistic wavefroms in Kerr spacetime and environmental effects.  The ultimate goal remains to generalize these waveforms to include generic orbits and broader environmental effects.

A key future direction in the waveform development is to address missing physics such as transient resonances in EMRI waveforms~\cite{Speri_2021,Gupta_2022,lynch2024fastinspiralstreatmentorbital}, followed by incorporating post-adiabatic effects~\cite{Burke:2023lno,Wardell_2023,van_de_Meent_2018}. The properties of the secondary, including its spin~\cite{Piovano_2021,lisa1Drummond_2022,lisa2_Drummond_2022,LISA_Scott_drummond2023extrememassratioinspiralwaveforms, Mathews_2022, piovano2024spinningparticlesnearkerr}
, must also be considered for more accurate predictions. These missing elements, along with the need to account for realistic astrophysical environmental effects, are essential for accurately performing parameter estimation and making predictions for any EMRI-related analyses with future space-based detectors.
\newline
\acknowledgments
We are deeply grateful to Scott Hughes for his invaluable guidance throughout this project, particularly on waveform modeling and the complexities of solving Teukolsky equation. His constant advice and insights have been instrumental in shaping this work. \\
H. K. acknowledges Alessandra Buonanno and the ACR group at the Max Planck Institute for Gravitational Physics (Albert Einstein Institute) for supporting his visit and providing an engaging research environment. H. K. also acknowledges Enrico Barausse and the Astroparticle Physics group at SISSA for supporting his visit and thanks Enrico for insightful discussions on the impact of accretion physics on EMRIs. \\
We thank Ollie Burke for his valuable advice on data analysis and Beatrice Bonga for stimulating discussions on the fundamentals of EMRIs. We acknowledge Michael Katz for his assistance with technical challenges related to the \texttt{FEW} package. We thank Rodrigo Vicente and Giovanni Maria Tomaselli for fruitful discussions on scalar clouds. \\
L. S. acknowledges the Perimeter Institute for Theoretical Physics for supporting his visit.
Research at Perimeter Institute is supported in part by the Government of Canada through the Department of Innovation, Science and Economic Development and by the Province of Ontario through the Ministry of Colleges and Universities.\\
Additionally, we performed our numerical simulations on the "Symmetry" HPC at the Perimeter Institute. Our MCMC runs have been performed on NVIDIA A100 GPUs through the ``Lakshmi'' facility at the Max Planck Institute for Gravitational Physics. \\
R.B. acknowledges financial support provided by FCT – Fundação para a Ciência e a Tecnologia, I.P., under the Scientific Employment Stimulus -- Individual Call -- Grant No. \href{https://doi.org/10.54499/2020.00470.CEECIND/CP1587/CT0010}{2020.00470.CEECIND}, the Project No.\href{https://doi.org/10.54499/2022.01324.PTDC}{2022.01324.PTDC}, and the Project ``GravNewFields'' funded under the ERC-Portugal program.

\bibliography{ref}

\begin{thebibliography}{134}%
\makeatletter
\providecommand \@ifxundefined [1]{%
 \@ifx{#1\undefined}
}%
\providecommand \@ifnum [1]{%
 \ifnum #1\expandafter \@firstoftwo
 \else \expandafter \@secondoftwo
 \fi
}%
\providecommand \@ifx [1]{%
 \ifx #1\expandafter \@firstoftwo
 \else \expandafter \@secondoftwo
 \fi
}%
\providecommand \natexlab [1]{#1}%
\providecommand \enquote  [1]{``#1''}%
\providecommand \bibnamefont  [1]{#1}%
\providecommand \bibfnamefont [1]{#1}%
\providecommand \citenamefont [1]{#1}%
\providecommand \href@noop [0]{\@secondoftwo}%
\providecommand \href [0]{\begingroup \@sanitize@url \@href}%
\providecommand \@href[1]{\@@startlink{#1}\@@href}%
\providecommand \@@href[1]{\endgroup#1\@@endlink}%
\providecommand \@sanitize@url [0]{\catcode `\\12\catcode `\$12\catcode
  `\&12\catcode `\#12\catcode `\^12\catcode `\_12\catcode `\%12\relax}%
\providecommand \@@startlink[1]{}%
\providecommand \@@endlink[0]{}%
\providecommand \url  [0]{\begingroup\@sanitize@url \@url }%
\providecommand \@url [1]{\endgroup\@href {#1}{\urlprefix }}%
\providecommand \urlprefix  [0]{URL }%
\providecommand \Eprint [0]{\href }%
\providecommand \doibase [0]{https://doi.org/}%
\providecommand \selectlanguage [0]{\@gobble}%
\providecommand \bibinfo  [0]{\@secondoftwo}%
\providecommand \bibfield  [0]{\@secondoftwo}%
\providecommand \translation [1]{[#1]}%
\providecommand \BibitemOpen [0]{}%
\providecommand \bibitemStop [0]{}%
\providecommand \bibitemNoStop [0]{.\EOS\space}%
\providecommand \EOS [0]{\spacefactor3000\relax}%
\providecommand \BibitemShut  [1]{\csname bibitem#1\endcsname}%
\let\auto@bib@innerbib\@empty
\bibitem [{\citenamefont {Drasco}\ and\ \citenamefont
  {Hughes}(2006{\natexlab{a}})}]{Hughes_generic_PhysRevD.73.024027}%
  \BibitemOpen
  \bibfield  {author} {\bibinfo {author} {\bibfnamefont {S.}~\bibnamefont
  {Drasco}}\ and\ \bibinfo {author} {\bibfnamefont {S.~A.}\ \bibnamefont
  {Hughes}},\ }\bibfield  {title} {\bibinfo {title} {Gravitational wave
  snapshots of generic extreme mass ratio inspirals},\ }\href
  {https://doi.org/10.1103/PhysRevD.73.024027} {\bibfield  {journal} {\bibinfo
  {journal} {Phys. Rev. D}\ }\textbf {\bibinfo {volume} {73}},\ \bibinfo
  {pages} {024027} (\bibinfo {year} {2006}{\natexlab{a}})}\BibitemShut
  {NoStop}%
\bibitem [{\citenamefont {Gair}\ \emph {et~al.}(2017)\citenamefont {Gair},
  \citenamefont {Babak}, \citenamefont {Sesana}, \citenamefont {Amaro-Seoane},
  \citenamefont {Barausse}, \citenamefont {Berry}, \citenamefont {Berti},\ and\
  \citenamefont {Sopuerta}}]{Gair_2017}%
  \BibitemOpen
  \bibfield  {author} {\bibinfo {author} {\bibfnamefont {J.~R.}\ \bibnamefont
  {Gair}}, \bibinfo {author} {\bibfnamefont {S.}~\bibnamefont {Babak}},
  \bibinfo {author} {\bibfnamefont {A.}~\bibnamefont {Sesana}}, \bibinfo
  {author} {\bibfnamefont {P.}~\bibnamefont {Amaro-Seoane}}, \bibinfo {author}
  {\bibfnamefont {E.}~\bibnamefont {Barausse}}, \bibinfo {author}
  {\bibfnamefont {C.~P.~L.}\ \bibnamefont {Berry}}, \bibinfo {author}
  {\bibfnamefont {E.}~\bibnamefont {Berti}},\ and\ \bibinfo {author}
  {\bibfnamefont {C.}~\bibnamefont {Sopuerta}},\ }\bibfield  {title} {\bibinfo
  {title} {Prospects for observing extreme-mass-ratio inspirals with lisa},\
  }\href {https://doi.org/10.1088/1742-6596/840/1/012021} {\bibfield  {journal}
  {\bibinfo  {journal} {Journal of Physics: Conference Series}\ }\textbf
  {\bibinfo {volume} {840}},\ \bibinfo {pages} {012021} (\bibinfo {year}
  {2017})}\BibitemShut {NoStop}%
\bibitem [{\citenamefont {Colpi}\ \emph {et~al.}(2024)\citenamefont {Colpi}
  \emph {et~al.}}]{redbook_colpi2024lisadefinitionstudyreport}%
  \BibitemOpen
  \bibfield  {author} {\bibinfo {author} {\bibfnamefont {M.}~\bibnamefont
  {Colpi}} \emph {et~al.},\ }\href {https://doi.org/10.48550/arXiv.2402.07571}
  {\bibinfo {title} {Lisa definition study report}} (\bibinfo {year} {2024}),\
  \Eprint {https://arxiv.org/abs/2402.07571} {arXiv:2402.07571 [astro-ph.CO]}
  \BibitemShut {NoStop}%
\bibitem [{\citenamefont {Babak}\ \emph {et~al.}(2017)\citenamefont {Babak},
  \citenamefont {Gair}, \citenamefont {Sesana}, \citenamefont {Barausse},
  \citenamefont {Sopuerta}, \citenamefont {Berry}, \citenamefont {Berti},
  \citenamefont {Amaro-Seoane}, \citenamefont {Petiteau},\ and\ \citenamefont
  {Klein}}]{Babak:2017tow}%
  \BibitemOpen
  \bibfield  {author} {\bibinfo {author} {\bibfnamefont {S.}~\bibnamefont
  {Babak}}, \bibinfo {author} {\bibfnamefont {J.}~\bibnamefont {Gair}},
  \bibinfo {author} {\bibfnamefont {A.}~\bibnamefont {Sesana}}, \bibinfo
  {author} {\bibfnamefont {E.}~\bibnamefont {Barausse}}, \bibinfo {author}
  {\bibfnamefont {C.~F.}\ \bibnamefont {Sopuerta}}, \bibinfo {author}
  {\bibfnamefont {C.~P.~L.}\ \bibnamefont {Berry}}, \bibinfo {author}
  {\bibfnamefont {E.}~\bibnamefont {Berti}}, \bibinfo {author} {\bibfnamefont
  {P.}~\bibnamefont {Amaro-Seoane}}, \bibinfo {author} {\bibfnamefont
  {A.}~\bibnamefont {Petiteau}},\ and\ \bibinfo {author} {\bibfnamefont
  {A.}~\bibnamefont {Klein}},\ }\bibfield  {title} {\bibinfo {title} {{Science
  with the space-based interferometer LISA. V: Extreme mass-ratio inspirals}},\
  }\href {https://doi.org/10.1103/PhysRevD.95.103012} {\bibfield  {journal}
  {\bibinfo  {journal} {Phys. Rev. D}\ }\textbf {\bibinfo {volume} {95}},\
  \bibinfo {pages} {103012} (\bibinfo {year} {2017})},\ \Eprint
  {https://arxiv.org/abs/1703.09722} {arXiv:1703.09722 [gr-qc]} \BibitemShut
  {NoStop}%
\bibitem [{\citenamefont {Pan}\ and\ \citenamefont {Yang}(2021)}]{Pan_2021}%
  \BibitemOpen
  \bibfield  {author} {\bibinfo {author} {\bibfnamefont {Z.}~\bibnamefont
  {Pan}}\ and\ \bibinfo {author} {\bibfnamefont {H.}~\bibnamefont {Yang}},\
  }\bibfield  {title} {\bibinfo {title} {Formation rate of extreme mass ratio
  inspirals in active galactic nuclei},\ }\bibfield  {journal} {\bibinfo
  {journal} {Physical Review D}\ }\textbf {\bibinfo {volume} {103}},\ \href
  {https://doi.org/10.1103/physrevd.103.103018} {10.1103/physrevd.103.103018}
  (\bibinfo {year} {2021})\BibitemShut {NoStop}%
\bibitem [{\citenamefont {Pan}\ \emph {et~al.}(2021)\citenamefont {Pan},
  \citenamefont {Lyu},\ and\ \citenamefont {Yang}}]{Pan2_2021}%
  \BibitemOpen
  \bibfield  {author} {\bibinfo {author} {\bibfnamefont {Z.}~\bibnamefont
  {Pan}}, \bibinfo {author} {\bibfnamefont {Z.}~\bibnamefont {Lyu}},\ and\
  \bibinfo {author} {\bibfnamefont {H.}~\bibnamefont {Yang}},\ }\bibfield
  {title} {\bibinfo {title} {Wet extreme mass ratio inspirals may be more
  common for spaceborne gravitational wave detection},\ }\bibfield  {journal}
  {\bibinfo  {journal} {Physical Review D}\ }\textbf {\bibinfo {volume}
  {104}},\ \href {https://doi.org/10.1103/physrevd.104.063007}
  {10.1103/physrevd.104.063007} (\bibinfo {year} {2021})\BibitemShut {NoStop}%
\bibitem [{\citenamefont {Broggi}\ \emph {et~al.}(2022)\citenamefont {Broggi},
  \citenamefont {Bortolas}, \citenamefont {Bonetti}, \citenamefont {Sesana},\
  and\ \citenamefont {Dotti}}]{Broggi:2022udp}%
  \BibitemOpen
  \bibfield  {author} {\bibinfo {author} {\bibfnamefont {L.}~\bibnamefont
  {Broggi}}, \bibinfo {author} {\bibfnamefont {E.}~\bibnamefont {Bortolas}},
  \bibinfo {author} {\bibfnamefont {M.}~\bibnamefont {Bonetti}}, \bibinfo
  {author} {\bibfnamefont {A.}~\bibnamefont {Sesana}},\ and\ \bibinfo {author}
  {\bibfnamefont {M.}~\bibnamefont {Dotti}},\ }\bibfield  {title} {\bibinfo
  {title} {{Extreme mass ratio inspirals and tidal disruption events in nuclear
  clusters \textendash{} I. Time-dependent rates}},\ }\href
  {https://doi.org/10.1093/mnras/stac1453} {\bibfield  {journal} {\bibinfo
  {journal} {Mon. Not. Roy. Astron. Soc.}\ }\textbf {\bibinfo {volume} {514}},\
  \bibinfo {pages} {3270} (\bibinfo {year} {2022})},\ \Eprint
  {https://arxiv.org/abs/2205.06277} {arXiv:2205.06277 [astro-ph.GA]}
  \BibitemShut {NoStop}%
\bibitem [{\citenamefont {Barack}\ and\ \citenamefont
  {Cutler}(2007)}]{Barack_2007}%
  \BibitemOpen
  \bibfield  {author} {\bibinfo {author} {\bibfnamefont {L.}~\bibnamefont
  {Barack}}\ and\ \bibinfo {author} {\bibfnamefont {C.}~\bibnamefont
  {Cutler}},\ }\bibfield  {title} {\bibinfo {title} {Using lisa
  extreme-mass-ratio inspiral sources to test off-kerr deviations in the
  geometry of massive black holes},\ }\bibfield  {journal} {\bibinfo  {journal}
  {Physical Review D}\ }\textbf {\bibinfo {volume} {75}},\ \href
  {https://doi.org/10.1103/physrevd.75.042003} {10.1103/physrevd.75.042003}
  (\bibinfo {year} {2007})\BibitemShut {NoStop}%
\bibitem [{\citenamefont {Tahura}\ \emph {et~al.}(2024)\citenamefont {Tahura},
  \citenamefont {Khalvati},\ and\ \citenamefont {Yang}}]{Tahura_2024}%
  \BibitemOpen
  \bibfield  {author} {\bibinfo {author} {\bibfnamefont {S.}~\bibnamefont
  {Tahura}}, \bibinfo {author} {\bibfnamefont {H.}~\bibnamefont {Khalvati}},\
  and\ \bibinfo {author} {\bibfnamefont {H.}~\bibnamefont {Yang}},\ }\bibfield
  {title} {\bibinfo {title} {Vacuum spacetime with multipole moments: The
  minimal size conjecture, black hole shadow, and gravitational wave
  observables},\ }\bibfield  {journal} {\bibinfo  {journal} {Physical Review
  D}\ }\textbf {\bibinfo {volume} {109}},\ \href
  {https://doi.org/10.1103/physrevd.109.124025} {10.1103/physrevd.109.124025}
  (\bibinfo {year} {2024})\BibitemShut {NoStop}%
\bibitem [{\citenamefont {Barausse}\ \emph {et~al.}(2015)\citenamefont
  {Barausse}, \citenamefont {Cardoso},\ and\ \citenamefont
  {Pani}}]{Barausse_2015}%
  \BibitemOpen
  \bibfield  {author} {\bibinfo {author} {\bibfnamefont {E.}~\bibnamefont
  {Barausse}}, \bibinfo {author} {\bibfnamefont {V.}~\bibnamefont {Cardoso}},\
  and\ \bibinfo {author} {\bibfnamefont {P.}~\bibnamefont {Pani}},\ }\bibfield
  {title} {\bibinfo {title} {Environmental effects for gravitational-wave
  astrophysics},\ }\href {https://doi.org/10.1088/1742-6596/610/1/012044}
  {\bibfield  {journal} {\bibinfo  {journal} {Journal of Physics: Conference
  Series}\ }\textbf {\bibinfo {volume} {610}},\ \bibinfo {pages} {012044}
  (\bibinfo {year} {2015})}\BibitemShut {NoStop}%
\bibitem [{\citenamefont {Kocsis}\ \emph {et~al.}(2011)\citenamefont {Kocsis},
  \citenamefont {Yunes},\ and\ \citenamefont {Loeb}}]{Kocsis_2011}%
  \BibitemOpen
  \bibfield  {author} {\bibinfo {author} {\bibfnamefont {B.}~\bibnamefont
  {Kocsis}}, \bibinfo {author} {\bibfnamefont {N.}~\bibnamefont {Yunes}},\ and\
  \bibinfo {author} {\bibfnamefont {A.}~\bibnamefont {Loeb}},\ }\bibfield
  {title} {\bibinfo {title} {Observable signatures of extreme mass-ratio
  inspiral black hole binaries embedded in thin accretion disks},\ }\bibfield
  {journal} {\bibinfo  {journal} {Physical Review D}\ }\textbf {\bibinfo
  {volume} {84}},\ \href {https://doi.org/10.1103/physrevd.84.024032}
  {10.1103/physrevd.84.024032} (\bibinfo {year} {2011})\BibitemShut {NoStop}%
\bibitem [{\citenamefont {Levin}(2003)}]{levin2003formationmassivestarsblack}%
  \BibitemOpen
  \bibfield  {author} {\bibinfo {author} {\bibfnamefont {Y.}~\bibnamefont
  {Levin}},\ }\href {https://arxiv.org/abs/astro-ph/0307084} {\bibinfo {title}
  {Formation of massive stars and black holes in self-gravitating agn discs,
  and gravitational waves in lisa band}} (\bibinfo {year} {2003}),\ \Eprint
  {https://arxiv.org/abs/astro-ph/0307084} {arXiv:astro-ph/0307084 [astro-ph]}
  \BibitemShut {NoStop}%
\bibitem [{\citenamefont {Gair}\ \emph {et~al.}(2013)\citenamefont {Gair},
  \citenamefont {Vallisneri}, \citenamefont {Larson},\ and\ \citenamefont
  {Baker}}]{Gair_2013}%
  \BibitemOpen
  \bibfield  {author} {\bibinfo {author} {\bibfnamefont {J.~R.}\ \bibnamefont
  {Gair}}, \bibinfo {author} {\bibfnamefont {M.}~\bibnamefont {Vallisneri}},
  \bibinfo {author} {\bibfnamefont {S.~L.}\ \bibnamefont {Larson}},\ and\
  \bibinfo {author} {\bibfnamefont {J.~G.}\ \bibnamefont {Baker}},\ }\bibfield
  {title} {\bibinfo {title} {Testing general relativity with low-frequency,
  space-based gravitational-wave detectors},\ }\bibfield  {journal} {\bibinfo
  {journal} {Living Reviews in Relativity}\ }\textbf {\bibinfo {volume} {16}},\
  \href {https://doi.org/10.12942/lrr-2013-7} {10.12942/lrr-2013-7} (\bibinfo
  {year} {2013})\BibitemShut {NoStop}%
\bibitem [{\citenamefont {Tahura}\ \emph {et~al.}(2022)\citenamefont {Tahura},
  \citenamefont {Pan},\ and\ \citenamefont {Yang}}]{Tahura_2022}%
  \BibitemOpen
  \bibfield  {author} {\bibinfo {author} {\bibfnamefont {S.}~\bibnamefont
  {Tahura}}, \bibinfo {author} {\bibfnamefont {Z.}~\bibnamefont {Pan}},\ and\
  \bibinfo {author} {\bibfnamefont {H.}~\bibnamefont {Yang}},\ }\bibfield
  {title} {\bibinfo {title} {Science potential for stellar-mass black holes as
  neighbors of sgr a*},\ }\bibfield  {journal} {\bibinfo  {journal} {Physical
  Review D}\ }\textbf {\bibinfo {volume} {105}},\ \href
  {https://doi.org/10.1103/physrevd.105.123018} {10.1103/physrevd.105.123018}
  (\bibinfo {year} {2022})\BibitemShut {NoStop}%
\bibitem [{\citenamefont {{Speri}}\ \emph {et~al.}(2024)\citenamefont
  {{Speri}}, \citenamefont {{Barsanti}}, \citenamefont {{Maselli}},
  \citenamefont {{Sotiriou}}, \citenamefont {{Warburton}}, \citenamefont {{van
  de Meent}}, \citenamefont {{Chua}}, \citenamefont {{Burke}},\ and\
  \citenamefont {{Gair}}}]{Speri_fundamental_2024arXiv240607607S}%
  \BibitemOpen
  \bibfield  {author} {\bibinfo {author} {\bibfnamefont {L.}~\bibnamefont
  {{Speri}}}, \bibinfo {author} {\bibfnamefont {S.}~\bibnamefont {{Barsanti}}},
  \bibinfo {author} {\bibfnamefont {A.}~\bibnamefont {{Maselli}}}, \bibinfo
  {author} {\bibfnamefont {T.~P.}\ \bibnamefont {{Sotiriou}}}, \bibinfo
  {author} {\bibfnamefont {N.}~\bibnamefont {{Warburton}}}, \bibinfo {author}
  {\bibfnamefont {M.}~\bibnamefont {{van de Meent}}}, \bibinfo {author}
  {\bibfnamefont {A.~J.~K.}\ \bibnamefont {{Chua}}}, \bibinfo {author}
  {\bibfnamefont {O.}~\bibnamefont {{Burke}}},\ and\ \bibinfo {author}
  {\bibfnamefont {J.}~\bibnamefont {{Gair}}},\ }\bibfield  {title} {\bibinfo
  {title} {{Probing fundamental physics with Extreme Mass Ratio Inspirals: a
  full Bayesian inference for scalar charge}},\ }\href
  {https://doi.org/10.48550/arXiv.2406.07607} {\bibfield  {journal} {\bibinfo
  {journal} {arXiv e-prints}\ ,\ \bibinfo {eid} {arXiv:2406.07607}} (\bibinfo
  {year} {2024})},\ \Eprint {https://arxiv.org/abs/2406.07607}
  {arXiv:2406.07607 [gr-qc]} \BibitemShut {NoStop}%
\bibitem [{\citenamefont {Khalvati}\ \emph {et~al.}(2024)\citenamefont
  {Khalvati}, \citenamefont {Lynch}, \citenamefont {Burke}, \citenamefont
  {Meent}, \citenamefont {Speri},\ and\ \citenamefont
  {Nasipak}}]{khalvati2024flux}%
  \BibitemOpen
  \bibfield  {author} {\bibinfo {author} {\bibfnamefont {H.}~\bibnamefont
  {Khalvati}}, \bibinfo {author} {\bibfnamefont {P.}~\bibnamefont {Lynch}},
  \bibinfo {author} {\bibfnamefont {O.}~\bibnamefont {Burke}}, \bibinfo
  {author} {\bibfnamefont {M.~V.~D.}\ \bibnamefont {Meent}}, \bibinfo {author}
  {\bibfnamefont {L.}~\bibnamefont {Speri}},\ and\ \bibinfo {author}
  {\bibfnamefont {Z.}~\bibnamefont {Nasipak}},\ }\bibfield  {title} {\bibinfo
  {title} {Systematics in adiabatic relativistic waveforms for extreme mass
  ratio inspirals}} (\bibinfo {year} {2024}),\ \bibinfo {note} {manuscript in
  preparation}\BibitemShut {NoStop}%
\bibitem [{\citenamefont {Burke}\ \emph {et~al.}(2024)\citenamefont {Burke},
  \citenamefont {Piovano}, \citenamefont {Warburton}, \citenamefont {Lynch},
  \citenamefont {Speri}, \citenamefont {Kavanagh}, \citenamefont {Wardell},
  \citenamefont {Pound}, \citenamefont {Durkan},\ and\ \citenamefont
  {Miller}}]{Burke:2023lno}%
  \BibitemOpen
  \bibfield  {author} {\bibinfo {author} {\bibfnamefont {O.}~\bibnamefont
  {Burke}}, \bibinfo {author} {\bibfnamefont {G.~A.}\ \bibnamefont {Piovano}},
  \bibinfo {author} {\bibfnamefont {N.}~\bibnamefont {Warburton}}, \bibinfo
  {author} {\bibfnamefont {P.}~\bibnamefont {Lynch}}, \bibinfo {author}
  {\bibfnamefont {L.}~\bibnamefont {Speri}}, \bibinfo {author} {\bibfnamefont
  {C.}~\bibnamefont {Kavanagh}}, \bibinfo {author} {\bibfnamefont
  {B.}~\bibnamefont {Wardell}}, \bibinfo {author} {\bibfnamefont
  {A.}~\bibnamefont {Pound}}, \bibinfo {author} {\bibfnamefont
  {L.}~\bibnamefont {Durkan}},\ and\ \bibinfo {author} {\bibfnamefont
  {J.}~\bibnamefont {Miller}},\ }\bibfield  {title} {\bibinfo {title}
  {{Assessing the importance of first postadiabatic terms for small-mass-ratio
  binaries}},\ }\href {https://doi.org/10.1103/PhysRevD.109.124048} {\bibfield
  {journal} {\bibinfo  {journal} {Phys. Rev. D}\ }\textbf {\bibinfo {volume}
  {109}},\ \bibinfo {pages} {124048} (\bibinfo {year} {2024})},\ \Eprint
  {https://arxiv.org/abs/2310.08927} {arXiv:2310.08927 [gr-qc]} \BibitemShut
  {NoStop}%
\bibitem [{\citenamefont {Gupta}\ \emph {et~al.}(2024)\citenamefont {Gupta}
  \emph {et~al.}}]{Gupta:2024gun}%
  \BibitemOpen
  \bibfield  {author} {\bibinfo {author} {\bibfnamefont {A.}~\bibnamefont
  {Gupta}} \emph {et~al.},\ }\href@noop {} {\bibinfo {title} {{Possible Causes
  of False General Relativity Violations in Gravitational Wave Observations}}}
  (\bibinfo {year} {2024}),\ \Eprint {https://arxiv.org/abs/2405.02197}
  {arXiv:2405.02197 [gr-qc]} \BibitemShut {NoStop}%
\bibitem [{\citenamefont {Chua}\ and\ \citenamefont
  {Cutler}(2022)}]{Chua:2021aah}%
  \BibitemOpen
  \bibfield  {author} {\bibinfo {author} {\bibfnamefont {A.~J.~K.}\
  \bibnamefont {Chua}}\ and\ \bibinfo {author} {\bibfnamefont {C.~J.}\
  \bibnamefont {Cutler}},\ }\bibfield  {title} {\bibinfo {title} {{Nonlocal
  parameter degeneracy in the intrinsic space of gravitational-wave signals
  from extreme-mass-ratio inspirals}},\ }\href
  {https://doi.org/10.1103/PhysRevD.106.124046} {\bibfield  {journal} {\bibinfo
   {journal} {Phys. Rev. D}\ }\textbf {\bibinfo {volume} {106}},\ \bibinfo
  {pages} {124046} (\bibinfo {year} {2022})},\ \Eprint
  {https://arxiv.org/abs/2109.14254} {arXiv:2109.14254 [gr-qc]} \BibitemShut
  {NoStop}%
\bibitem [{\citenamefont {Speri}\ \emph {et~al.}(2024)\citenamefont {Speri},
  \citenamefont {Katz}, \citenamefont {Chua}, \citenamefont {Hughes},
  \citenamefont {Warburton}, \citenamefont {Thompson}, \citenamefont
  {Chapman-Bird},\ and\ \citenamefont {Gair}}]{Speri_2024}%
  \BibitemOpen
  \bibfield  {author} {\bibinfo {author} {\bibfnamefont {L.}~\bibnamefont
  {Speri}}, \bibinfo {author} {\bibfnamefont {M.~L.}\ \bibnamefont {Katz}},
  \bibinfo {author} {\bibfnamefont {A.~J.~K.}\ \bibnamefont {Chua}}, \bibinfo
  {author} {\bibfnamefont {S.~A.}\ \bibnamefont {Hughes}}, \bibinfo {author}
  {\bibfnamefont {N.}~\bibnamefont {Warburton}}, \bibinfo {author}
  {\bibfnamefont {J.~E.}\ \bibnamefont {Thompson}}, \bibinfo {author}
  {\bibfnamefont {C.~E.~A.}\ \bibnamefont {Chapman-Bird}},\ and\ \bibinfo
  {author} {\bibfnamefont {J.~R.}\ \bibnamefont {Gair}},\ }\bibfield  {title}
  {\bibinfo {title} {Fast and fourier: extreme mass ratio inspiral waveforms in
  the frequency domain},\ }\bibfield  {journal} {\bibinfo  {journal} {Frontiers
  in Applied Mathematics and Statistics}\ }\textbf {\bibinfo {volume} {9}},\
  \href {https://doi.org/10.3389/fams.2023.1266739} {10.3389/fams.2023.1266739}
  (\bibinfo {year} {2024})\BibitemShut {NoStop}%
\bibitem [{\citenamefont {Nasipak}(2024)}]{Nasipak_2024}%
  \BibitemOpen
  \bibfield  {author} {\bibinfo {author} {\bibfnamefont {Z.}~\bibnamefont
  {Nasipak}},\ }\bibfield  {title} {\bibinfo {title} {Adiabatic gravitational
  waveform model for compact objects undergoing quasicircular inspirals into
  rotating massive black holes},\ }\bibfield  {journal} {\bibinfo  {journal}
  {Physical Review D}\ }\textbf {\bibinfo {volume} {109}},\ \href
  {https://doi.org/10.1103/physrevd.109.044020} {10.1103/physrevd.109.044020}
  (\bibinfo {year} {2024})\BibitemShut {NoStop}%
\bibitem [{\citenamefont {Afshordi}\ \emph {et~al.}(2023)\citenamefont
  {Afshordi} \emph
  {et~al.}}]{lisaconsortiumwaveformworkinggroup2023waveformmodellinglaserinterferometer}%
  \BibitemOpen
  \bibfield  {author} {\bibinfo {author} {\bibfnamefont {N.}~\bibnamefont
  {Afshordi}} \emph {et~al.},\ }\href {https://arxiv.org/abs/2311.01300}
  {\bibinfo {title} {Waveform modelling for the laser interferometer space
  antenna}} (\bibinfo {year} {2023}),\ \Eprint
  {https://arxiv.org/abs/2311.01300} {arXiv:2311.01300 [gr-qc]} \BibitemShut
  {NoStop}%
\bibitem [{\citenamefont {Lynch}\ \emph {et~al.}(2024)\citenamefont {Lynch},
  \citenamefont {Witzany}, \citenamefont {van~de Meent},\ and\ \citenamefont
  {Warburton}}]{lynch2024fastinspiralstreatmentorbital}%
  \BibitemOpen
  \bibfield  {author} {\bibinfo {author} {\bibfnamefont {P.}~\bibnamefont
  {Lynch}}, \bibinfo {author} {\bibfnamefont {V.}~\bibnamefont {Witzany}},
  \bibinfo {author} {\bibfnamefont {M.}~\bibnamefont {van~de Meent}},\ and\
  \bibinfo {author} {\bibfnamefont {N.}~\bibnamefont {Warburton}},\ }\href
  {https://arxiv.org/abs/2405.21072} {\bibinfo {title} {Fast inspirals and the
  treatment of orbital resonances}} (\bibinfo {year} {2024}),\ \Eprint
  {https://arxiv.org/abs/2405.21072} {arXiv:2405.21072 [gr-qc]} \BibitemShut
  {NoStop}%
\bibitem [{\citenamefont {Drummond}\ \emph
  {et~al.}(2023{\natexlab{a}})\citenamefont {Drummond}, \citenamefont {Lynch},
  \citenamefont {Hanselman}, \citenamefont {Becker},\ and\ \citenamefont
  {Hughes}}]{drummond2023extrememassratioinspiralwaveforms}%
  \BibitemOpen
  \bibfield  {author} {\bibinfo {author} {\bibfnamefont {L.~V.}\ \bibnamefont
  {Drummond}}, \bibinfo {author} {\bibfnamefont {P.}~\bibnamefont {Lynch}},
  \bibinfo {author} {\bibfnamefont {A.~G.}\ \bibnamefont {Hanselman}}, \bibinfo
  {author} {\bibfnamefont {D.~R.}\ \bibnamefont {Becker}},\ and\ \bibinfo
  {author} {\bibfnamefont {S.~A.}\ \bibnamefont {Hughes}},\ }\href
  {https://arxiv.org/abs/2310.08438} {\bibinfo {title} {Extreme mass-ratio
  inspiral and waveforms for a spinning body into a kerr black hole via
  osculating geodesics and near-identity transformations}} (\bibinfo {year}
  {2023}{\natexlab{a}}),\ \Eprint {https://arxiv.org/abs/2310.08438}
  {arXiv:2310.08438 [gr-qc]} \BibitemShut {NoStop}%
\bibitem [{\citenamefont {Rink}\ \emph {et~al.}(2024)\citenamefont {Rink},
  \citenamefont {Bachhar}, \citenamefont {Islam}, \citenamefont {Rifat},
  \citenamefont {Gonzalez-Quesada}, \citenamefont {Field}, \citenamefont
  {Khanna}, \citenamefont {Hughes},\ and\ \citenamefont
  {Varma}}]{rink2024gravitationalwavesurrogatemodel}%
  \BibitemOpen
  \bibfield  {author} {\bibinfo {author} {\bibfnamefont {K.}~\bibnamefont
  {Rink}}, \bibinfo {author} {\bibfnamefont {R.}~\bibnamefont {Bachhar}},
  \bibinfo {author} {\bibfnamefont {T.}~\bibnamefont {Islam}}, \bibinfo
  {author} {\bibfnamefont {N.~E.~M.}\ \bibnamefont {Rifat}}, \bibinfo {author}
  {\bibfnamefont {K.}~\bibnamefont {Gonzalez-Quesada}}, \bibinfo {author}
  {\bibfnamefont {S.~E.}\ \bibnamefont {Field}}, \bibinfo {author}
  {\bibfnamefont {G.}~\bibnamefont {Khanna}}, \bibinfo {author} {\bibfnamefont
  {S.~A.}\ \bibnamefont {Hughes}},\ and\ \bibinfo {author} {\bibfnamefont
  {V.}~\bibnamefont {Varma}},\ }\href {https://arxiv.org/abs/2407.18319}
  {\bibinfo {title} {Gravitational wave surrogate model for spinning,
  intermediate mass ratio binaries based on perturbation theory and numerical
  relativity}} (\bibinfo {year} {2024}),\ \Eprint
  {https://arxiv.org/abs/2407.18319} {arXiv:2407.18319 [gr-qc]} \BibitemShut
  {NoStop}%
\bibitem [{\citenamefont {Katz}\ \emph {et~al.}(2021)\citenamefont {Katz},
  \citenamefont {Chua}, \citenamefont {Speri}, \citenamefont {Warburton},\ and\
  \citenamefont {Hughes}}]{Katz_2021}%
  \BibitemOpen
  \bibfield  {author} {\bibinfo {author} {\bibfnamefont {M.~L.}\ \bibnamefont
  {Katz}}, \bibinfo {author} {\bibfnamefont {A.~J.~K.}\ \bibnamefont {Chua}},
  \bibinfo {author} {\bibfnamefont {L.}~\bibnamefont {Speri}}, \bibinfo
  {author} {\bibfnamefont {N.}~\bibnamefont {Warburton}},\ and\ \bibinfo
  {author} {\bibfnamefont {S.~A.}\ \bibnamefont {Hughes}},\ }\bibfield  {title}
  {\bibinfo {title} {Fast extreme-mass-ratio-inspiral waveforms: New tools for
  millihertz gravitational-wave data analysis},\ }\bibfield  {journal}
  {\bibinfo  {journal} {Physical Review D}\ }\textbf {\bibinfo {volume}
  {104}},\ \href {https://doi.org/10.1103/physrevd.104.064047}
  {10.1103/physrevd.104.064047} (\bibinfo {year} {2021})\BibitemShut {NoStop}%
\bibitem [{\citenamefont {Chua}\ \emph {et~al.}(2021)\citenamefont {Chua},
  \citenamefont {Katz}, \citenamefont {Warburton},\ and\ \citenamefont
  {Hughes}}]{few_Chua_2021}%
  \BibitemOpen
  \bibfield  {author} {\bibinfo {author} {\bibfnamefont {A.~J.}\ \bibnamefont
  {Chua}}, \bibinfo {author} {\bibfnamefont {M.~L.}\ \bibnamefont {Katz}},
  \bibinfo {author} {\bibfnamefont {N.}~\bibnamefont {Warburton}},\ and\
  \bibinfo {author} {\bibfnamefont {S.~A.}\ \bibnamefont {Hughes}},\ }\bibfield
   {title} {\bibinfo {title} {Rapid generation of fully relativistic
  extreme-mass-ratio-inspiral waveform templates for lisa data analysis},\
  }\bibfield  {journal} {\bibinfo  {journal} {Physical Review Letters}\
  }\textbf {\bibinfo {volume} {126}},\ \href
  {https://doi.org/10.1103/physrevlett.126.051102}
  {10.1103/physrevlett.126.051102} (\bibinfo {year} {2021})\BibitemShut
  {NoStop}%
\bibitem [{\citenamefont {Pound}\ and\ \citenamefont
  {Wardell}(2021)}]{Pound:2021qin}%
  \BibitemOpen
  \bibfield  {author} {\bibinfo {author} {\bibfnamefont {A.}~\bibnamefont
  {Pound}}\ and\ \bibinfo {author} {\bibfnamefont {B.}~\bibnamefont
  {Wardell}},\ }\href {https://doi.org/10.1007/978-981-15-4702-7\_38-1}
  {\bibinfo {title} {{Black hole perturbation theory and gravitational
  self-force}}} (\bibinfo {year} {2021}),\ \Eprint
  {https://arxiv.org/abs/2101.04592} {arXiv:2101.04592 [gr-qc]} \BibitemShut
  {NoStop}%
\bibitem [{\citenamefont {Speri}\ \emph {et~al.}(2022)\citenamefont {Speri},
  \citenamefont {Antonelli}, \citenamefont {Sberna}, \citenamefont {Babak},
  \citenamefont {Barausse}, \citenamefont {Gair},\ and\ \citenamefont
  {Katz}}]{Speri}%
  \BibitemOpen
  \bibfield  {author} {\bibinfo {author} {\bibfnamefont {L.}~\bibnamefont
  {Speri}}, \bibinfo {author} {\bibfnamefont {A.}~\bibnamefont {Antonelli}},
  \bibinfo {author} {\bibfnamefont {L.}~\bibnamefont {Sberna}}, \bibinfo
  {author} {\bibfnamefont {S.}~\bibnamefont {Babak}}, \bibinfo {author}
  {\bibfnamefont {E.}~\bibnamefont {Barausse}}, \bibinfo {author}
  {\bibfnamefont {J.~R.}\ \bibnamefont {Gair}},\ and\ \bibinfo {author}
  {\bibfnamefont {M.~L.}\ \bibnamefont {Katz}},\ }\href
  {https://doi.org/10.48550/ARXIV.2207.10086} {\bibinfo {title} {Measuring
  accretion-disk effects with gravitational waves from extreme mass ratio
  inspirals}} (\bibinfo {year} {2022})\BibitemShut {NoStop}%
\bibitem [{\citenamefont {Brito}\ and\ \citenamefont
  {Shah}(2023)}]{Brito_2023}%
  \BibitemOpen
  \bibfield  {author} {\bibinfo {author} {\bibfnamefont {R.}~\bibnamefont
  {Brito}}\ and\ \bibinfo {author} {\bibfnamefont {S.}~\bibnamefont {Shah}},\
  }\bibfield  {title} {\bibinfo {title} {Extreme mass-ratio inspirals into
  black holes surrounded by scalar clouds},\ }\bibfield  {journal} {\bibinfo
  {journal} {Physical Review D}\ }\textbf {\bibinfo {volume} {108}},\ \href
  {https://doi.org/10.1103/physrevd.108.084019} {10.1103/physrevd.108.084019}
  (\bibinfo {year} {2023})\BibitemShut {NoStop}%
\bibitem [{Ker(2024)}]{KerrCircNonvac}%
  \BibitemOpen
  \href@noop {} {\bibinfo {title} {{FastEMRIWaveforms: Relativistic Kerr
  Circular Non-Vacuum Extension}}},\ \bibinfo {howpublished}
  {\url{https://github.com/Hassankh92/FastEMRIWaveforms_KerrCircNonvac}}
  (\bibinfo {year} {2024}),\ \bibinfo {note} {gitHub repository}\BibitemShut
  {NoStop}%
\bibitem [{\citenamefont {Khalvati}(2024)}]{KerrCircNonvac_Data}%
  \BibitemOpen
  \bibfield  {author} {\bibinfo {author} {\bibfnamefont {H.}~\bibnamefont
  {Khalvati}},\ }\href@noop {} {\bibinfo {title} {{Dataset for
  FastEMRIWaveforms: Relativistic Kerr Circular Non-Vacuum Extension}}},\
  \bibinfo {howpublished} {\url{https://doi.org/10.5281/zenodo.15041149}}
  (\bibinfo {year} {2024}),\ \bibinfo {note} {zenodo repository}\BibitemShut
  {NoStop}%
\bibitem [{\citenamefont {Barack}\ and\ \citenamefont
  {Cutler}(2004)}]{AK_Barack_2004}%
  \BibitemOpen
  \bibfield  {author} {\bibinfo {author} {\bibfnamefont {L.}~\bibnamefont
  {Barack}}\ and\ \bibinfo {author} {\bibfnamefont {C.}~\bibnamefont
  {Cutler}},\ }\bibfield  {title} {\bibinfo {title} {Lisa capture sources:
  Approximate waveforms, signal-to-noise ratios, and parameter estimation
  accuracy},\ }\bibfield  {journal} {\bibinfo  {journal} {Physical Review D}\
  }\textbf {\bibinfo {volume} {69}},\ \href
  {https://doi.org/10.1103/physrevd.69.082005} {10.1103/physrevd.69.082005}
  (\bibinfo {year} {2004})\BibitemShut {NoStop}%
\bibitem [{\citenamefont {Babak}\ \emph {et~al.}(2007)\citenamefont {Babak},
  \citenamefont {Fang}, \citenamefont {Gair}, \citenamefont {Glampedakis},\
  and\ \citenamefont {Hughes}}]{Kludge}%
  \BibitemOpen
  \bibfield  {author} {\bibinfo {author} {\bibfnamefont {S.}~\bibnamefont
  {Babak}}, \bibinfo {author} {\bibfnamefont {H.}~\bibnamefont {Fang}},
  \bibinfo {author} {\bibfnamefont {J.~R.}\ \bibnamefont {Gair}}, \bibinfo
  {author} {\bibfnamefont {K.}~\bibnamefont {Glampedakis}},\ and\ \bibinfo
  {author} {\bibfnamefont {S.~A.}\ \bibnamefont {Hughes}},\ }\bibfield  {title}
  {\bibinfo {title} {``kludge'' gravitational waveforms for a test-body
  orbiting a kerr black hole},\ }\href
  {https://doi.org/10.1103/PhysRevD.75.024005} {\bibfield  {journal} {\bibinfo
  {journal} {Phys. Rev. D}\ }\textbf {\bibinfo {volume} {75}},\ \bibinfo
  {pages} {024005} (\bibinfo {year} {2007})}\BibitemShut {NoStop}%
\bibitem [{\citenamefont {Chua}\ \emph {et~al.}(2017)\citenamefont {Chua},
  \citenamefont {Moore},\ and\ \citenamefont {Gair}}]{AAK_Chua_2017}%
  \BibitemOpen
  \bibfield  {author} {\bibinfo {author} {\bibfnamefont {A.~J.~K.}\
  \bibnamefont {Chua}}, \bibinfo {author} {\bibfnamefont {C.~J.}\ \bibnamefont
  {Moore}},\ and\ \bibinfo {author} {\bibfnamefont {J.~R.}\ \bibnamefont
  {Gair}},\ }\bibfield  {title} {\bibinfo {title} {Augmented kludge waveforms
  for detecting extreme-mass-ratio inspirals},\ }\bibfield  {journal} {\bibinfo
   {journal} {Physical Review D}\ }\textbf {\bibinfo {volume} {96}},\ \href
  {https://doi.org/10.1103/physrevd.96.044005} {10.1103/physrevd.96.044005}
  (\bibinfo {year} {2017})\BibitemShut {NoStop}%
\bibitem [{\citenamefont {Peters}\ and\ \citenamefont
  {Mathews}(1963)}]{Peters_Math_PhysRev.131.435}%
  \BibitemOpen
  \bibfield  {author} {\bibinfo {author} {\bibfnamefont {P.~C.}\ \bibnamefont
  {Peters}}\ and\ \bibinfo {author} {\bibfnamefont {J.}~\bibnamefont
  {Mathews}},\ }\bibfield  {title} {\bibinfo {title} {Gravitational radiation
  from point masses in a keplerian orbit},\ }\href
  {https://doi.org/10.1103/PhysRev.131.435} {\bibfield  {journal} {\bibinfo
  {journal} {Phys. Rev.}\ }\textbf {\bibinfo {volume} {131}},\ \bibinfo {pages}
  {435} (\bibinfo {year} {1963})}\BibitemShut {NoStop}%
\bibitem [{\citenamefont {Gair}\ and\ \citenamefont
  {Glampedakis}(2006)}]{Gair_NK_PhysRevD.73.064037}%
  \BibitemOpen
  \bibfield  {author} {\bibinfo {author} {\bibfnamefont {J.~R.}\ \bibnamefont
  {Gair}}\ and\ \bibinfo {author} {\bibfnamefont {K.}~\bibnamefont
  {Glampedakis}},\ }\bibfield  {title} {\bibinfo {title} {Improved approximate
  inspirals of test bodies into kerr black holes},\ }\href
  {https://doi.org/10.1103/PhysRevD.73.064037} {\bibfield  {journal} {\bibinfo
  {journal} {Phys. Rev. D}\ }\textbf {\bibinfo {volume} {73}},\ \bibinfo
  {pages} {064037} (\bibinfo {year} {2006})}\BibitemShut {NoStop}%
\bibitem [{\citenamefont {Drasco}\ and\ \citenamefont
  {Hughes}(2006{\natexlab{b}})}]{Drasco_2006}%
  \BibitemOpen
  \bibfield  {author} {\bibinfo {author} {\bibfnamefont {S.}~\bibnamefont
  {Drasco}}\ and\ \bibinfo {author} {\bibfnamefont {S.~A.}\ \bibnamefont
  {Hughes}},\ }\bibfield  {title} {\bibinfo {title} {Gravitational wave
  snapshots of generic extreme mass ratio inspirals},\ }\bibfield  {journal}
  {\bibinfo  {journal} {Physical Review D}\ }\textbf {\bibinfo {volume} {73}},\
  \href {https://doi.org/10.1103/physrevd.73.024027}
  {10.1103/physrevd.73.024027} (\bibinfo {year}
  {2006}{\natexlab{b}})\BibitemShut {NoStop}%
\bibitem [{\citenamefont {Yunes}\ and\ \citenamefont
  {Berti}(2008)}]{Yunes_berti_3PN_PhysRevD.77.124006}%
  \BibitemOpen
  \bibfield  {author} {\bibinfo {author} {\bibfnamefont {N.}~\bibnamefont
  {Yunes}}\ and\ \bibinfo {author} {\bibfnamefont {E.}~\bibnamefont {Berti}},\
  }\bibfield  {title} {\bibinfo {title} {Accuracy of the post-newtonian
  approximation: Optimal asymptotic expansion for quasicircular, extreme-mass
  ratio inspirals},\ }\href {https://doi.org/10.1103/PhysRevD.77.124006}
  {\bibfield  {journal} {\bibinfo  {journal} {Phys. Rev. D}\ }\textbf {\bibinfo
  {volume} {77}},\ \bibinfo {pages} {124006} (\bibinfo {year}
  {2008})}\BibitemShut {NoStop}%
\bibitem [{\citenamefont {Hughes}\ \emph {et~al.}(2005)\citenamefont {Hughes},
  \citenamefont {Drasco}, \citenamefont {Flanagan},\ and\ \citenamefont
  {Franklin}}]{Hughes_adiabatic_PhysRevLett.94.221101}%
  \BibitemOpen
  \bibfield  {author} {\bibinfo {author} {\bibfnamefont {S.~A.}\ \bibnamefont
  {Hughes}}, \bibinfo {author} {\bibfnamefont {S.}~\bibnamefont {Drasco}},
  \bibinfo {author} {\bibfnamefont {E.~E.}\ \bibnamefont {Flanagan}},\ and\
  \bibinfo {author} {\bibfnamefont {J.}~\bibnamefont {Franklin}},\ }\bibfield
  {title} {\bibinfo {title} {Gravitational radiation reaction and inspiral
  waveforms in the adiabatic limit},\ }\href
  {https://doi.org/10.1103/PhysRevLett.94.221101} {\bibfield  {journal}
  {\bibinfo  {journal} {Phys. Rev. Lett.}\ }\textbf {\bibinfo {volume} {94}},\
  \bibinfo {pages} {221101} (\bibinfo {year} {2005})}\BibitemShut {NoStop}%
\bibitem [{\citenamefont {Hinderer}\ and\ \citenamefont
  {Flanagan}(2008)}]{Hinderer_2008}%
  \BibitemOpen
  \bibfield  {author} {\bibinfo {author} {\bibfnamefont {T.}~\bibnamefont
  {Hinderer}}\ and\ \bibinfo {author} {\bibfnamefont {E.~E.}\ \bibnamefont
  {Flanagan}},\ }\bibfield  {title} {\bibinfo {title} {Two-timescale analysis
  of extreme mass ratio inspirals in kerr spacetime: Orbital motion},\
  }\bibfield  {journal} {\bibinfo  {journal} {Physical Review D}\ }\textbf
  {\bibinfo {volume} {78}},\ \href {https://doi.org/10.1103/physrevd.78.064028}
  {10.1103/physrevd.78.064028} (\bibinfo {year} {2008})\BibitemShut {NoStop}%
\bibitem [{\citenamefont {Miller}\ and\ \citenamefont
  {Pound}(2021)}]{Miller:2020bft}%
  \BibitemOpen
  \bibfield  {author} {\bibinfo {author} {\bibfnamefont {J.}~\bibnamefont
  {Miller}}\ and\ \bibinfo {author} {\bibfnamefont {A.}~\bibnamefont {Pound}},\
  }\bibfield  {title} {\bibinfo {title} {{Two-timescale evolution of
  extreme-mass-ratio inspirals: waveform generation scheme for quasicircular
  orbits in Schwarzschild spacetime}},\ }\href
  {https://doi.org/10.1103/PhysRevD.103.064048} {\bibfield  {journal} {\bibinfo
   {journal} {Phys. Rev. D}\ }\textbf {\bibinfo {volume} {103}},\ \bibinfo
  {pages} {064048} (\bibinfo {year} {2021})},\ \Eprint
  {https://arxiv.org/abs/2006.11263} {arXiv:2006.11263 [gr-qc]} \BibitemShut
  {NoStop}%
\bibitem [{\citenamefont {Hughes}\ \emph {et~al.}(2021)\citenamefont {Hughes},
  \citenamefont {Warburton}, \citenamefont {Khanna}, \citenamefont {Chua},\
  and\ \citenamefont {Katz}}]{Hughes_2021}%
  \BibitemOpen
  \bibfield  {author} {\bibinfo {author} {\bibfnamefont {S.~A.}\ \bibnamefont
  {Hughes}}, \bibinfo {author} {\bibfnamefont {N.}~\bibnamefont {Warburton}},
  \bibinfo {author} {\bibfnamefont {G.}~\bibnamefont {Khanna}}, \bibinfo
  {author} {\bibfnamefont {A.~J.~K.}\ \bibnamefont {Chua}},\ and\ \bibinfo
  {author} {\bibfnamefont {M.~L.}\ \bibnamefont {Katz}},\ }\bibfield  {title}
  {\bibinfo {title} {Adiabatic waveforms for extreme mass-ratio inspirals via
  multivoice decomposition in time and frequency},\ }\bibfield  {journal}
  {\bibinfo  {journal} {Physical Review D}\ }\textbf {\bibinfo {volume}
  {103}},\ \href {https://doi.org/10.1103/physrevd.103.104014}
  {10.1103/physrevd.103.104014} (\bibinfo {year} {2021})\BibitemShut {NoStop}%
\bibitem [{\citenamefont {Barack}\ and\ \citenamefont
  {Pound}(2019)}]{Barack:2018yvs}%
  \BibitemOpen
  \bibfield  {author} {\bibinfo {author} {\bibfnamefont {L.}~\bibnamefont
  {Barack}}\ and\ \bibinfo {author} {\bibfnamefont {A.}~\bibnamefont {Pound}},\
  }\bibfield  {title} {\bibinfo {title} {{Self-force and radiation reaction in
  general relativity}},\ }\href {https://doi.org/10.1088/1361-6633/aae552}
  {\bibfield  {journal} {\bibinfo  {journal} {Rept. Prog. Phys.}\ }\textbf
  {\bibinfo {volume} {82}},\ \bibinfo {pages} {016904} (\bibinfo {year}
  {2019})},\ \Eprint {https://arxiv.org/abs/1805.10385} {arXiv:1805.10385
  [gr-qc]} \BibitemShut {NoStop}%
\bibitem [{\citenamefont {Cutler}\ \emph {et~al.}(1994)\citenamefont {Cutler},
  \citenamefont {Kennefick},\ and\ \citenamefont
  {Poisson}}]{cutler_eric_PhysRevD.50.3816}%
  \BibitemOpen
  \bibfield  {author} {\bibinfo {author} {\bibfnamefont {C.}~\bibnamefont
  {Cutler}}, \bibinfo {author} {\bibfnamefont {D.}~\bibnamefont {Kennefick}},\
  and\ \bibinfo {author} {\bibfnamefont {E.}~\bibnamefont {Poisson}},\
  }\bibfield  {title} {\bibinfo {title} {Gravitational radiation reaction for
  bound motion around a schwarzschild black hole},\ }\href
  {https://doi.org/10.1103/PhysRevD.50.3816} {\bibfield  {journal} {\bibinfo
  {journal} {Phys. Rev. D}\ }\textbf {\bibinfo {volume} {50}},\ \bibinfo
  {pages} {3816} (\bibinfo {year} {1994})}\BibitemShut {NoStop}%
\bibitem [{\citenamefont {Hughes}(2000)}]{Hughes2000_PhysRevD.61.084004}%
  \BibitemOpen
  \bibfield  {author} {\bibinfo {author} {\bibfnamefont {S.~A.}\ \bibnamefont
  {Hughes}},\ }\bibfield  {title} {\bibinfo {title} {Evolution of circular,
  nonequatorial orbits of kerr black holes due to gravitational-wave
  emission},\ }\href {https://doi.org/10.1103/PhysRevD.61.084004} {\bibfield
  {journal} {\bibinfo  {journal} {Phys. Rev. D}\ }\textbf {\bibinfo {volume}
  {61}},\ \bibinfo {pages} {084004} (\bibinfo {year} {2000})}\BibitemShut
  {NoStop}%
\bibitem [{\citenamefont {Teukolsky}(1972)}]{Teukolsky}%
  \BibitemOpen
  \bibfield  {author} {\bibinfo {author} {\bibfnamefont {S.~A.}\ \bibnamefont
  {Teukolsky}},\ }\bibfield  {title} {\bibinfo {title} {Rotating black holes:
  Separable wave equations for gravitational and electromagnetic
  perturbations},\ }\href {https://doi.org/10.1103/PhysRevLett.29.1114}
  {\bibfield  {journal} {\bibinfo  {journal} {Phys. Rev. Lett.}\ }\textbf
  {\bibinfo {volume} {29}},\ \bibinfo {pages} {1114} (\bibinfo {year}
  {1972})}\BibitemShut {NoStop}%
\bibitem [{\citenamefont {{Teukolsky}}(1973)}]{Teuk0_1973ApJ...185..635T}%
  \BibitemOpen
  \bibfield  {author} {\bibinfo {author} {\bibfnamefont {S.~A.}\ \bibnamefont
  {{Teukolsky}}},\ }\bibfield  {title} {\bibinfo {title} {{Perturbations of a
  Rotating Black Hole. I. Fundamental Equations for Gravitational,
  Electromagnetic, and Neutrino-Field Perturbations}},\ }\href
  {https://doi.org/10.1086/152444} {\bibfield  {journal} {\bibinfo  {journal}
  {\apj}\ }\textbf {\bibinfo {volume} {185}},\ \bibinfo {pages} {635} (\bibinfo
  {year} {1973})}\BibitemShut {NoStop}%
\bibitem [{\citenamefont {O'Sullivan}\ and\ \citenamefont
  {Hughes}(2014)}]{Hughes_Sullivan_PhysRevD.90.124039}%
  \BibitemOpen
  \bibfield  {author} {\bibinfo {author} {\bibfnamefont {S.}~\bibnamefont
  {O'Sullivan}}\ and\ \bibinfo {author} {\bibfnamefont {S.~A.}\ \bibnamefont
  {Hughes}},\ }\bibfield  {title} {\bibinfo {title} {Strong-field tidal
  distortions of rotating black holes: Formalism and results for circular,
  equatorial orbits},\ }\href {https://doi.org/10.1103/PhysRevD.90.124039}
  {\bibfield  {journal} {\bibinfo  {journal} {Phys. Rev. D}\ }\textbf {\bibinfo
  {volume} {90}},\ \bibinfo {pages} {124039} (\bibinfo {year}
  {2014})}\BibitemShut {NoStop}%
\bibitem [{\citenamefont {Schmidt}(2002)}]{Schmidt_2002}%
  \BibitemOpen
  \bibfield  {author} {\bibinfo {author} {\bibfnamefont {W.}~\bibnamefont
  {Schmidt}},\ }\bibfield  {title} {\bibinfo {title} {Celestial mechanics in
  kerr spacetime},\ }\href {https://doi.org/10.1088/0264-9381/19/10/314}
  {\bibfield  {journal} {\bibinfo  {journal} {Classical and Quantum Gravity}\
  }\textbf {\bibinfo {volume} {19}},\ \bibinfo {pages} {2743–2764} (\bibinfo
  {year} {2002})}\BibitemShut {NoStop}%
\bibitem [{\citenamefont {Isaacson}(1968)}]{Isaacson_PhysRev.166.1272}%
  \BibitemOpen
  \bibfield  {author} {\bibinfo {author} {\bibfnamefont {R.~A.}\ \bibnamefont
  {Isaacson}},\ }\bibfield  {title} {\bibinfo {title} {Gravitational radiation
  in the limit of high frequency. ii. nonlinear terms and the effective stress
  tensor},\ }\href {https://doi.org/10.1103/PhysRev.166.1272} {\bibfield
  {journal} {\bibinfo  {journal} {Phys. Rev.}\ }\textbf {\bibinfo {volume}
  {166}},\ \bibinfo {pages} {1272} (\bibinfo {year} {1968})}\BibitemShut
  {NoStop}%
\bibitem [{\citenamefont {Wardell}\ \emph {et~al.}(2023)\citenamefont
  {Wardell}, \citenamefont {Pound}, \citenamefont {Warburton}, \citenamefont
  {Miller}, \citenamefont {Durkan},\ and\ \citenamefont
  {Le~Tiec}}]{Wardell_2023}%
  \BibitemOpen
  \bibfield  {author} {\bibinfo {author} {\bibfnamefont {B.}~\bibnamefont
  {Wardell}}, \bibinfo {author} {\bibfnamefont {A.}~\bibnamefont {Pound}},
  \bibinfo {author} {\bibfnamefont {N.}~\bibnamefont {Warburton}}, \bibinfo
  {author} {\bibfnamefont {J.}~\bibnamefont {Miller}}, \bibinfo {author}
  {\bibfnamefont {L.}~\bibnamefont {Durkan}},\ and\ \bibinfo {author}
  {\bibfnamefont {A.}~\bibnamefont {Le~Tiec}},\ }\bibfield  {title} {\bibinfo
  {title} {Gravitational waveforms for compact binaries from second-order
  self-force theory},\ }\href {https://doi.org/10.1103/PhysRevLett.130.241402}
  {\bibfield  {journal} {\bibinfo  {journal} {Phys. Rev. Lett.}\ }\textbf
  {\bibinfo {volume} {130}},\ \bibinfo {pages} {241402} (\bibinfo {year}
  {2023})}\BibitemShut {NoStop}%
\bibitem [{\citenamefont {Katz}\ \emph {et~al.}(2020)\citenamefont {Katz},
  \citenamefont {Chua}, \citenamefont {Warburton},\ and\ \citenamefont
  {Hughes.}}]{michael_l_katz_2020_4005001}%
  \BibitemOpen
  \bibfield  {author} {\bibinfo {author} {\bibfnamefont {M.~L.}\ \bibnamefont
  {Katz}}, \bibinfo {author} {\bibfnamefont {A.~J.~K.}\ \bibnamefont {Chua}},
  \bibinfo {author} {\bibfnamefont {N.}~\bibnamefont {Warburton}},\ and\
  \bibinfo {author} {\bibfnamefont {S.~A.}\ \bibnamefont {Hughes.}},\ }\href
  {https://doi.org/10.5281/zenodo.4005001} {\bibinfo {title}
  {{BlackHolePerturbationToolkit/FastEMRIWaveforms: Official Release}}}
  (\bibinfo {year} {2020})\BibitemShut {NoStop}%
\bibitem [{BHP()}]{BHPToolkit}%
  \BibitemOpen
  \href@noop {} {\bibinfo {title} {{Black Hole Perturbation Toolkit}}},\
  \bibinfo {howpublished}
  {(\href{http://bhptoolkit.org/}{bhptoolkit.org})}\BibitemShut {NoStop}%
\bibitem [{\citenamefont
  {Hughes}(2024)}]{hughes2024parameterizingblackholeorbits}%
  \BibitemOpen
  \bibfield  {author} {\bibinfo {author} {\bibfnamefont {S.~A.}\ \bibnamefont
  {Hughes}},\ }\href {https://arxiv.org/abs/2401.09577} {\bibinfo {title}
  {Parameterizing black hole orbits for adiabatic inspiral}} (\bibinfo {year}
  {2024}),\ \Eprint {https://arxiv.org/abs/2401.09577} {arXiv:2401.09577
  [gr-qc]} \BibitemShut {NoStop}%
\bibitem [{\citenamefont {Yunes}\ \emph
  {et~al.}(2011{\natexlab{a}})\citenamefont {Yunes}, \citenamefont {Buonanno},
  \citenamefont {Hughes}, \citenamefont {Pan}, \citenamefont {Barausse},
  \citenamefont {Miller},\ and\ \citenamefont {Throwe}}]{Yunes_2011}%
  \BibitemOpen
  \bibfield  {author} {\bibinfo {author} {\bibfnamefont {N.}~\bibnamefont
  {Yunes}}, \bibinfo {author} {\bibfnamefont {A.}~\bibnamefont {Buonanno}},
  \bibinfo {author} {\bibfnamefont {S.~A.}\ \bibnamefont {Hughes}}, \bibinfo
  {author} {\bibfnamefont {Y.}~\bibnamefont {Pan}}, \bibinfo {author}
  {\bibfnamefont {E.}~\bibnamefont {Barausse}}, \bibinfo {author}
  {\bibfnamefont {M.~C.}\ \bibnamefont {Miller}},\ and\ \bibinfo {author}
  {\bibfnamefont {W.}~\bibnamefont {Throwe}},\ }\bibfield  {title} {\bibinfo
  {title} {Extreme mass-ratio inspirals in the effective-one-body approach:
  Quasicircular, equatorial orbits around a spinning black hole},\ }\bibfield
  {journal} {\bibinfo  {journal} {Physical Review D}\ }\textbf {\bibinfo
  {volume} {83}},\ \href {https://doi.org/10.1103/physrevd.83.044044}
  {10.1103/physrevd.83.044044} (\bibinfo {year}
  {2011}{\natexlab{a}})\BibitemShut {NoStop}%
\bibitem [{\citenamefont {Hirata}(2011{\natexlab{a}})}]{Hirata2:2010vp}%
  \BibitemOpen
  \bibfield  {author} {\bibinfo {author} {\bibfnamefont {C.~M.}\ \bibnamefont
  {Hirata}},\ }\bibfield  {title} {\bibinfo {title} {{Lindblad resonance
  torques in relativistic discs: II. Computation of resonance strengths}},\
  }\href {https://doi.org/10.1111/j.1365-2966.2011.18619.x} {\bibfield
  {journal} {\bibinfo  {journal} {Mon. Not. Roy. Astron. Soc.}\ }\textbf
  {\bibinfo {volume} {414}},\ \bibinfo {pages} {3212} (\bibinfo {year}
  {2011}{\natexlab{a}})},\ \Eprint {https://arxiv.org/abs/1010.0759}
  {arXiv:1010.0759 [astro-ph.HE]} \BibitemShut {NoStop}%
\bibitem [{\citenamefont {{Press}}\ and\ \citenamefont
  {{Teukolsky}}(1973)}]{press_teuk_1973ApJ...185..649P}%
  \BibitemOpen
  \bibfield  {author} {\bibinfo {author} {\bibfnamefont {W.~H.}\ \bibnamefont
  {{Press}}}\ and\ \bibinfo {author} {\bibfnamefont {S.~A.}\ \bibnamefont
  {{Teukolsky}}},\ }\bibfield  {title} {\bibinfo {title} {{Perturbations of a
  Rotating Black Hole. II. Dynamical Stability of the Kerr Metric}},\ }\href
  {https://doi.org/10.1086/152445} {\bibfield  {journal} {\bibinfo  {journal}
  {\apj}\ }\textbf {\bibinfo {volume} {185}},\ \bibinfo {pages} {649} (\bibinfo
  {year} {1973})}\BibitemShut {NoStop}%
\bibitem [{\citenamefont {{Karnesis}}\ \emph {et~al.}(2023)\citenamefont
  {{Karnesis}}, \citenamefont {{Katz}}, \citenamefont {{Korsakova}},
  \citenamefont {{Gair}},\ and\ \citenamefont
  {{Stergioulas}}}]{2023MNRAS.526.4814K}%
  \BibitemOpen
  \bibfield  {author} {\bibinfo {author} {\bibfnamefont {N.}~\bibnamefont
  {{Karnesis}}}, \bibinfo {author} {\bibfnamefont {M.~L.}\ \bibnamefont
  {{Katz}}}, \bibinfo {author} {\bibfnamefont {N.}~\bibnamefont {{Korsakova}}},
  \bibinfo {author} {\bibfnamefont {J.~R.}\ \bibnamefont {{Gair}}},\ and\
  \bibinfo {author} {\bibfnamefont {N.}~\bibnamefont {{Stergioulas}}},\
  }\bibfield  {title} {\bibinfo {title} {{Eryn: a multipurpose sampler for
  Bayesian inference}},\ }\href {https://doi.org/10.1093/mnras/stad2939}
  {\bibfield  {journal} {\bibinfo  {journal} {\mnras}\ }\textbf {\bibinfo
  {volume} {526}},\ \bibinfo {pages} {4814} (\bibinfo {year} {2023})},\ \Eprint
  {https://arxiv.org/abs/2303.02164} {arXiv:2303.02164 [astro-ph.IM]}
  \BibitemShut {NoStop}%
\bibitem [{\citenamefont {{Goodman}}\ and\ \citenamefont
  {{Weare}}(2010)}]{2010CAMCS...5...65G}%
  \BibitemOpen
  \bibfield  {author} {\bibinfo {author} {\bibfnamefont {J.}~\bibnamefont
  {{Goodman}}}\ and\ \bibinfo {author} {\bibfnamefont {J.}~\bibnamefont
  {{Weare}}},\ }\bibfield  {title} {\bibinfo {title} {{Ensemble samplers with
  affine invariance}},\ }\href {https://doi.org/10.2140/camcos.2010.5.65}
  {\bibfield  {journal} {\bibinfo  {journal} {Communications in Applied
  Mathematics and Computational Science}\ }\textbf {\bibinfo {volume} {5}},\
  \bibinfo {pages} {65} (\bibinfo {year} {2010})}\BibitemShut {NoStop}%
\bibitem [{\citenamefont {{Del Pozzo}}\ \emph {et~al.}(2011)\citenamefont {{Del
  Pozzo}}, \citenamefont {{Veitch}},\ and\ \citenamefont
  {{Vecchio}}}]{2011PhRvD..83h2002D}%
  \BibitemOpen
  \bibfield  {author} {\bibinfo {author} {\bibfnamefont {W.}~\bibnamefont {{Del
  Pozzo}}}, \bibinfo {author} {\bibfnamefont {J.}~\bibnamefont {{Veitch}}},\
  and\ \bibinfo {author} {\bibfnamefont {A.}~\bibnamefont {{Vecchio}}},\
  }\bibfield  {title} {\bibinfo {title} {{Testing general relativity using
  Bayesian model selection: Applications to observations of gravitational waves
  from compact binary systems}},\ }\href
  {https://doi.org/10.1103/PhysRevD.83.082002} {\bibfield  {journal} {\bibinfo
  {journal} {\prd}\ }\textbf {\bibinfo {volume} {83}},\ \bibinfo {eid} {082002}
  (\bibinfo {year} {2011})},\ \Eprint {https://arxiv.org/abs/1101.1391}
  {arXiv:1101.1391 [gr-qc]} \BibitemShut {NoStop}%
\bibitem [{\citenamefont {{Toubiana}}\ \emph {et~al.}(2021)\citenamefont
  {{Toubiana}}, \citenamefont {{Wong}}, \citenamefont {{Babak}}, \citenamefont
  {{Barausse}}, \citenamefont {{Berti}}, \citenamefont {{Gair}}, \citenamefont
  {{Marsat}},\ and\ \citenamefont {{Taylor}}}]{2021PhRvD.104h3027T}%
  \BibitemOpen
  \bibfield  {author} {\bibinfo {author} {\bibfnamefont {A.}~\bibnamefont
  {{Toubiana}}}, \bibinfo {author} {\bibfnamefont {K.~W.~K.}\ \bibnamefont
  {{Wong}}}, \bibinfo {author} {\bibfnamefont {S.}~\bibnamefont {{Babak}}},
  \bibinfo {author} {\bibfnamefont {E.}~\bibnamefont {{Barausse}}}, \bibinfo
  {author} {\bibfnamefont {E.}~\bibnamefont {{Berti}}}, \bibinfo {author}
  {\bibfnamefont {J.~R.}\ \bibnamefont {{Gair}}}, \bibinfo {author}
  {\bibfnamefont {S.}~\bibnamefont {{Marsat}}},\ and\ \bibinfo {author}
  {\bibfnamefont {S.~R.}\ \bibnamefont {{Taylor}}},\ }\bibfield  {title}
  {\bibinfo {title} {{Discriminating between different scenarios for the
  formation and evolution of massive black holes with LISA}},\ }\href
  {https://doi.org/10.1103/PhysRevD.104.083027} {\bibfield  {journal} {\bibinfo
   {journal} {\prd}\ }\textbf {\bibinfo {volume} {104}},\ \bibinfo {eid}
  {083027} (\bibinfo {year} {2021})},\ \Eprint
  {https://arxiv.org/abs/2106.13819} {arXiv:2106.13819 [gr-qc]} \BibitemShut
  {NoStop}%
\bibitem [{\citenamefont {{Toubiana}}\ \emph {et~al.}(2024)\citenamefont
  {{Toubiana}}, \citenamefont {{Pompili}}, \citenamefont {{Buonanno}},
  \citenamefont {{Gair}},\ and\ \citenamefont {{Katz}}}]{2024PhRvD.109j4019T}%
  \BibitemOpen
  \bibfield  {author} {\bibinfo {author} {\bibfnamefont {A.}~\bibnamefont
  {{Toubiana}}}, \bibinfo {author} {\bibfnamefont {L.}~\bibnamefont
  {{Pompili}}}, \bibinfo {author} {\bibfnamefont {A.}~\bibnamefont
  {{Buonanno}}}, \bibinfo {author} {\bibfnamefont {J.~R.}\ \bibnamefont
  {{Gair}}},\ and\ \bibinfo {author} {\bibfnamefont {M.~L.}\ \bibnamefont
  {{Katz}}},\ }\bibfield  {title} {\bibinfo {title} {{Measuring source
  properties and quasinormal mode frequencies of heavy massive black-hole
  binaries with LISA}},\ }\href {https://doi.org/10.1103/PhysRevD.109.104019}
  {\bibfield  {journal} {\bibinfo  {journal} {\prd}\ }\textbf {\bibinfo
  {volume} {109}},\ \bibinfo {eid} {104019} (\bibinfo {year} {2024})},\ \Eprint
  {https://arxiv.org/abs/2307.15086} {arXiv:2307.15086 [gr-qc]} \BibitemShut
  {NoStop}%
\bibitem [{\citenamefont {Lartillot}\ and\ \citenamefont
  {Philippe}(2006)}]{10.1080/10635150500433722}%
  \BibitemOpen
  \bibfield  {author} {\bibinfo {author} {\bibfnamefont {N.}~\bibnamefont
  {Lartillot}}\ and\ \bibinfo {author} {\bibfnamefont {H.}~\bibnamefont
  {Philippe}},\ }\bibfield  {title} {\bibinfo {title} {{Computing Bayes Factors
  Using Thermodynamic Integration}},\ }\href
  {https://doi.org/10.1080/10635150500433722} {\bibfield  {journal} {\bibinfo
  {journal} {Systematic Biology}\ }\textbf {\bibinfo {volume} {55}},\ \bibinfo
  {pages} {195} (\bibinfo {year} {2006})},\ \Eprint
  {https://arxiv.org/abs/https://academic.oup.com/sysbio/article-pdf/55/2/195/26557316/10635150500433722.pdf}
  {https://academic.oup.com/sysbio/article-pdf/55/2/195/26557316/10635150500433722.pdf}
  \BibitemShut {NoStop}%
\bibitem [{\citenamefont {Xie}\ \emph {et~al.}(2010)\citenamefont {Xie},
  \citenamefont {Lewis}, \citenamefont {Fan}, \citenamefont {Kuo},\ and\
  \citenamefont {Chen}}]{10.1093/sysbio/syq085}%
  \BibitemOpen
  \bibfield  {author} {\bibinfo {author} {\bibfnamefont {W.}~\bibnamefont
  {Xie}}, \bibinfo {author} {\bibfnamefont {P.~O.}\ \bibnamefont {Lewis}},
  \bibinfo {author} {\bibfnamefont {Y.}~\bibnamefont {Fan}}, \bibinfo {author}
  {\bibfnamefont {L.}~\bibnamefont {Kuo}},\ and\ \bibinfo {author}
  {\bibfnamefont {M.-H.}\ \bibnamefont {Chen}},\ }\bibfield  {title} {\bibinfo
  {title} {{Improving Marginal Likelihood Estimation for Bayesian Phylogenetic
  Model Selection}},\ }\href {https://doi.org/10.1093/sysbio/syq085} {\bibfield
   {journal} {\bibinfo  {journal} {Systematic Biology}\ }\textbf {\bibinfo
  {volume} {60}},\ \bibinfo {pages} {150} (\bibinfo {year} {2010})},\ \Eprint
  {https://arxiv.org/abs/https://academic.oup.com/sysbio/article-pdf/60/2/150/24552358/syq085.pdf}
  {https://academic.oup.com/sysbio/article-pdf/60/2/150/24552358/syq085.pdf}
  \BibitemShut {NoStop}%
\bibitem [{\citenamefont {Srinivasan}\ \emph {et~al.}(2024)\citenamefont
  {Srinivasan}, \citenamefont {Crisostomi}, \citenamefont {Trotta},
  \citenamefont {Barausse},\ and\ \citenamefont
  {Breschi}}]{Srinivasan:2024uax}%
  \BibitemOpen
  \bibfield  {author} {\bibinfo {author} {\bibfnamefont {R.}~\bibnamefont
  {Srinivasan}}, \bibinfo {author} {\bibfnamefont {M.}~\bibnamefont
  {Crisostomi}}, \bibinfo {author} {\bibfnamefont {R.}~\bibnamefont {Trotta}},
  \bibinfo {author} {\bibfnamefont {E.}~\bibnamefont {Barausse}},\ and\
  \bibinfo {author} {\bibfnamefont {M.}~\bibnamefont {Breschi}},\ }\href
  {https://arxiv.org/abs/2404.12294} {\bibinfo {title} {$floz$: Improved
  bayesian evidence estimation from posterior samples with normalizing flows}}
  (\bibinfo {year} {2024}),\ \Eprint {https://arxiv.org/abs/2404.12294}
  {arXiv:2404.12294 [stat.ML]} \BibitemShut {NoStop}%
\bibitem [{\citenamefont {{Ivezić}}\ \emph {et~al.}(2019)\citenamefont
  {{Ivezić}}, \citenamefont {{Connolly}}, \citenamefont {{VanderPlas}},\ and\
  \citenamefont {{Gray}}}]{ivezic}%
  \BibitemOpen
  \bibfield  {author} {\bibinfo {author} {\bibfnamefont {Z.}~\bibnamefont
  {{Ivezić}}}, \bibinfo {author} {\bibfnamefont {A.~J.}\ \bibnamefont
  {{Connolly}}}, \bibinfo {author} {\bibfnamefont {J.~T.}\ \bibnamefont
  {{VanderPlas}}},\ and\ \bibinfo {author} {\bibfnamefont {A.}~\bibnamefont
  {{Gray}}},\ }\href
  {https://press.princeton.edu/books/hardcover/9780691198309/statistics-data-mining-and-machine-learning-in-astronomy-pdf}
  {\emph {\bibinfo {title} {Statistics, Data Mining, and Machine Learning in
  Astronomy: A Practical Python Guide for the Analysis of Survey Data}}}\
  (\bibinfo  {publisher} {Princeton Series in Modern Observational Astronomy},\
  \bibinfo {year} {2019})\BibitemShut {NoStop}%
\bibitem [{\citenamefont {Papamakarios}\ \emph {et~al.}(2017)\citenamefont
  {Papamakarios}, \citenamefont {Pavlakou},\ and\ \citenamefont
  {Murray}}]{MAFs}%
  \BibitemOpen
  \bibfield  {author} {\bibinfo {author} {\bibfnamefont {G.}~\bibnamefont
  {Papamakarios}}, \bibinfo {author} {\bibfnamefont {T.}~\bibnamefont
  {Pavlakou}},\ and\ \bibinfo {author} {\bibfnamefont {I.}~\bibnamefont
  {Murray}},\ }\bibfield  {title} {\bibinfo {title} {Masked autoregressive flow
  for density estimation},\ }\href@noop {} {\bibfield  {journal} {\bibinfo
  {journal} {Advances in neural information processing systems}\ }\textbf
  {\bibinfo {volume} {30}} (\bibinfo {year} {2017})}\BibitemShut {NoStop}%
\bibitem [{\citenamefont {Durkan}\ \emph {et~al.}(2020)\citenamefont {Durkan},
  \citenamefont {Bekasov}, \citenamefont {Murray},\ and\ \citenamefont
  {Papamakarios}}]{nflows}%
  \BibitemOpen
  \bibfield  {author} {\bibinfo {author} {\bibfnamefont {C.}~\bibnamefont
  {Durkan}}, \bibinfo {author} {\bibfnamefont {A.}~\bibnamefont {Bekasov}},
  \bibinfo {author} {\bibfnamefont {I.}~\bibnamefont {Murray}},\ and\ \bibinfo
  {author} {\bibfnamefont {G.}~\bibnamefont {Papamakarios}},\ }\href
  {https://doi.org/10.5281/zenodo.4296287} {\bibinfo {title} {{nflows}:
  normalizing flows in {PyTorch}}} (\bibinfo {year} {2020})\BibitemShut
  {NoStop}%
\bibitem [{\citenamefont {Team}(2018)}]{scirdv}%
  \BibitemOpen
  \bibfield  {author} {\bibinfo {author} {\bibfnamefont {L.~S.~S.}\
  \bibnamefont {Team}},\ }\bibfield  {title} {\bibinfo {title} {{LISA Science
  Requirements Document}},\ }\href {https://doi.org/10.2140/camcos.2010.5.65}
  {\bibfield  {journal} {\bibinfo  {journal} {ESA-L3-EST-SCI-RS-001}\ }\textbf
  {\bibinfo {volume} {5}},\ \bibinfo {pages} {65} (\bibinfo {year}
  {2018})}\BibitemShut {NoStop}%
\bibitem [{\citenamefont {Karnesis}\ \emph {et~al.}(2021)\citenamefont
  {Karnesis}, \citenamefont {Babak}, \citenamefont {Pieroni}, \citenamefont
  {Cornish},\ and\ \citenamefont {Littenberg}}]{PhysRevD.104.043019}%
  \BibitemOpen
  \bibfield  {author} {\bibinfo {author} {\bibfnamefont {N.}~\bibnamefont
  {Karnesis}}, \bibinfo {author} {\bibfnamefont {S.}~\bibnamefont {Babak}},
  \bibinfo {author} {\bibfnamefont {M.}~\bibnamefont {Pieroni}}, \bibinfo
  {author} {\bibfnamefont {N.}~\bibnamefont {Cornish}},\ and\ \bibinfo {author}
  {\bibfnamefont {T.}~\bibnamefont {Littenberg}},\ }\bibfield  {title}
  {\bibinfo {title} {Characterization of the stochastic signal originating from
  compact binary populations as measured by lisa},\ }\href
  {https://doi.org/10.1103/PhysRevD.104.043019} {\bibfield  {journal} {\bibinfo
   {journal} {Phys. Rev. D}\ }\textbf {\bibinfo {volume} {104}},\ \bibinfo
  {pages} {043019} (\bibinfo {year} {2021})}\BibitemShut {NoStop}%
\bibitem [{\citenamefont {Katz}\ \emph {et~al.}(2024)\citenamefont {Katz},
  \citenamefont {CChapmanbird}, \citenamefont {Speri}, \citenamefont
  {Karnesis},\ and\ \citenamefont {Korsakova}}]{lisatools}%
  \BibitemOpen
  \bibfield  {author} {\bibinfo {author} {\bibfnamefont {M.}~\bibnamefont
  {Katz}}, \bibinfo {author} {\bibnamefont {CChapmanbird}}, \bibinfo {author}
  {\bibfnamefont {L.}~\bibnamefont {Speri}}, \bibinfo {author} {\bibfnamefont
  {N.}~\bibnamefont {Karnesis}},\ and\ \bibinfo {author} {\bibfnamefont
  {N.}~\bibnamefont {Korsakova}},\ }\href
  {https://doi.org/10.5281/zenodo.10930980} {\bibinfo {title}
  {mikekatz04/lisaanalysistools: First main release.}} (\bibinfo {year}
  {2024})\BibitemShut {NoStop}%
\bibitem [{\citenamefont {{Katz}}\ \emph {et~al.}(2022)\citenamefont {{Katz}},
  \citenamefont {{Bayle}}, \citenamefont {{Chua}},\ and\ \citenamefont
  {{Vallisneri}}}]{2022PhRvD.106j3001K}%
  \BibitemOpen
  \bibfield  {author} {\bibinfo {author} {\bibfnamefont {M.~L.}\ \bibnamefont
  {{Katz}}}, \bibinfo {author} {\bibfnamefont {J.-B.}\ \bibnamefont {{Bayle}}},
  \bibinfo {author} {\bibfnamefont {A.~J.~K.}\ \bibnamefont {{Chua}}},\ and\
  \bibinfo {author} {\bibfnamefont {M.}~\bibnamefont {{Vallisneri}}},\
  }\bibfield  {title} {\bibinfo {title} {{Assessing the data-analysis impact of
  LISA orbit approximations using a GPU-accelerated response model}},\ }\href
  {https://doi.org/10.1103/PhysRevD.106.103001} {\bibfield  {journal} {\bibinfo
   {journal} {\prd}\ }\textbf {\bibinfo {volume} {106}},\ \bibinfo {eid}
  {103001} (\bibinfo {year} {2022})},\ \Eprint
  {https://arxiv.org/abs/2204.06633} {arXiv:2204.06633 [gr-qc]} \BibitemShut
  {NoStop}%
\bibitem [{\citenamefont {{Aghanim}}\ \emph {et~al.}(2020)\citenamefont
  {{Aghanim}} \emph {et~al.}}]{2020A&A...641A...6P}%
  \BibitemOpen
  \bibfield  {author} {\bibinfo {author} {\bibfnamefont {N.}~\bibnamefont
  {{Aghanim}}} \emph {et~al.},\ }\bibfield  {title} {\bibinfo {title} {{Planck
  2018 results. VI. Cosmological parameters}},\ }\href
  {https://doi.org/10.1051/0004-6361/201833910} {\bibfield  {journal} {\bibinfo
   {journal} {\aap}\ }\textbf {\bibinfo {volume} {641}},\ \bibinfo {eid} {A6}
  (\bibinfo {year} {2020})},\ \Eprint {https://arxiv.org/abs/1807.06209}
  {arXiv:1807.06209 [astro-ph.CO]} \BibitemShut {NoStop}%
\bibitem [{\citenamefont {Comp\`ere}\ and\ \citenamefont
  {K\"uchler}(2021)}]{Geoffrey_PhysRevLett.126.241106}%
  \BibitemOpen
  \bibfield  {author} {\bibinfo {author} {\bibfnamefont {G.}~\bibnamefont
  {Comp\`ere}}\ and\ \bibinfo {author} {\bibfnamefont {L.}~\bibnamefont
  {K\"uchler}},\ }\bibfield  {title} {\bibinfo {title} {Self-consistent
  adiabatic inspiral and transition motion},\ }\href
  {https://doi.org/10.1103/PhysRevLett.126.241106} {\bibfield  {journal}
  {\bibinfo  {journal} {Phys. Rev. Lett.}\ }\textbf {\bibinfo {volume} {126}},\
  \bibinfo {pages} {241106} (\bibinfo {year} {2021})}\BibitemShut {NoStop}%
\bibitem [{\citenamefont {Küchler}\ \emph {et~al.}(2024)\citenamefont
  {Küchler}, \citenamefont {Compère}, \citenamefont {Durkan},\ and\
  \citenamefont {Pound}}]{K_chler_2024}%
  \BibitemOpen
  \bibfield  {author} {\bibinfo {author} {\bibfnamefont {L.}~\bibnamefont
  {Küchler}}, \bibinfo {author} {\bibfnamefont {G.}~\bibnamefont {Compère}},
  \bibinfo {author} {\bibfnamefont {L.}~\bibnamefont {Durkan}},\ and\ \bibinfo
  {author} {\bibfnamefont {A.}~\bibnamefont {Pound}},\ }\bibfield  {title}
  {\bibinfo {title} {Self-force framework for transition-to-plunge waveforms},\
  }\bibfield  {journal} {\bibinfo  {journal} {SciPost Physics}\ }\textbf
  {\bibinfo {volume} {17}},\ \href
  {https://doi.org/10.21468/scipostphys.17.2.056}
  {10.21468/scipostphys.17.2.056} (\bibinfo {year} {2024})\BibitemShut
  {NoStop}%
\bibitem [{\citenamefont {Dittmann}\ and\ \citenamefont
  {Miller}(2020)}]{Dittmann:2019sbm}%
  \BibitemOpen
  \bibfield  {author} {\bibinfo {author} {\bibfnamefont {A.~J.}\ \bibnamefont
  {Dittmann}}\ and\ \bibinfo {author} {\bibfnamefont {M.~C.}\ \bibnamefont
  {Miller}},\ }\bibfield  {title} {\bibinfo {title} {{Star formation in
  accretion discs and SMBH growth}},\ }\href
  {https://doi.org/10.1093/mnras/staa463} {\bibfield  {journal} {\bibinfo
  {journal} {Mon. Not. Roy. Astron. Soc.}\ }\textbf {\bibinfo {volume} {493}},\
  \bibinfo {pages} {3732} (\bibinfo {year} {2020})},\ \Eprint
  {https://arxiv.org/abs/1911.08685} {arXiv:1911.08685 [astro-ph.HE]}
  \BibitemShut {NoStop}%
\bibitem [{\citenamefont {Yang}\ \emph {et~al.}(2021)\citenamefont {Yang},
  \citenamefont {Wang}, \citenamefont {Fan}, \citenamefont {Barth},
  \citenamefont {Hennawi}, \citenamefont {Nanni}, \citenamefont {Bian},
  \citenamefont {Davies}, \citenamefont {Farina}, \citenamefont {Schindler},
  \citenamefont {Bañados}, \citenamefont {Decarli}, \citenamefont {Eilers},
  \citenamefont {Green}, \citenamefont {Guo}, \citenamefont {Jiang},
  \citenamefont {Li}, \citenamefont {Venemans}, \citenamefont {Walter},
  \citenamefont {Wu},\ and\ \citenamefont {Yue}}]{Yang_2021}%
  \BibitemOpen
  \bibfield  {author} {\bibinfo {author} {\bibfnamefont {J.}~\bibnamefont
  {Yang}}, \bibinfo {author} {\bibfnamefont {F.}~\bibnamefont {Wang}}, \bibinfo
  {author} {\bibfnamefont {X.}~\bibnamefont {Fan}}, \bibinfo {author}
  {\bibfnamefont {A.~J.}\ \bibnamefont {Barth}}, \bibinfo {author}
  {\bibfnamefont {J.~F.}\ \bibnamefont {Hennawi}}, \bibinfo {author}
  {\bibfnamefont {R.}~\bibnamefont {Nanni}}, \bibinfo {author} {\bibfnamefont
  {F.}~\bibnamefont {Bian}}, \bibinfo {author} {\bibfnamefont {F.~B.}\
  \bibnamefont {Davies}}, \bibinfo {author} {\bibfnamefont {E.~P.}\
  \bibnamefont {Farina}}, \bibinfo {author} {\bibfnamefont {J.-T.}\
  \bibnamefont {Schindler}}, \bibinfo {author} {\bibfnamefont {E.}~\bibnamefont
  {Bañados}}, \bibinfo {author} {\bibfnamefont {R.}~\bibnamefont {Decarli}},
  \bibinfo {author} {\bibfnamefont {A.-C.}\ \bibnamefont {Eilers}}, \bibinfo
  {author} {\bibfnamefont {R.}~\bibnamefont {Green}}, \bibinfo {author}
  {\bibfnamefont {H.}~\bibnamefont {Guo}}, \bibinfo {author} {\bibfnamefont
  {L.}~\bibnamefont {Jiang}}, \bibinfo {author} {\bibfnamefont {J.-T.}\
  \bibnamefont {Li}}, \bibinfo {author} {\bibfnamefont {B.}~\bibnamefont
  {Venemans}}, \bibinfo {author} {\bibfnamefont {F.}~\bibnamefont {Walter}},
  \bibinfo {author} {\bibfnamefont {X.-B.}\ \bibnamefont {Wu}},\ and\ \bibinfo
  {author} {\bibfnamefont {M.}~\bibnamefont {Yue}},\ }\bibfield  {title}
  {\bibinfo {title} {Probing early supermassive black hole growth and quasar
  evolution with near-infrared spectroscopy of 37 reionization-era quasars at
  $6.3 \le z \le 7.64$},\ }\href {https://doi.org/10.3847/1538-4357/ac2b32}
  {\bibfield  {journal} {\bibinfo  {journal} {The Astrophysical Journal}\
  }\textbf {\bibinfo {volume} {923}},\ \bibinfo {pages} {262} (\bibinfo {year}
  {2021})}\BibitemShut {NoStop}%
\bibitem [{\citenamefont {{Macuga}}\ \emph {et~al.}(2019)\citenamefont
  {{Macuga}}, \citenamefont {{Martini}}, \citenamefont {{Miller}},
  \citenamefont {{Brodwin}}, \citenamefont {{Hayashi}}, \citenamefont
  {{Kodama}}, \citenamefont {{Koyama}}, \citenamefont {{Overzier}},
  \citenamefont {{Shimakawa}}, \citenamefont {{Tadaki}},\ and\ \citenamefont
  {{Tanaka}}}]{2019ApJ...874...54M}%
  \BibitemOpen
  \bibfield  {author} {\bibinfo {author} {\bibfnamefont {M.}~\bibnamefont
  {{Macuga}}}, \bibinfo {author} {\bibfnamefont {P.}~\bibnamefont {{Martini}}},
  \bibinfo {author} {\bibfnamefont {E.~D.}\ \bibnamefont {{Miller}}}, \bibinfo
  {author} {\bibfnamefont {M.}~\bibnamefont {{Brodwin}}}, \bibinfo {author}
  {\bibfnamefont {M.}~\bibnamefont {{Hayashi}}}, \bibinfo {author}
  {\bibfnamefont {T.}~\bibnamefont {{Kodama}}}, \bibinfo {author}
  {\bibfnamefont {Y.}~\bibnamefont {{Koyama}}}, \bibinfo {author}
  {\bibfnamefont {R.~A.}\ \bibnamefont {{Overzier}}}, \bibinfo {author}
  {\bibfnamefont {R.}~\bibnamefont {{Shimakawa}}}, \bibinfo {author}
  {\bibfnamefont {K.-i.}\ \bibnamefont {{Tadaki}}},\ and\ \bibinfo {author}
  {\bibfnamefont {I.}~\bibnamefont {{Tanaka}}},\ }\bibfield  {title} {\bibinfo
  {title} {{The Fraction of Active Galactic Nuclei in the USS 1558-003
  Protocluster at z = 2.53}},\ }\href
  {https://doi.org/10.3847/1538-4357/ab0746} {\bibfield  {journal} {\bibinfo
  {journal} {\apj}\ }\textbf {\bibinfo {volume} {874}},\ \bibinfo {eid} {54}
  (\bibinfo {year} {2019})},\ \Eprint {https://arxiv.org/abs/1805.06569}
  {arXiv:1805.06569 [astro-ph.GA]} \BibitemShut {NoStop}%
\bibitem [{\citenamefont {Madau}\ and\ \citenamefont
  {Dickinson}(2014)}]{Madau:2014bja}%
  \BibitemOpen
  \bibfield  {author} {\bibinfo {author} {\bibfnamefont {P.}~\bibnamefont
  {Madau}}\ and\ \bibinfo {author} {\bibfnamefont {M.}~\bibnamefont
  {Dickinson}},\ }\bibfield  {title} {\bibinfo {title} {{Cosmic Star Formation
  History}},\ }\href {https://doi.org/10.1146/annurev-astro-081811-125615}
  {\bibfield  {journal} {\bibinfo  {journal} {Ann. Rev. Astron. Astrophys.}\
  }\textbf {\bibinfo {volume} {52}},\ \bibinfo {pages} {415} (\bibinfo {year}
  {2014})},\ \Eprint {https://arxiv.org/abs/1403.0007} {arXiv:1403.0007
  [astro-ph.CO]} \BibitemShut {NoStop}%
\bibitem [{\citenamefont {Yunes}\ \emph
  {et~al.}(2011{\natexlab{b}})\citenamefont {Yunes}, \citenamefont {Kocsis},
  \citenamefont {Loeb},\ and\ \citenamefont {Haiman}}]{Yunes_disk}%
  \BibitemOpen
  \bibfield  {author} {\bibinfo {author} {\bibfnamefont {N.}~\bibnamefont
  {Yunes}}, \bibinfo {author} {\bibfnamefont {B.}~\bibnamefont {Kocsis}},
  \bibinfo {author} {\bibfnamefont {A.}~\bibnamefont {Loeb}},\ and\ \bibinfo
  {author} {\bibfnamefont {Z.}~\bibnamefont {Haiman}},\ }\bibfield  {title}
  {\bibinfo {title} {Imprint of accretion disk-induced migration on
  gravitational waves from extreme mass ratio inspirals},\ }\bibfield
  {journal} {\bibinfo  {journal} {Physical Review Letters}\ }\textbf {\bibinfo
  {volume} {107}},\ \href {https://doi.org/10.1103/physrevlett.107.171103}
  {10.1103/physrevlett.107.171103} (\bibinfo {year}
  {2011}{\natexlab{b}})\BibitemShut {NoStop}%
\bibitem [{\citenamefont {Derdzinski}\ \emph {et~al.}(2021)\citenamefont
  {Derdzinski}, \citenamefont {D'Orazio}, \citenamefont {Duffell},
  \citenamefont {Haiman},\ and\ \citenamefont
  {MacFadyen}}]{Derdzinski:2020wlw}%
  \BibitemOpen
  \bibfield  {author} {\bibinfo {author} {\bibfnamefont {A.}~\bibnamefont
  {Derdzinski}}, \bibinfo {author} {\bibfnamefont {D.}~\bibnamefont
  {D'Orazio}}, \bibinfo {author} {\bibfnamefont {P.}~\bibnamefont {Duffell}},
  \bibinfo {author} {\bibfnamefont {Z.}~\bibnamefont {Haiman}},\ and\ \bibinfo
  {author} {\bibfnamefont {A.}~\bibnamefont {MacFadyen}},\ }\bibfield  {title}
  {\bibinfo {title} {{Evolution of gas disc\textendash{}embedded intermediate
  mass ratio inspirals in the $LISA$ band}},\ }\href
  {https://doi.org/10.1093/mnras/staa3976} {\bibfield  {journal} {\bibinfo
  {journal} {Mon. Not. Roy. Astron. Soc.}\ }\textbf {\bibinfo {volume} {501}},\
  \bibinfo {pages} {3540} (\bibinfo {year} {2021})},\ \Eprint
  {https://arxiv.org/abs/2005.11333} {arXiv:2005.11333 [astro-ph.HE]}
  \BibitemShut {NoStop}%
\bibitem [{\citenamefont {Zwick}\ \emph {et~al.}(2021)\citenamefont {Zwick},
  \citenamefont {Derdzinski}, \citenamefont {Garg}, \citenamefont {Capelo},\
  and\ \citenamefont {Mayer}}]{Zwick:2021dlg}%
  \BibitemOpen
  \bibfield  {author} {\bibinfo {author} {\bibfnamefont {L.}~\bibnamefont
  {Zwick}}, \bibinfo {author} {\bibfnamefont {A.}~\bibnamefont {Derdzinski}},
  \bibinfo {author} {\bibfnamefont {M.}~\bibnamefont {Garg}}, \bibinfo {author}
  {\bibfnamefont {P.~R.}\ \bibnamefont {Capelo}},\ and\ \bibinfo {author}
  {\bibfnamefont {L.}~\bibnamefont {Mayer}},\ }\href@noop {} {\bibinfo {title}
  {{Dirty waveforms: multiband harmonic content of gas-embedded gravitational
  wave sources}}} (\bibinfo {year} {2021}),\ \Eprint
  {https://arxiv.org/abs/2110.09097} {arXiv:2110.09097 [astro-ph.HE]}
  \BibitemShut {NoStop}%
\bibitem [{\citenamefont {Garg}\ \emph {et~al.}(2022)\citenamefont {Garg},
  \citenamefont {Derdzinski}, \citenamefont {Zwick}, \citenamefont {Capelo},\
  and\ \citenamefont {Mayer}}]{Garg:2022nko}%
  \BibitemOpen
  \bibfield  {author} {\bibinfo {author} {\bibfnamefont {M.}~\bibnamefont
  {Garg}}, \bibinfo {author} {\bibfnamefont {A.}~\bibnamefont {Derdzinski}},
  \bibinfo {author} {\bibfnamefont {L.}~\bibnamefont {Zwick}}, \bibinfo
  {author} {\bibfnamefont {P.~R.}\ \bibnamefont {Capelo}},\ and\ \bibinfo
  {author} {\bibfnamefont {L.}~\bibnamefont {Mayer}},\ }\href
  {https://doi.org/10.1093/mnras/stac2711} {\bibinfo {title} {{The imprint of
  gas on gravitational waves from LISA intermediate-mass black hole binaries}}}
  (\bibinfo {year} {2022}),\ \Eprint {https://arxiv.org/abs/2206.05292}
  {arXiv:2206.05292 [astro-ph.GA]} \BibitemShut {NoStop}%
\bibitem [{\citenamefont {{Levin}}(2007)}]{Levin_2007MNRAS.374..515L}%
  \BibitemOpen
  \bibfield  {author} {\bibinfo {author} {\bibfnamefont {Y.}~\bibnamefont
  {{Levin}}},\ }\bibfield  {title} {\bibinfo {title} {{Starbursts near
  supermassive black holes: young stars in the Galactic Centre, and
  gravitational waves in LISA band}},\ }\href
  {https://doi.org/10.1111/j.1365-2966.2006.11155.x} {\bibfield  {journal}
  {\bibinfo  {journal} {\mnras}\ }\textbf {\bibinfo {volume} {374}},\ \bibinfo
  {pages} {515} (\bibinfo {year} {2007})},\ \Eprint
  {https://arxiv.org/abs/astro-ph/0603583} {arXiv:astro-ph/0603583 [astro-ph]}
  \BibitemShut {NoStop}%
\bibitem [{\citenamefont {Tanaka}\ \emph {et~al.}(2002)\citenamefont {Tanaka},
  \citenamefont {Takeuchi},\ and\ \citenamefont {Ward}}]{Tanaka_2002}%
  \BibitemOpen
  \bibfield  {author} {\bibinfo {author} {\bibfnamefont {H.}~\bibnamefont
  {Tanaka}}, \bibinfo {author} {\bibfnamefont {T.}~\bibnamefont {Takeuchi}},\
  and\ \bibinfo {author} {\bibfnamefont {W.~R.}\ \bibnamefont {Ward}},\
  }\bibfield  {title} {\bibinfo {title} {Three-dimensional interaction between
  a planet and an isothermal gaseous disk. i. corotation and lindblad torques
  and planet migration},\ }\href {https://doi.org/10.1086/324713} {\bibfield
  {journal} {\bibinfo  {journal} {The Astrophysical Journal}\ }\textbf
  {\bibinfo {volume} {565}},\ \bibinfo {pages} {1257} (\bibinfo {year}
  {2002})}\BibitemShut {NoStop}%
\bibitem [{\citenamefont {Tanaka}\ and\ \citenamefont
  {Ward}(2004)}]{Tanaka_2004}%
  \BibitemOpen
  \bibfield  {author} {\bibinfo {author} {\bibfnamefont {H.}~\bibnamefont
  {Tanaka}}\ and\ \bibinfo {author} {\bibfnamefont {W.~R.}\ \bibnamefont
  {Ward}},\ }\bibfield  {title} {\bibinfo {title} {Three-dimensional
  interaction between a planet and an isothermal gaseous disk. ii. eccentricity
  waves and bending waves},\ }\href {https://doi.org/10.1086/380992} {\bibfield
   {journal} {\bibinfo  {journal} {The Astrophysical Journal}\ }\textbf
  {\bibinfo {volume} {602}},\ \bibinfo {pages} {388} (\bibinfo {year}
  {2004})}\BibitemShut {NoStop}%
\bibitem [{\citenamefont {Hirata}(2011{\natexlab{b}})}]{Hirata:2010vn}%
  \BibitemOpen
  \bibfield  {author} {\bibinfo {author} {\bibfnamefont {C.~M.}\ \bibnamefont
  {Hirata}},\ }\bibfield  {title} {\bibinfo {title} {{Lindblad resonance
  torques in relativistic discs: I. Basic equations}},\ }\href
  {https://doi.org/10.1111/j.1365-2966.2011.18617.x} {\bibfield  {journal}
  {\bibinfo  {journal} {Mon. Not. Roy. Astron. Soc.}\ }\textbf {\bibinfo
  {volume} {414}},\ \bibinfo {pages} {3198} (\bibinfo {year}
  {2011}{\natexlab{b}})},\ \Eprint {https://arxiv.org/abs/1010.0758}
  {arXiv:1010.0758 [astro-ph.HE]} \BibitemShut {NoStop}%
\bibitem [{Note1()}]{Note1}%
  \BibitemOpen
  \bibinfo {note} {For a short but more detailed description of this model, we
  refer the reader to\cite {Shakura_1973A&A....24..337S}}\BibitemShut {NoStop}%
\bibitem [{\citenamefont {Dhani}\ \emph {et~al.}(2024)\citenamefont {Dhani},
  \citenamefont {V\"olkel}, \citenamefont {Buonanno}, \citenamefont {Estelles},
  \citenamefont {Gair}, \citenamefont {Pfeiffer}, \citenamefont {Pompili},\
  and\ \citenamefont {Toubiana}}]{Dhani:2024jja}%
  \BibitemOpen
  \bibfield  {author} {\bibinfo {author} {\bibfnamefont {A.}~\bibnamefont
  {Dhani}}, \bibinfo {author} {\bibfnamefont {S.}~\bibnamefont {V\"olkel}},
  \bibinfo {author} {\bibfnamefont {A.}~\bibnamefont {Buonanno}}, \bibinfo
  {author} {\bibfnamefont {H.}~\bibnamefont {Estelles}}, \bibinfo {author}
  {\bibfnamefont {J.}~\bibnamefont {Gair}}, \bibinfo {author} {\bibfnamefont
  {H.~P.}\ \bibnamefont {Pfeiffer}}, \bibinfo {author} {\bibfnamefont
  {L.}~\bibnamefont {Pompili}},\ and\ \bibinfo {author} {\bibfnamefont
  {A.}~\bibnamefont {Toubiana}},\ }\href@noop {} {\bibinfo {title} {{Systematic
  Biases in Estimating the Properties of Black Holes Due to Inaccurate
  Gravitational-Wave Models}}} (\bibinfo {year} {2024}),\ \Eprint
  {https://arxiv.org/abs/2404.05811} {arXiv:2404.05811 [gr-qc]} \BibitemShut
  {NoStop}%
\bibitem [{\citenamefont {Kass}\ and\ \citenamefont
  {Raftery}(1995)}]{doi:10.1080/01621459.1995.10476572}%
  \BibitemOpen
  \bibfield  {author} {\bibinfo {author} {\bibfnamefont {R.~E.}\ \bibnamefont
  {Kass}}\ and\ \bibinfo {author} {\bibfnamefont {A.~E.}\ \bibnamefont
  {Raftery}},\ }\bibfield  {title} {\bibinfo {title} {Bayes factors},\ }\href
  {https://doi.org/10.1080/01621459.1995.10476572} {\bibfield  {journal}
  {\bibinfo  {journal} {Journal of the American Statistical Association}\
  }\textbf {\bibinfo {volume} {90}},\ \bibinfo {pages} {773} (\bibinfo {year}
  {1995})}\BibitemShut {NoStop}%
\bibitem [{\citenamefont {Peccei}\ and\ \citenamefont
  {Quinn}(1977)}]{PhysRevLett.38.1440}%
  \BibitemOpen
  \bibfield  {author} {\bibinfo {author} {\bibfnamefont {R.~D.}\ \bibnamefont
  {Peccei}}\ and\ \bibinfo {author} {\bibfnamefont {H.~R.}\ \bibnamefont
  {Quinn}},\ }\bibfield  {title} {\bibinfo {title} {$\mathrm{CP}$ conservation
  in the presence of pseudoparticles},\ }\href
  {https://doi.org/10.1103/PhysRevLett.38.1440} {\bibfield  {journal} {\bibinfo
   {journal} {Phys. Rev. Lett.}\ }\textbf {\bibinfo {volume} {38}},\ \bibinfo
  {pages} {1440} (\bibinfo {year} {1977})}\BibitemShut {NoStop}%
\bibitem [{\citenamefont {Weinberg}(1978)}]{PhysRevLett.40.223}%
  \BibitemOpen
  \bibfield  {author} {\bibinfo {author} {\bibfnamefont {S.}~\bibnamefont
  {Weinberg}},\ }\bibfield  {title} {\bibinfo {title} {A new light boson?},\
  }\href {https://doi.org/10.1103/PhysRevLett.40.223} {\bibfield  {journal}
  {\bibinfo  {journal} {Phys. Rev. Lett.}\ }\textbf {\bibinfo {volume} {40}},\
  \bibinfo {pages} {223} (\bibinfo {year} {1978})}\BibitemShut {NoStop}%
\bibitem [{\citenamefont {Arvanitaki}\ \emph {et~al.}(2010)\citenamefont
  {Arvanitaki}, \citenamefont {Dimopoulos}, \citenamefont {Dubovsky},
  \citenamefont {Kaloper},\ and\ \citenamefont
  {March-Russell}}]{Arvanitaki_2010}%
  \BibitemOpen
  \bibfield  {author} {\bibinfo {author} {\bibfnamefont {A.}~\bibnamefont
  {Arvanitaki}}, \bibinfo {author} {\bibfnamefont {S.}~\bibnamefont
  {Dimopoulos}}, \bibinfo {author} {\bibfnamefont {S.}~\bibnamefont
  {Dubovsky}}, \bibinfo {author} {\bibfnamefont {N.}~\bibnamefont {Kaloper}},\
  and\ \bibinfo {author} {\bibfnamefont {J.}~\bibnamefont {March-Russell}},\
  }\bibfield  {title} {\bibinfo {title} {String axiverse},\ }\bibfield
  {journal} {\bibinfo  {journal} {Physical Review D}\ }\textbf {\bibinfo
  {volume} {81}},\ \href {https://doi.org/10.1103/physrevd.81.123530}
  {10.1103/physrevd.81.123530} (\bibinfo {year} {2010})\BibitemShut {NoStop}%
\bibitem [{\citenamefont {Fabbrichesi}\ \emph {et~al.}(2020)\citenamefont
  {Fabbrichesi}, \citenamefont {Gabrielli},\ and\ \citenamefont
  {Lanfranchi}}]{Fabbrichesi:2020wbt}%
  \BibitemOpen
  \bibfield  {author} {\bibinfo {author} {\bibfnamefont {M.}~\bibnamefont
  {Fabbrichesi}}, \bibinfo {author} {\bibfnamefont {E.}~\bibnamefont
  {Gabrielli}},\ and\ \bibinfo {author} {\bibfnamefont {G.}~\bibnamefont
  {Lanfranchi}},\ }\href {https://doi.org/10.1007/978-3-030-62519-1} {\bibinfo
  {title} {{The Dark Photon}}} (\bibinfo {year} {2020}),\ \Eprint
  {https://arxiv.org/abs/2005.01515} {arXiv:2005.01515 [hep-ph]} \BibitemShut
  {NoStop}%
\bibitem [{\citenamefont {Ferreira}(2021)}]{Ferreira:2020fam}%
  \BibitemOpen
  \bibfield  {author} {\bibinfo {author} {\bibfnamefont {E.~G.~M.}\
  \bibnamefont {Ferreira}},\ }\bibfield  {title} {\bibinfo {title}
  {{Ultra-light dark matter}},\ }\href
  {https://doi.org/10.1007/s00159-021-00135-6} {\bibfield  {journal} {\bibinfo
  {journal} {Astron. Astrophys. Rev.}\ }\textbf {\bibinfo {volume} {29}},\
  \bibinfo {pages} {7} (\bibinfo {year} {2021})},\ \Eprint
  {https://arxiv.org/abs/2005.03254} {arXiv:2005.03254 [astro-ph.CO]}
  \BibitemShut {NoStop}%
\bibitem [{\citenamefont {Hui}(2021)}]{Hui:2021tkt}%
  \BibitemOpen
  \bibfield  {author} {\bibinfo {author} {\bibfnamefont {L.}~\bibnamefont
  {Hui}},\ }\bibfield  {title} {\bibinfo {title} {{Wave Dark Matter}},\ }\href
  {https://doi.org/10.1146/annurev-astro-120920-010024} {\bibfield  {journal}
  {\bibinfo  {journal} {Ann. Rev. Astron. Astrophys.}\ }\textbf {\bibinfo
  {volume} {59}},\ \bibinfo {pages} {247} (\bibinfo {year} {2021})},\ \Eprint
  {https://arxiv.org/abs/2101.11735} {arXiv:2101.11735 [astro-ph.CO]}
  \BibitemShut {NoStop}%
\bibitem [{\citenamefont {Antypas}\ \emph {et~al.}(2022)\citenamefont {Antypas}
  \emph {et~al.}}]{Antypas:2022asj}%
  \BibitemOpen
  \bibfield  {author} {\bibinfo {author} {\bibfnamefont {D.}~\bibnamefont
  {Antypas}} \emph {et~al.},\ }\href@noop {} {\bibinfo {title} {{New Horizons:
  Scalar and Vector Ultralight Dark Matter}}} (\bibinfo {year} {2022}),\
  \Eprint {https://arxiv.org/abs/2203.14915} {arXiv:2203.14915 [hep-ex]}
  \BibitemShut {NoStop}%
\bibitem [{\citenamefont {Seidel}\ and\ \citenamefont
  {Suen}(1994)}]{Seidel:1993zk}%
  \BibitemOpen
  \bibfield  {author} {\bibinfo {author} {\bibfnamefont {E.}~\bibnamefont
  {Seidel}}\ and\ \bibinfo {author} {\bibfnamefont {W.-M.}\ \bibnamefont
  {Suen}},\ }\bibfield  {title} {\bibinfo {title} {{Formation of solitonic
  stars through gravitational cooling}},\ }\href
  {https://doi.org/10.1103/PhysRevLett.72.2516} {\bibfield  {journal} {\bibinfo
   {journal} {Phys. Rev. Lett.}\ }\textbf {\bibinfo {volume} {72}},\ \bibinfo
  {pages} {2516} (\bibinfo {year} {1994})},\ \Eprint
  {https://arxiv.org/abs/gr-qc/9309015} {arXiv:gr-qc/9309015} \BibitemShut
  {NoStop}%
\bibitem [{\citenamefont {Liebling}\ and\ \citenamefont
  {Palenzuela}(2012)}]{Liebling:2012fv}%
  \BibitemOpen
  \bibfield  {author} {\bibinfo {author} {\bibfnamefont {S.~L.}\ \bibnamefont
  {Liebling}}\ and\ \bibinfo {author} {\bibfnamefont {C.}~\bibnamefont
  {Palenzuela}},\ }\bibfield  {title} {\bibinfo {title} {{Dynamical Boson
  Stars}},\ }\href {https://doi.org/10.12942/lrr-2012-6} {\bibfield  {journal}
  {\bibinfo  {journal} {Living Rev. Rel.}\ }\textbf {\bibinfo {volume} {15}},\
  \bibinfo {pages} {6} (\bibinfo {year} {2012})},\ \Eprint
  {https://arxiv.org/abs/1202.5809} {arXiv:1202.5809 [gr-qc]} \BibitemShut
  {NoStop}%
\bibitem [{\citenamefont {Schive}\ \emph
  {et~al.}(2014{\natexlab{a}})\citenamefont {Schive}, \citenamefont {Liao},
  \citenamefont {Woo}, \citenamefont {Wong}, \citenamefont {Chiueh},
  \citenamefont {Broadhurst},\ and\ \citenamefont {Hwang}}]{Schive:2014hza}%
  \BibitemOpen
  \bibfield  {author} {\bibinfo {author} {\bibfnamefont {H.-Y.}\ \bibnamefont
  {Schive}}, \bibinfo {author} {\bibfnamefont {M.-H.}\ \bibnamefont {Liao}},
  \bibinfo {author} {\bibfnamefont {T.-P.}\ \bibnamefont {Woo}}, \bibinfo
  {author} {\bibfnamefont {S.-K.}\ \bibnamefont {Wong}}, \bibinfo {author}
  {\bibfnamefont {T.}~\bibnamefont {Chiueh}}, \bibinfo {author} {\bibfnamefont
  {T.}~\bibnamefont {Broadhurst}},\ and\ \bibinfo {author} {\bibfnamefont
  {W.~Y.~P.}\ \bibnamefont {Hwang}},\ }\bibfield  {title} {\bibinfo {title}
  {{Understanding the Core-Halo Relation of Quantum Wave Dark Matter from 3D
  Simulations}},\ }\href {https://doi.org/10.1103/PhysRevLett.113.261302}
  {\bibfield  {journal} {\bibinfo  {journal} {Phys. Rev. Lett.}\ }\textbf
  {\bibinfo {volume} {113}},\ \bibinfo {pages} {261302} (\bibinfo {year}
  {2014}{\natexlab{a}})},\ \Eprint {https://arxiv.org/abs/1407.7762}
  {arXiv:1407.7762 [astro-ph.GA]} \BibitemShut {NoStop}%
\bibitem [{\citenamefont {Schive}\ \emph
  {et~al.}(2014{\natexlab{b}})\citenamefont {Schive}, \citenamefont {Chiueh},\
  and\ \citenamefont {Broadhurst}}]{Schive:2014dra}%
  \BibitemOpen
  \bibfield  {author} {\bibinfo {author} {\bibfnamefont {H.-Y.}\ \bibnamefont
  {Schive}}, \bibinfo {author} {\bibfnamefont {T.}~\bibnamefont {Chiueh}},\
  and\ \bibinfo {author} {\bibfnamefont {T.}~\bibnamefont {Broadhurst}},\
  }\bibfield  {title} {\bibinfo {title} {{Cosmic Structure as the Quantum
  Interference of a Coherent Dark Wave}},\ }\href
  {https://doi.org/10.1038/nphys2996} {\bibfield  {journal} {\bibinfo
  {journal} {Nature Phys.}\ }\textbf {\bibinfo {volume} {10}},\ \bibinfo
  {pages} {496} (\bibinfo {year} {2014}{\natexlab{b}})},\ \Eprint
  {https://arxiv.org/abs/1406.6586} {arXiv:1406.6586 [astro-ph.GA]}
  \BibitemShut {NoStop}%
\bibitem [{\citenamefont {Cardoso}\ \emph {et~al.}(2022)\citenamefont
  {Cardoso}, \citenamefont {Ikeda}, \citenamefont {Vicente},\ and\
  \citenamefont {Zilh\~ao}}]{Cardoso:2022nzc}%
  \BibitemOpen
  \bibfield  {author} {\bibinfo {author} {\bibfnamefont {V.}~\bibnamefont
  {Cardoso}}, \bibinfo {author} {\bibfnamefont {T.}~\bibnamefont {Ikeda}},
  \bibinfo {author} {\bibfnamefont {R.}~\bibnamefont {Vicente}},\ and\ \bibinfo
  {author} {\bibfnamefont {M.}~\bibnamefont {Zilh\~ao}},\ }\bibfield  {title}
  {\bibinfo {title} {{Parasitic black holes: The swallowing of a fuzzy dark
  matter soliton}},\ }\href {https://doi.org/10.1103/PhysRevD.106.L121302}
  {\bibfield  {journal} {\bibinfo  {journal} {Phys. Rev. D}\ }\textbf {\bibinfo
  {volume} {106}},\ \bibinfo {pages} {L121302} (\bibinfo {year} {2022})},\
  \Eprint {https://arxiv.org/abs/2207.09469} {arXiv:2207.09469 [gr-qc]}
  \BibitemShut {NoStop}%
\bibitem [{\citenamefont {Brito}\ \emph {et~al.}(2015)\citenamefont {Brito},
  \citenamefont {Cardoso},\ and\ \citenamefont {Pani}}]{Brito:2015oca}%
  \BibitemOpen
  \bibfield  {author} {\bibinfo {author} {\bibfnamefont {R.}~\bibnamefont
  {Brito}}, \bibinfo {author} {\bibfnamefont {V.}~\bibnamefont {Cardoso}},\
  and\ \bibinfo {author} {\bibfnamefont {P.}~\bibnamefont {Pani}},\ }\bibfield
  {title} {\bibinfo {title} {{Superradiance}: {New Frontiers in Black Hole
  Physics}},\ }\href {https://doi.org/10.1007/978-3-319-19000-6} {\bibfield
  {journal} {\bibinfo  {journal} {Lect. Notes Phys.}\ }\textbf {\bibinfo
  {volume} {906}},\ \bibinfo {pages} {pp.1} (\bibinfo {year} {2015})},\ \Eprint
  {https://arxiv.org/abs/1501.06570} {arXiv:1501.06570 [gr-qc]} \BibitemShut
  {NoStop}%
\bibitem [{\citenamefont {Yoshino}\ and\ \citenamefont
  {Kodama}(2014)}]{Yoshino:2013ofa}%
  \BibitemOpen
  \bibfield  {author} {\bibinfo {author} {\bibfnamefont {H.}~\bibnamefont
  {Yoshino}}\ and\ \bibinfo {author} {\bibfnamefont {H.}~\bibnamefont
  {Kodama}},\ }\bibfield  {title} {\bibinfo {title} {{Gravitational radiation
  from an axion cloud around a black hole: Superradiant phase}},\ }\href
  {https://doi.org/10.1093/ptep/ptu029} {\bibfield  {journal} {\bibinfo
  {journal} {PTEP}\ }\textbf {\bibinfo {volume} {2014}},\ \bibinfo {pages}
  {043E02} (\bibinfo {year} {2014})},\ \Eprint
  {https://arxiv.org/abs/1312.2326} {arXiv:1312.2326 [gr-qc]} \BibitemShut
  {NoStop}%
\bibitem [{\citenamefont {Arvanitaki}\ \emph {et~al.}(2015)\citenamefont
  {Arvanitaki}, \citenamefont {Baryakhtar},\ and\ \citenamefont
  {Huang}}]{Arvanitaki:2014wva}%
  \BibitemOpen
  \bibfield  {author} {\bibinfo {author} {\bibfnamefont {A.}~\bibnamefont
  {Arvanitaki}}, \bibinfo {author} {\bibfnamefont {M.}~\bibnamefont
  {Baryakhtar}},\ and\ \bibinfo {author} {\bibfnamefont {X.}~\bibnamefont
  {Huang}},\ }\bibfield  {title} {\bibinfo {title} {{Discovering the QCD Axion
  with Black Holes and Gravitational Waves}},\ }\href
  {https://doi.org/10.1103/PhysRevD.91.084011} {\bibfield  {journal} {\bibinfo
  {journal} {Phys. Rev. D}\ }\textbf {\bibinfo {volume} {91}},\ \bibinfo
  {pages} {084011} (\bibinfo {year} {2015})},\ \Eprint
  {https://arxiv.org/abs/1411.2263} {arXiv:1411.2263 [hep-ph]} \BibitemShut
  {NoStop}%
\bibitem [{\citenamefont {Brito}\ \emph
  {et~al.}(2017{\natexlab{a}})\citenamefont {Brito}, \citenamefont {Ghosh},
  \citenamefont {Barausse}, \citenamefont {Berti}, \citenamefont {Cardoso},
  \citenamefont {Dvorkin}, \citenamefont {Klein},\ and\ \citenamefont
  {Pani}}]{Brito:2017zvb}%
  \BibitemOpen
  \bibfield  {author} {\bibinfo {author} {\bibfnamefont {R.}~\bibnamefont
  {Brito}}, \bibinfo {author} {\bibfnamefont {S.}~\bibnamefont {Ghosh}},
  \bibinfo {author} {\bibfnamefont {E.}~\bibnamefont {Barausse}}, \bibinfo
  {author} {\bibfnamefont {E.}~\bibnamefont {Berti}}, \bibinfo {author}
  {\bibfnamefont {V.}~\bibnamefont {Cardoso}}, \bibinfo {author} {\bibfnamefont
  {I.}~\bibnamefont {Dvorkin}}, \bibinfo {author} {\bibfnamefont
  {A.}~\bibnamefont {Klein}},\ and\ \bibinfo {author} {\bibfnamefont
  {P.}~\bibnamefont {Pani}},\ }\bibfield  {title} {\bibinfo {title}
  {{Gravitational wave searches for ultralight bosons with LIGO and LISA}},\
  }\href {https://doi.org/10.1103/PhysRevD.96.064050} {\bibfield  {journal}
  {\bibinfo  {journal} {Phys. Rev. D}\ }\textbf {\bibinfo {volume} {96}},\
  \bibinfo {pages} {064050} (\bibinfo {year} {2017}{\natexlab{a}})},\ \Eprint
  {https://arxiv.org/abs/1706.06311} {arXiv:1706.06311 [gr-qc]} \BibitemShut
  {NoStop}%
\bibitem [{\citenamefont {Siemonsen}\ and\ \citenamefont
  {East}(2020)}]{Siemonsen:2019ebd}%
  \BibitemOpen
  \bibfield  {author} {\bibinfo {author} {\bibfnamefont {N.}~\bibnamefont
  {Siemonsen}}\ and\ \bibinfo {author} {\bibfnamefont {W.~E.}\ \bibnamefont
  {East}},\ }\bibfield  {title} {\bibinfo {title} {{Gravitational wave
  signatures of ultralight vector bosons from black hole superradiance}},\
  }\href {https://doi.org/10.1103/PhysRevD.101.024019} {\bibfield  {journal}
  {\bibinfo  {journal} {Phys. Rev. D}\ }\textbf {\bibinfo {volume} {101}},\
  \bibinfo {pages} {024019} (\bibinfo {year} {2020})},\ \Eprint
  {https://arxiv.org/abs/1910.09476} {arXiv:1910.09476 [gr-qc]} \BibitemShut
  {NoStop}%
\bibitem [{\citenamefont {Siemonsen}\ \emph {et~al.}(2023)\citenamefont
  {Siemonsen}, \citenamefont {May},\ and\ \citenamefont
  {East}}]{Siemonsen:2022yyf}%
  \BibitemOpen
  \bibfield  {author} {\bibinfo {author} {\bibfnamefont {N.}~\bibnamefont
  {Siemonsen}}, \bibinfo {author} {\bibfnamefont {T.}~\bibnamefont {May}},\
  and\ \bibinfo {author} {\bibfnamefont {W.~E.}\ \bibnamefont {East}},\
  }\bibfield  {title} {\bibinfo {title} {{Modeling the black hole superradiance
  gravitational waveform}},\ }\href
  {https://doi.org/10.1103/PhysRevD.107.104003} {\bibfield  {journal} {\bibinfo
   {journal} {Phys. Rev. D}\ }\textbf {\bibinfo {volume} {107}},\ \bibinfo
  {pages} {104003} (\bibinfo {year} {2023})},\ \Eprint
  {https://arxiv.org/abs/2211.03845} {arXiv:2211.03845 [gr-qc]} \BibitemShut
  {NoStop}%
\bibitem [{\citenamefont {Zhang}\ and\ \citenamefont
  {Yang}(2020)}]{Zhang_2020}%
  \BibitemOpen
  \bibfield  {author} {\bibinfo {author} {\bibfnamefont {J.}~\bibnamefont
  {Zhang}}\ and\ \bibinfo {author} {\bibfnamefont {H.}~\bibnamefont {Yang}},\
  }\bibfield  {title} {\bibinfo {title} {Dynamic signatures of black hole
  binaries with superradiant clouds},\ }\bibfield  {journal} {\bibinfo
  {journal} {Physical Review D}\ }\textbf {\bibinfo {volume} {101}},\ \href
  {https://doi.org/10.1103/physrevd.101.043020} {10.1103/physrevd.101.043020}
  (\bibinfo {year} {2020})\BibitemShut {NoStop}%
\bibitem [{\citenamefont {Baumann}\ \emph
  {et~al.}(2022{\natexlab{a}})\citenamefont {Baumann}, \citenamefont {Bertone},
  \citenamefont {Stout},\ and\ \citenamefont {Tomaselli}}]{Baumann:2021fkf}%
  \BibitemOpen
  \bibfield  {author} {\bibinfo {author} {\bibfnamefont {D.}~\bibnamefont
  {Baumann}}, \bibinfo {author} {\bibfnamefont {G.}~\bibnamefont {Bertone}},
  \bibinfo {author} {\bibfnamefont {J.}~\bibnamefont {Stout}},\ and\ \bibinfo
  {author} {\bibfnamefont {G.~M.}\ \bibnamefont {Tomaselli}},\ }\bibfield
  {title} {\bibinfo {title} {{Ionization of gravitational atoms}},\ }\href
  {https://doi.org/10.1103/PhysRevD.105.115036} {\bibfield  {journal} {\bibinfo
   {journal} {Phys. Rev. D}\ }\textbf {\bibinfo {volume} {105}},\ \bibinfo
  {pages} {115036} (\bibinfo {year} {2022}{\natexlab{a}})},\ \Eprint
  {https://arxiv.org/abs/2112.14777} {arXiv:2112.14777 [gr-qc]} \BibitemShut
  {NoStop}%
\bibitem [{\citenamefont {Tomaselli}\ \emph {et~al.}(2023)\citenamefont
  {Tomaselli}, \citenamefont {Spieksma},\ and\ \citenamefont
  {Bertone}}]{Tomaselli:2023ysb}%
  \BibitemOpen
  \bibfield  {author} {\bibinfo {author} {\bibfnamefont {G.~M.}\ \bibnamefont
  {Tomaselli}}, \bibinfo {author} {\bibfnamefont {T.~F.~M.}\ \bibnamefont
  {Spieksma}},\ and\ \bibinfo {author} {\bibfnamefont {G.}~\bibnamefont
  {Bertone}},\ }\bibfield  {title} {\bibinfo {title} {{Dynamical friction in
  gravitational atoms}},\ }\href
  {https://doi.org/10.1088/1475-7516/2023/07/070} {\bibfield  {journal}
  {\bibinfo  {journal} {JCAP}\ }\textbf {\bibinfo {volume} {07}},\ \bibinfo
  {pages} {070}},\ \Eprint {https://arxiv.org/abs/2305.15460} {arXiv:2305.15460
  [gr-qc]} \BibitemShut {NoStop}%
\bibitem [{\citenamefont {Baumann}\ \emph {et~al.}(2019)\citenamefont
  {Baumann}, \citenamefont {Chia},\ and\ \citenamefont
  {Porto}}]{Baumann:2018vus}%
  \BibitemOpen
  \bibfield  {author} {\bibinfo {author} {\bibfnamefont {D.}~\bibnamefont
  {Baumann}}, \bibinfo {author} {\bibfnamefont {H.~S.}\ \bibnamefont {Chia}},\
  and\ \bibinfo {author} {\bibfnamefont {R.~A.}\ \bibnamefont {Porto}},\
  }\bibfield  {title} {\bibinfo {title} {{Probing Ultralight Bosons with Binary
  Black Holes}},\ }\href {https://doi.org/10.1103/PhysRevD.99.044001}
  {\bibfield  {journal} {\bibinfo  {journal} {Phys. Rev. D}\ }\textbf {\bibinfo
  {volume} {99}},\ \bibinfo {pages} {044001} (\bibinfo {year} {2019})},\
  \Eprint {https://arxiv.org/abs/1804.03208} {arXiv:1804.03208 [gr-qc]}
  \BibitemShut {NoStop}%
\bibitem [{\citenamefont {Baumann}\ \emph {et~al.}(2020)\citenamefont
  {Baumann}, \citenamefont {Chia}, \citenamefont {Porto},\ and\ \citenamefont
  {Stout}}]{Baumann:2019ztm}%
  \BibitemOpen
  \bibfield  {author} {\bibinfo {author} {\bibfnamefont {D.}~\bibnamefont
  {Baumann}}, \bibinfo {author} {\bibfnamefont {H.~S.}\ \bibnamefont {Chia}},
  \bibinfo {author} {\bibfnamefont {R.~A.}\ \bibnamefont {Porto}},\ and\
  \bibinfo {author} {\bibfnamefont {J.}~\bibnamefont {Stout}},\ }\bibfield
  {title} {\bibinfo {title} {{Gravitational Collider Physics}},\ }\href
  {https://doi.org/10.1103/PhysRevD.101.083019} {\bibfield  {journal} {\bibinfo
   {journal} {Phys. Rev. D}\ }\textbf {\bibinfo {volume} {101}},\ \bibinfo
  {pages} {083019} (\bibinfo {year} {2020})},\ \Eprint
  {https://arxiv.org/abs/1912.04932} {arXiv:1912.04932 [gr-qc]} \BibitemShut
  {NoStop}%
\bibitem [{\citenamefont {Zhang}\ and\ \citenamefont
  {Yang}(2019)}]{Zhang:2018kib}%
  \BibitemOpen
  \bibfield  {author} {\bibinfo {author} {\bibfnamefont {J.}~\bibnamefont
  {Zhang}}\ and\ \bibinfo {author} {\bibfnamefont {H.}~\bibnamefont {Yang}},\
  }\bibfield  {title} {\bibinfo {title} {{Gravitational floating orbits around
  hairy black holes}},\ }\href {https://doi.org/10.1103/PhysRevD.99.064018}
  {\bibfield  {journal} {\bibinfo  {journal} {Phys. Rev. D}\ }\textbf {\bibinfo
  {volume} {99}},\ \bibinfo {pages} {064018} (\bibinfo {year} {2019})},\
  \Eprint {https://arxiv.org/abs/1808.02905} {arXiv:1808.02905 [gr-qc]}
  \BibitemShut {NoStop}%
\bibitem [{\citenamefont {Baumann}\ \emph
  {et~al.}(2022{\natexlab{b}})\citenamefont {Baumann}, \citenamefont {Bertone},
  \citenamefont {Stout},\ and\ \citenamefont {Tomaselli}}]{Baumann:2022pkl}%
  \BibitemOpen
  \bibfield  {author} {\bibinfo {author} {\bibfnamefont {D.}~\bibnamefont
  {Baumann}}, \bibinfo {author} {\bibfnamefont {G.}~\bibnamefont {Bertone}},
  \bibinfo {author} {\bibfnamefont {J.}~\bibnamefont {Stout}},\ and\ \bibinfo
  {author} {\bibfnamefont {G.~M.}\ \bibnamefont {Tomaselli}},\ }\bibfield
  {title} {\bibinfo {title} {{Sharp Signals of Boson Clouds in Black Hole
  Binary Inspirals}},\ }\href {https://doi.org/10.1103/PhysRevLett.128.221102}
  {\bibfield  {journal} {\bibinfo  {journal} {Phys. Rev. Lett.}\ }\textbf
  {\bibinfo {volume} {128}},\ \bibinfo {pages} {221102} (\bibinfo {year}
  {2022}{\natexlab{b}})},\ \Eprint {https://arxiv.org/abs/2206.01212}
  {arXiv:2206.01212 [gr-qc]} \BibitemShut {NoStop}%
\bibitem [{\citenamefont {Tomaselli}\ \emph
  {et~al.}(2024{\natexlab{a}})\citenamefont {Tomaselli}, \citenamefont
  {Spieksma},\ and\ \citenamefont {Bertone}}]{Tomaselli:2024dbw}%
  \BibitemOpen
  \bibfield  {author} {\bibinfo {author} {\bibfnamefont {G.~M.}\ \bibnamefont
  {Tomaselli}}, \bibinfo {author} {\bibfnamefont {T.~F.~M.}\ \bibnamefont
  {Spieksma}},\ and\ \bibinfo {author} {\bibfnamefont {G.}~\bibnamefont
  {Bertone}},\ }\href@noop {} {\bibinfo {title} {{The legacy of boson clouds on
  black hole binaries}}} (\bibinfo {year} {2024}{\natexlab{a}}),\ \Eprint
  {https://arxiv.org/abs/2407.12908} {arXiv:2407.12908 [gr-qc]} \BibitemShut
  {NoStop}%
\bibitem [{\citenamefont {Tomaselli}\ \emph
  {et~al.}(2024{\natexlab{b}})\citenamefont {Tomaselli}, \citenamefont
  {Spieksma},\ and\ \citenamefont {Bertone}}]{Tomaselli:2024bdd}%
  \BibitemOpen
  \bibfield  {author} {\bibinfo {author} {\bibfnamefont {G.~M.}\ \bibnamefont
  {Tomaselli}}, \bibinfo {author} {\bibfnamefont {T.~F.~M.}\ \bibnamefont
  {Spieksma}},\ and\ \bibinfo {author} {\bibfnamefont {G.}~\bibnamefont
  {Bertone}},\ }\href@noop {} {\bibinfo {title} {{The resonant history of
  gravitational atoms in black hole binaries}}} (\bibinfo {year}
  {2024}{\natexlab{b}}),\ \Eprint {https://arxiv.org/abs/2403.03147}
  {arXiv:2403.03147 [gr-qc]} \BibitemShut {NoStop}%
\bibitem [{\citenamefont {Cole}\ \emph {et~al.}(2023)\citenamefont {Cole},
  \citenamefont {Bertone}, \citenamefont {Coogan}, \citenamefont {Gaggero},
  \citenamefont {Karydas}, \citenamefont {Kavanagh}, \citenamefont {Spieksma},\
  and\ \citenamefont {Tomaselli}}]{Cole:2022yzw}%
  \BibitemOpen
  \bibfield  {author} {\bibinfo {author} {\bibfnamefont {P.~S.}\ \bibnamefont
  {Cole}}, \bibinfo {author} {\bibfnamefont {G.}~\bibnamefont {Bertone}},
  \bibinfo {author} {\bibfnamefont {A.}~\bibnamefont {Coogan}}, \bibinfo
  {author} {\bibfnamefont {D.}~\bibnamefont {Gaggero}}, \bibinfo {author}
  {\bibfnamefont {T.}~\bibnamefont {Karydas}}, \bibinfo {author} {\bibfnamefont
  {B.~J.}\ \bibnamefont {Kavanagh}}, \bibinfo {author} {\bibfnamefont
  {T.~F.~M.}\ \bibnamefont {Spieksma}},\ and\ \bibinfo {author} {\bibfnamefont
  {G.~M.}\ \bibnamefont {Tomaselli}},\ }\bibfield  {title} {\bibinfo {title}
  {{Distinguishing environmental effects on binary black hole gravitational
  waveforms}},\ }\href {https://doi.org/10.1038/s41550-023-01990-2} {\bibfield
  {journal} {\bibinfo  {journal} {Nature Astron.}\ }\textbf {\bibinfo {volume}
  {7}},\ \bibinfo {pages} {943} (\bibinfo {year} {2023})},\ \Eprint
  {https://arxiv.org/abs/2211.01362} {arXiv:2211.01362 [gr-qc]} \BibitemShut
  {NoStop}%
\bibitem [{\citenamefont {Duque}\ \emph {et~al.}(2023)\citenamefont {Duque},
  \citenamefont {Macedo}, \citenamefont {Vicente},\ and\ \citenamefont
  {Cardoso}}]{Duque:2023cac}%
  \BibitemOpen
  \bibfield  {author} {\bibinfo {author} {\bibfnamefont {F.}~\bibnamefont
  {Duque}}, \bibinfo {author} {\bibfnamefont {C.~F.~B.}\ \bibnamefont
  {Macedo}}, \bibinfo {author} {\bibfnamefont {R.}~\bibnamefont {Vicente}},\
  and\ \bibinfo {author} {\bibfnamefont {V.}~\bibnamefont {Cardoso}},\
  }\href@noop {} {\bibinfo {title} {{Axion Weak Leaks: extreme mass-ratio
  inspirals in ultra-light dark matter}}} (\bibinfo {year} {2023}),\ \Eprint
  {https://arxiv.org/abs/2312.06767} {arXiv:2312.06767 [gr-qc]} \BibitemShut
  {NoStop}%
\bibitem [{\citenamefont {Clough}(2021)}]{Clough:2021qlv}%
  \BibitemOpen
  \bibfield  {author} {\bibinfo {author} {\bibfnamefont {K.}~\bibnamefont
  {Clough}},\ }\bibfield  {title} {\bibinfo {title} {{Continuity equations for
  general matter: applications in numerical relativity}},\ }\href
  {https://doi.org/10.1088/1361-6382/ac10ee} {\bibfield  {journal} {\bibinfo
  {journal} {Class. Quant. Grav.}\ }\textbf {\bibinfo {volume} {38}},\ \bibinfo
  {pages} {167001} (\bibinfo {year} {2021})},\ \Eprint
  {https://arxiv.org/abs/2104.13420} {arXiv:2104.13420 [gr-qc]} \BibitemShut
  {NoStop}%
\bibitem [{\citenamefont {Dyson}\ \emph {et~al.}(2025)\citenamefont {Dyson},
  \citenamefont {Spieksma}, \citenamefont {Brito}, \citenamefont {van~de
  Meent},\ and\ \citenamefont
  {Dolan}}]{dyson2025environmentaleffectsextrememass}%
  \BibitemOpen
  \bibfield  {author} {\bibinfo {author} {\bibfnamefont {C.}~\bibnamefont
  {Dyson}}, \bibinfo {author} {\bibfnamefont {T.~F.~M.}\ \bibnamefont
  {Spieksma}}, \bibinfo {author} {\bibfnamefont {R.}~\bibnamefont {Brito}},
  \bibinfo {author} {\bibfnamefont {M.}~\bibnamefont {van~de Meent}},\ and\
  \bibinfo {author} {\bibfnamefont {S.}~\bibnamefont {Dolan}},\ }\href
  {https://arxiv.org/abs/2501.09806} {\bibinfo {title} {Environmental effects
  in extreme mass ratio inspirals: perturbations to the environment in kerr}}
  (\bibinfo {year} {2025}),\ \Eprint {https://arxiv.org/abs/2501.09806}
  {arXiv:2501.09806 [gr-qc]} \BibitemShut {NoStop}%
\bibitem [{\citenamefont {Cardoso}\ \emph {et~al.}(2018)\citenamefont
  {Cardoso}, \citenamefont {Dias}, \citenamefont {Hartnett}, \citenamefont
  {Middleton}, \citenamefont {Pani},\ and\ \citenamefont
  {Santos}}]{Cardoso:2018tly}%
  \BibitemOpen
  \bibfield  {author} {\bibinfo {author} {\bibfnamefont {V.}~\bibnamefont
  {Cardoso}}, \bibinfo {author} {\bibfnamefont {O.~J.~C.}\ \bibnamefont
  {Dias}}, \bibinfo {author} {\bibfnamefont {G.~S.}\ \bibnamefont {Hartnett}},
  \bibinfo {author} {\bibfnamefont {M.}~\bibnamefont {Middleton}}, \bibinfo
  {author} {\bibfnamefont {P.}~\bibnamefont {Pani}},\ and\ \bibinfo {author}
  {\bibfnamefont {J.~E.}\ \bibnamefont {Santos}},\ }\bibfield  {title}
  {\bibinfo {title} {{Constraining the mass of dark photons and axion-like
  particles through black-hole superradiance}},\ }\href
  {https://doi.org/10.1088/1475-7516/2018/03/043} {\bibfield  {journal}
  {\bibinfo  {journal} {JCAP}\ }\textbf {\bibinfo {volume} {03}},\ \bibinfo
  {pages} {043}},\ \Eprint {https://arxiv.org/abs/1801.01420} {arXiv:1801.01420
  [gr-qc]} \BibitemShut {NoStop}%
\bibitem [{\citenamefont {Brito}\ \emph
  {et~al.}(2017{\natexlab{b}})\citenamefont {Brito}, \citenamefont {Ghosh},
  \citenamefont {Barausse}, \citenamefont {Berti}, \citenamefont {Cardoso},
  \citenamefont {Dvorkin}, \citenamefont {Klein},\ and\ \citenamefont
  {Pani}}]{Brito:2017wnc}%
  \BibitemOpen
  \bibfield  {author} {\bibinfo {author} {\bibfnamefont {R.}~\bibnamefont
  {Brito}}, \bibinfo {author} {\bibfnamefont {S.}~\bibnamefont {Ghosh}},
  \bibinfo {author} {\bibfnamefont {E.}~\bibnamefont {Barausse}}, \bibinfo
  {author} {\bibfnamefont {E.}~\bibnamefont {Berti}}, \bibinfo {author}
  {\bibfnamefont {V.}~\bibnamefont {Cardoso}}, \bibinfo {author} {\bibfnamefont
  {I.}~\bibnamefont {Dvorkin}}, \bibinfo {author} {\bibfnamefont
  {A.}~\bibnamefont {Klein}},\ and\ \bibinfo {author} {\bibfnamefont
  {P.}~\bibnamefont {Pani}},\ }\bibfield  {title} {\bibinfo {title}
  {{Stochastic and resolvable gravitational waves from ultralight bosons}},\
  }\href {https://doi.org/10.1103/PhysRevLett.119.131101} {\bibfield  {journal}
  {\bibinfo  {journal} {Phys. Rev. Lett.}\ }\textbf {\bibinfo {volume} {119}},\
  \bibinfo {pages} {131101} (\bibinfo {year} {2017}{\natexlab{b}})},\ \Eprint
  {https://arxiv.org/abs/1706.05097} {arXiv:1706.05097 [gr-qc]} \BibitemShut
  {NoStop}%
\bibitem [{\citenamefont {Kejriwal}\ \emph {et~al.}(2024)\citenamefont
  {Kejriwal}, \citenamefont {Speri},\ and\ \citenamefont
  {Chua}}]{Shubham_PhysRevD.110.084060}%
  \BibitemOpen
  \bibfield  {author} {\bibinfo {author} {\bibfnamefont {S.}~\bibnamefont
  {Kejriwal}}, \bibinfo {author} {\bibfnamefont {L.}~\bibnamefont {Speri}},\
  and\ \bibinfo {author} {\bibfnamefont {A.~J.~K.}\ \bibnamefont {Chua}},\
  }\bibfield  {title} {\bibinfo {title} {Impact of correlations on the modeling
  and inference of beyond vacuum--general relativistic effects in
  extreme-mass-ratio inspirals},\ }\href
  {https://doi.org/10.1103/PhysRevD.110.084060} {\bibfield  {journal} {\bibinfo
   {journal} {Phys. Rev. D}\ }\textbf {\bibinfo {volume} {110}},\ \bibinfo
  {pages} {084060} (\bibinfo {year} {2024})}\BibitemShut {NoStop}%
\bibitem [{\citenamefont {Speri}\ and\ \citenamefont
  {Gair}(2021)}]{Speri_2021}%
  \BibitemOpen
  \bibfield  {author} {\bibinfo {author} {\bibfnamefont {L.}~\bibnamefont
  {Speri}}\ and\ \bibinfo {author} {\bibfnamefont {J.~R.}\ \bibnamefont
  {Gair}},\ }\bibfield  {title} {\bibinfo {title} {Assessing the impact of
  transient orbital resonances},\ }\bibfield  {journal} {\bibinfo  {journal}
  {Physical Review D}\ }\textbf {\bibinfo {volume} {103}},\ \href
  {https://doi.org/10.1103/physrevd.103.124032} {10.1103/physrevd.103.124032}
  (\bibinfo {year} {2021})\BibitemShut {NoStop}%
\bibitem [{\citenamefont {Gupta}\ \emph {et~al.}(2022)\citenamefont {Gupta},
  \citenamefont {Speri}, \citenamefont {Bonga}, \citenamefont {Chua},\ and\
  \citenamefont {Tanaka}}]{Gupta_2022}%
  \BibitemOpen
  \bibfield  {author} {\bibinfo {author} {\bibfnamefont {P.}~\bibnamefont
  {Gupta}}, \bibinfo {author} {\bibfnamefont {L.}~\bibnamefont {Speri}},
  \bibinfo {author} {\bibfnamefont {B.}~\bibnamefont {Bonga}}, \bibinfo
  {author} {\bibfnamefont {A.~J.}\ \bibnamefont {Chua}},\ and\ \bibinfo
  {author} {\bibfnamefont {T.}~\bibnamefont {Tanaka}},\ }\bibfield  {title}
  {\bibinfo {title} {Modeling transient resonances in extreme-mass-ratio
  inspirals},\ }\bibfield  {journal} {\bibinfo  {journal} {Physical Review D}\
  }\textbf {\bibinfo {volume} {106}},\ \href
  {https://doi.org/10.1103/physrevd.106.104001} {10.1103/physrevd.106.104001}
  (\bibinfo {year} {2022})\BibitemShut {NoStop}%
\bibitem [{\citenamefont {van~de Meent}(2018)}]{van_de_Meent_2018}%
  \BibitemOpen
  \bibfield  {author} {\bibinfo {author} {\bibfnamefont {M.}~\bibnamefont
  {van~de Meent}},\ }\bibfield  {title} {\bibinfo {title} {Gravitational
  self-force on generic bound geodesics in kerr spacetime},\ }\bibfield
  {journal} {\bibinfo  {journal} {Physical Review D}\ }\textbf {\bibinfo
  {volume} {97}},\ \href {https://doi.org/10.1103/physrevd.97.104033}
  {10.1103/physrevd.97.104033} (\bibinfo {year} {2018})\BibitemShut {NoStop}%
\bibitem [{\citenamefont {Piovano}\ \emph {et~al.}(2021)\citenamefont
  {Piovano}, \citenamefont {Brito}, \citenamefont {Maselli},\ and\
  \citenamefont {Pani}}]{Piovano_2021}%
  \BibitemOpen
  \bibfield  {author} {\bibinfo {author} {\bibfnamefont {G.~A.}\ \bibnamefont
  {Piovano}}, \bibinfo {author} {\bibfnamefont {R.}~\bibnamefont {Brito}},
  \bibinfo {author} {\bibfnamefont {A.}~\bibnamefont {Maselli}},\ and\ \bibinfo
  {author} {\bibfnamefont {P.}~\bibnamefont {Pani}},\ }\bibfield  {title}
  {\bibinfo {title} {Assessing the detectability of the secondary spin in
  extreme mass-ratio inspirals with fully relativistic numerical waveforms},\
  }\bibfield  {journal} {\bibinfo  {journal} {Physical Review D}\ }\textbf
  {\bibinfo {volume} {104}},\ \href
  {https://doi.org/10.1103/physrevd.104.124019} {10.1103/physrevd.104.124019}
  (\bibinfo {year} {2021})\BibitemShut {NoStop}%
\bibitem [{\citenamefont {Drummond}\ and\ \citenamefont
  {Hughes}(2022{\natexlab{a}})}]{lisa1Drummond_2022}%
  \BibitemOpen
  \bibfield  {author} {\bibinfo {author} {\bibfnamefont {L.~V.}\ \bibnamefont
  {Drummond}}\ and\ \bibinfo {author} {\bibfnamefont {S.~A.}\ \bibnamefont
  {Hughes}},\ }\bibfield  {title} {\bibinfo {title} {Precisely computing bound
  orbits of spinning bodies around black holes. i. general framework and
  results for nearly equatorial orbits},\ }\bibfield  {journal} {\bibinfo
  {journal} {Physical Review D}\ }\textbf {\bibinfo {volume} {105}},\ \href
  {https://doi.org/10.1103/physrevd.105.124040} {10.1103/physrevd.105.124040}
  (\bibinfo {year} {2022}{\natexlab{a}})\BibitemShut {NoStop}%
\bibitem [{\citenamefont {Drummond}\ and\ \citenamefont
  {Hughes}(2022{\natexlab{b}})}]{lisa2_Drummond_2022}%
  \BibitemOpen
  \bibfield  {author} {\bibinfo {author} {\bibfnamefont {L.~V.}\ \bibnamefont
  {Drummond}}\ and\ \bibinfo {author} {\bibfnamefont {S.~A.}\ \bibnamefont
  {Hughes}},\ }\bibfield  {title} {\bibinfo {title} {Precisely computing bound
  orbits of spinning bodies around black holes. ii. generic orbits},\
  }\bibfield  {journal} {\bibinfo  {journal} {Physical Review D}\ }\textbf
  {\bibinfo {volume} {105}},\ \href
  {https://doi.org/10.1103/physrevd.105.124041} {10.1103/physrevd.105.124041}
  (\bibinfo {year} {2022}{\natexlab{b}})\BibitemShut {NoStop}%
\bibitem [{\citenamefont {Drummond}\ \emph
  {et~al.}(2023{\natexlab{b}})\citenamefont {Drummond}, \citenamefont {Lynch},
  \citenamefont {Hanselman}, \citenamefont {Becker},\ and\ \citenamefont
  {Hughes}}]{LISA_Scott_drummond2023extrememassratioinspiralwaveforms}%
  \BibitemOpen
  \bibfield  {author} {\bibinfo {author} {\bibfnamefont {L.~V.}\ \bibnamefont
  {Drummond}}, \bibinfo {author} {\bibfnamefont {P.}~\bibnamefont {Lynch}},
  \bibinfo {author} {\bibfnamefont {A.~G.}\ \bibnamefont {Hanselman}}, \bibinfo
  {author} {\bibfnamefont {D.~R.}\ \bibnamefont {Becker}},\ and\ \bibinfo
  {author} {\bibfnamefont {S.~A.}\ \bibnamefont {Hughes}},\ }\href
  {https://arxiv.org/abs/2310.08438} {\bibinfo {title} {Extreme mass-ratio
  inspiral and waveforms for a spinning body into a kerr black hole via
  osculating geodesics and near-identity transformations}} (\bibinfo {year}
  {2023}{\natexlab{b}}),\ \Eprint {https://arxiv.org/abs/2310.08438}
  {arXiv:2310.08438 [gr-qc]} \BibitemShut {NoStop}%
\bibitem [{\citenamefont {Mathews}\ \emph {et~al.}(2022)\citenamefont
  {Mathews}, \citenamefont {Pound},\ and\ \citenamefont
  {Wardell}}]{Mathews_2022}%
  \BibitemOpen
  \bibfield  {author} {\bibinfo {author} {\bibfnamefont {J.}~\bibnamefont
  {Mathews}}, \bibinfo {author} {\bibfnamefont {A.}~\bibnamefont {Pound}},\
  and\ \bibinfo {author} {\bibfnamefont {B.}~\bibnamefont {Wardell}},\
  }\bibfield  {title} {\bibinfo {title} {Self-force calculations with a
  spinning secondary},\ }\bibfield  {journal} {\bibinfo  {journal} {Physical
  Review D}\ }\textbf {\bibinfo {volume} {105}},\ \href
  {https://doi.org/10.1103/physrevd.105.084031} {10.1103/physrevd.105.084031}
  (\bibinfo {year} {2022})\BibitemShut {NoStop}%
\bibitem [{\citenamefont {Piovano}\ \emph {et~al.}(2024)\citenamefont
  {Piovano}, \citenamefont {Pantelidou}, \citenamefont {Uilliam},\ and\
  \citenamefont {Witzany}}]{piovano2024spinningparticlesnearkerr}%
  \BibitemOpen
  \bibfield  {author} {\bibinfo {author} {\bibfnamefont {G.~A.}\ \bibnamefont
  {Piovano}}, \bibinfo {author} {\bibfnamefont {C.}~\bibnamefont {Pantelidou}},
  \bibinfo {author} {\bibfnamefont {J.~M.}\ \bibnamefont {Uilliam}},\ and\
  \bibinfo {author} {\bibfnamefont {V.}~\bibnamefont {Witzany}},\ }\href
  {https://arxiv.org/abs/2410.05769} {\bibinfo {title} {Spinning particles near
  kerr black holes: Orbits and gravitational-wave fluxes through the
  hamilton-jacobi formalism}} (\bibinfo {year} {2024}),\ \Eprint
  {https://arxiv.org/abs/2410.05769} {arXiv:2410.05769 [gr-qc]} \BibitemShut
  {NoStop}%
\end{thebibliography}%
\clearpage
\onecolumngrid
\appendix
\section{MCMC details}
\label{sec:appendix}
In this Appendix, we report additional details regarding our PE studies, such as the priors used in our analysis, the injected parameters, and full posterior distributions for two of the beyond-vacuum GR systems studied.

\begin{table}[htbp]
\caption{Priors used for the MCMC runs presented in this work. Barred values correspond to the injected parameters (Table~\ref{tab:injection}).  For all the results described, we have set $\delta=0.01$. $U$ denotes the uniform distribution.}
\label{tab:priors}
\centering
\begin{tabular}{ccc}
\toprule
Parameter & \qquad & Prior distribution \\	
\midrule
$\ln M$ & \qquad & $U[(1 - \delta) \cdot \ln \bar{M}, \, (1 + \delta) \cdot \ln \bar{M}]$ \\
$\ln \mu$ & \qquad & $U[(1 - \delta) \cdot \ln \bar{\mu}, \, (1 + \delta) \cdot \ln \bar{\mu}]$ \\
$a$ & \qquad & $U[(1 - \delta) \cdot \bar{a}, \, (1 + \delta) \cdot \bar{a}]$ \\
$p_0$ & \qquad & $U[(1 - \delta) \cdot \bar{p_0}, \, (1 + \delta) \cdot \bar{p_0}]$ \\
$d_L$ & \qquad & $U[0.01, \, 20]$ \\
$\cos \theta_S$ & \qquad & $U[-0.99999, \, 0.99999]$ \\
$\phi_S$ & \qquad & $U[0, \, 2 \pi]$ \\
$\cos \theta_K$ & \qquad & $U[-0.99999, \, 0.99999]$ \\
$\phi_K$ & \qquad & $U[0, \, 2 \pi]$ \\
$\Phi_{\phi_0}$ & \qquad & $U[0, \, 2 \pi]$ \\
$A$ & \qquad & $U[-10^{-3}, \, 10^{-3}]$ \\
$n_r$ & \qquad & $U[-10, \, 10]$ \\
$M_c\, [M]$ & \qquad & $U[0.0, \,0.1]$\\
\bottomrule
\end{tabular}
\end{table}

\begin{table}[htbp]
\caption{Injected parameters for the three systems considered in this study. The different column names represent the different environmental effects included in the waveform. The values of $p_0$ and $d_L$ are different between the two "Disk" systems because we fix the SNR $\rho = 50$ and the observation time $T_{\rm obs} = 4 \rm yr$.}
\label{tab:injection}
\centering
\begin{tabular}{ccccc}
\toprule
Parameter & \qquad & Disk, $\mu = 10 M_\odot$ & Disk, $\mu = 50 M_\odot$ & Cloud \\	
\midrule
$M \, [M_\odot]$ & \qquad & $10^6$ & $10^6$ & $4 \times 10^5$\\
$\mu \, [M_\odot]$ & \qquad & $10$ & $50$ & $20$\\
$a$ & \qquad & $0.9$ & $0.9$ & $0.6$\\
$p_0\, [M]$ & \qquad & $10.227577$ & $15.483580$ & $19.936900$\\
$d_L\, [\rm Gpc]$ & \qquad & $1.849068$ & $3.227112$ & $4.594097$\\
$\theta_S$ & \qquad & $0.542088$ & $0.542088$ & $\pi / 4$\\
$\phi_S$ & \qquad & $5.357656$ & $5.357656$ & $\pi / 3$\\
$\theta_K$ & \qquad & $1.734812$ & $1.734812$ & $\pi / 6$\\
$\phi_K$ & \qquad & $3.200417$ & $3.200417$ & $\pi / 5$\\
$\Phi_{\phi_0}$ & \qquad & $3.0$ & $3.0$ & $\pi / 3$\\
$A$ & \qquad & $1.92 \times 10^{-5}$ & $1.92 \times 10^{-5}$ & // \\
$n_r$ & \qquad & $8.0$ & $8.0$ & // \\
$M_c\, [M]$ & \qquad & // & // & $0.05$ \\
\bottomrule
\end{tabular}
\end{table}

\renewcommand{\arraystretch}{1.5}
\begin{table}[htbp]
    \centering
    \caption{Recovered parameters for the three systems considered in this study. The different column names represent the different environmental effects included in the waveform. The values of $p_0$ and $d_L$ are different between the two "Disk" systems because we fix the SNR $\rho = 50$ and the observation time $T_{\rm obs} = 4 \rm yr$. The quoted values correspond to medians and $95\%$ credible intervals. 
    }
    \label{tab:recovery}
    \begin{tabular}{ccccc}
        \toprule
        Parameter & & $\mu = 10\, M_\odot$ & $\mu = 50\, M_\odot$ & Scalar cloud Template \\	
        \midrule
        $M \, [M_\odot]$ & & $999990^{+130}_{-170}$ & $1000020^{+300}_{-320}$ & $399990\pm 170$ \\ 
        $\mu \, [M_\odot]$ & & $\left( 100000.5^{+8.5}_{-7.3} \right) \times 10^{-4}$ & $\left( 49999.3^{+9.6}_{-8.9} \right) \times 10^{-3}$ & $\left( 20000.3^{+4.9}_{-5.1} \right) \times 10^{-3}$ \\ 
        $a$ & & $\left( 89999.8^{+1.5}_{-2.1} \right) \times 10^{-5}$ & $\left( 90000.1\pm 2.3 \right) \times 10^{-5}$ & $\left( 5999.9^{+2.0}_{-1.9} \right) \times 10^{-4}$ \\ 
        $p_0 \, [M]$ & & $\left( 102276.5^{+11.8}_{-9.5} \right) \times 10^{-4}$ & $\left( 15483.3^{+3.4}_{-3.2} \right) \times 10^{-3}$ & $\left( 19933.9^{+5.6}_{-5.9} \right) \times 10^{-3}$ \\ 
        $d_L \, \rm[Gpc]$ & & $1.86^{+0.16}_{-0.14}$ & $3.26^{+0.28}_{-0.25}$ & $4.42^{+0.23}_{-0.25}$ \\ 
        $\theta_S$ & & $\left( 542.4^{+7.3}_{-7.2} \right) \times 10^{-3}$ & $0.541^{+0.014}_{-0.015}$ & $0.785^{+0.014}_{-0.013}$ \\ 
        $\phi_S$ & & $5.358^{+0.011}_{-0.012}$ & $5.359^{+0.020}_{-0.021}$ & $1.047\pm 0.013$ \\ 
        $\theta_K$ & & $1.739^{+0.059}_{-0.052}$ & $1.740^{+0.064}_{-0.054}$ & $0.53^{+0.57}_{-0.11}$ \\ 
        $\phi_K$ & & $3.201^{+0.048}_{-0.055}$ & $3.198^{+0.050}_{-0.056}$ & $0.64^{+0.66}_{-0.23}$ \\ 
        $\Phi_{\phi_0}$ & & $3.00^{+0.17}_{-0.16}$ & $3.02\pm 0.54$ & $1.05^{+0.57}_{-0.56}$ \\ 
        $A$ & & $\left( 2.0^{+1.7}_{-1.4} \right) \times 10^{-5}$ & $\left( 19.3^{+3.1}_{-2.6} \right) \times 10^{-6}$ & // \\ 
        $n_r$ & & $7.0^{+2.8}_{-3.8}$ & $7.99^{+0.51}_{-0.58}$ & // \\ 
        $M_c \, [M]$ & & // & // & $\left( 499.9\pm 3.3 \right) \times 10^{-4}$ \\ 
        \bottomrule
    \end{tabular}
\end{table}

\begin{figure*}[htbp]
	\centering
	\includegraphics[width=\textwidth]{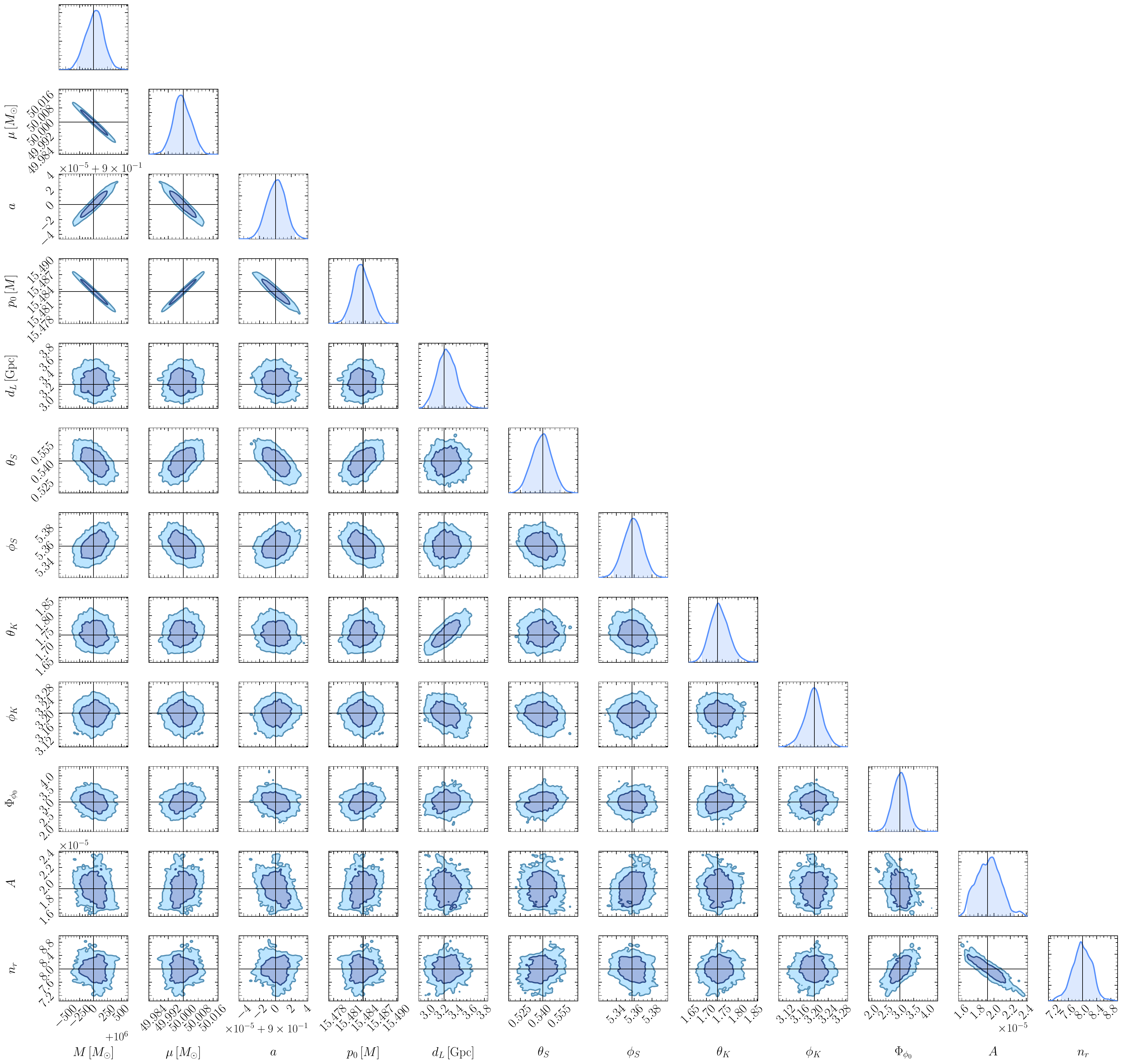}
	\caption{Full posterior for the ``Disk" injection and recovery with $\mu=50\, M_\odot$. Black solid lines represent the injected parameters, while shaded areas in the 2D (1D) histograms represent the $1-$ and $2-\sigma$ regions ($95\%$ credible regions).}
	\label{fig:mcmc_disk50}
\end{figure*}

\begin{figure*}[htbp]
	\centering
	\includegraphics[width=\textwidth]{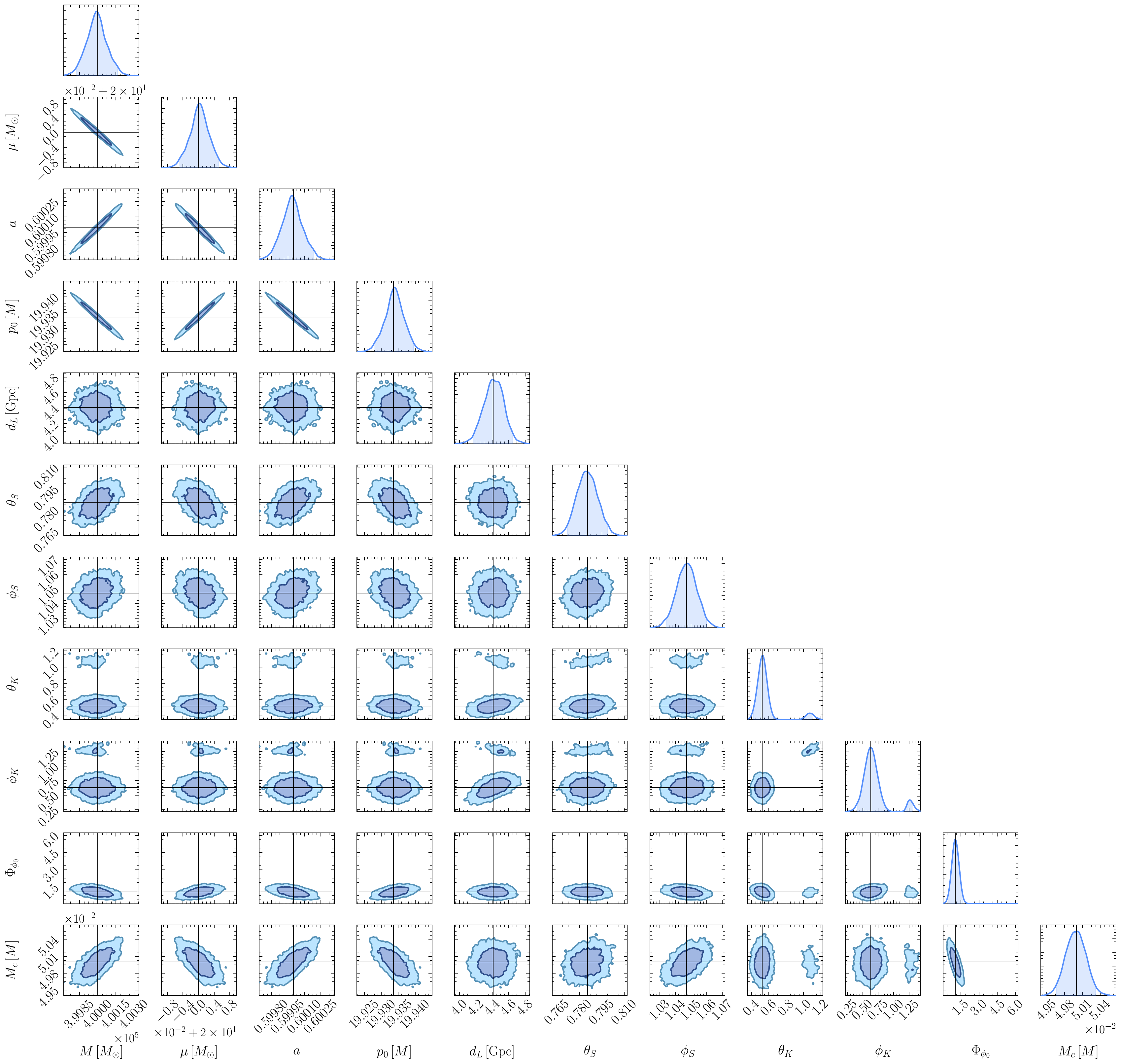}
	\caption{Full posterior for the ``Scalar Cloud" injection and recovery. Black solid lines represent the injected parameters, while shaded areas in the 2D (1D) histograms represent the $1-$ and $2-\sigma$ regions ($95\%$ credible regions).}
	\label{fig:mcmc_axion}
\end{figure*}

\end{document}